
\input phyzzx
\def\ee{\eqno\eq }
\def\abs#1{{\left\vert #1 \right\vert}}
\def\tr{{\rm tr}\,}
\def\bfalpha{{\vec{\alpha}}}
\def\bfdelta{{\bf \delta}}
\def\journal#1&#2(#3){\unskip, {\sl #1}{\bf #2}(19#3)}
\def\andjournal#1&#2(#3){{\sl #1}{\bf #2}(19#3)}
\def\andvol&#1(#2){{\bf #1}(19#2)}

\def\ahikko{$\alpha =p^+$ HIKKO\ }
\def\Nbar{\vbox{\ialign{##\crcr
          \hskip 1.5pt\hrulefill\hskip 1.1pt
          \crcr\noalign{\kern-1pt\vskip0.07cm\nointerlineskip}
          $\hfil\displaystyle{N}\hfil$\crcr}}}
\def\bra#1{\langle #1 \vert}
\def\ket#1{\vert #1 \rangle}
\def\VEV#1{\langle #1 \rangle}
\def\vac{\ket{0}}
\def\bravac{\bra{0}}
\def\QB{Q_{\rm B}}
\def\sfd{\ket{\Psi}}
\def\FP{\rm FP}
\def\bz#1{b_0^{-(#1)}}
\def\bm{b_0^-}
\def\bp{b_0^+}
\def\cp{c_0^+}
\def\cpr{c_0^{+(r)}}
\def\bpr{b_0^{+(r)}}
\def\cm{c_0^-}

\def\arn{\alpha_n^{i(r)}(E)}
\def\arz{\alpha_0^{i(r)}(E)}

\def\bon{b_n^{(1)}}
\def\btn{b_n^{(2)}}
\def\brn{b_n^{(r)}}

\def\con{c_n^{(1)}}
\def\ctn{c_n^{(2)}}
\def\crn{c_n^{(r)}}

\def\col{c_\ell^{(1)}}
\def\G{G_{ij}}
\def\bsub#1{\,{}_{\lower2pt\hbox{$\scriptstyle #1$} \!}}
\def\ah{{\rm a.h.}}
\def\Nrsnm{\Nbar^{rs}_{nm}}
\def\Nrsnz{\Nbar^{rs}_{n0}}
\def\Nrszz{\Nbar^{rs}_{00}}
\def\Nrn{\Nbar^r_n}
\def\O#1{{\cal O}^{(#1)}}
\def\calO{{\cal O}}
\def\calD{{\cal D}}
\def\calB{{\cal B}}

\def\calH{\hat{\cal H}}
\def\calA{{\cal A}}

\def\calM{{\cal M}}
\def\bfp{{\bf p}}
\def\U{{\cal U}}

\def\pder#1{{\partial \over \partial #1}}
\def\eulerder#1{ \left\{ #1,\pder{#1} \right\} }

\def\SUM#1#2{\sum_{\scriptstyle #1
                   \atop \lower2pt\hbox{$\scriptstyle #2$}}}
\def\SUMP#1#2{\mathop{{\sum}'}_{\scriptstyle #1
                   \atop \lower2pt\hbox{$\scriptstyle #2$}}}
\def\rtG{\sqrt G}
\def\DX{{\cal D}\!X}
\def\D#1{{\cal D}#1}
\def\opp{\hat p}
\def\opx{\hat x}
\def\opw{\hat w}
\def\opH{\hat H}
\def\Osp{$OSp(1,1\vert 2)$\ }
\def\INT#1#2{\int_{\scriptstyle #1
                   \atop \lower2pt\hbox{$\scriptstyle #2$}}}
\def\bfn{{\bf n}}
\def\bfm{{\bf m}}
\def\bfx{{\bf x}}

\def\intT{\int_{C}}
\def\tilde{\widetilde}

\def\mymatrix#1#2#3#4{\left( \matrix{#1  &  #2  \cr
                                      #3  &  #4  \cr} \right) }
\def\myvector#1#2{\pmatrix{#1 \cr #2  \cr}}
\def\Odd{$ O(d,d;R)$\ }
\def\Oddz{$ O(d,d;Z)$\ }

\pubnum{YITP/K-961 \cr
IASSNS-HEP-92/3 \cr
MIT-CTP-2058}
\date{January 1992}

\titlepage
\singlespace
\title{\bf
Target Space Duality as a Symmetry of String Field Theory}

\author{Taichiro Kugo}
\address{\null\hskip-2mm
Department of Physics\break
Kyoto University,~Kyoto 606,~Japan}

\andauthor{Barton Zwiebach
\foot{{\rm Present Address: Institute for Advanced Study, Princeton,
NJ 08540. Permanent Address: Center for Theoretical Physics, MIT,
Cambridge, Mass. 02139.}} }
\address{\null\hskip-2mm
Yukawa Institute for Theoretical Physics\break
Kyoto University,~Kyoto 606,~Japan}

\abstract{
Toroidal backgrounds for bosonic strings are used to understand
target space duality as a symmetry of string field theory and to study
explicitly issues in background independence. Our starting point
is the notion that the string field coordinates $X(\sigma)$
and the momenta $P(\sigma)$ are background independent objects
whose field algebra is always the same; backgrounds correspond
to inequivalent representations of this algebra. We propose
classical string field solutions relating any two toroidal
backgrounds and discuss the space where these solutions are defined.
String field theories formulated around dual backgrounds are shown
to be related by a homogeneous field redefinition, and are therefore
equivalent, if and only if their string field coupling constants
are identical. Using this discrete equivalence of backgrounds and
the classical solutions we find discrete symmetry
transformations of the string field leaving the string action
invariant. These symmetries, which are spontaneously broken
for generic backgrounds, are shown to generate the full group
of duality symmetries, and in general are seen to arise from
the string field gauge group.}
\endpage

\singlespace

\hsize=450pt
\hoffset=.5cm
\vsize=620pt
\voffset=-10pt

\REF\KIYA{K.Kikkawa and M. Yamasaki
\journal Phys. Lett. &B149 (84) 357;
\hfill\break
N. Sakai and I. Senda \journal Prog. Theor. Phys. &75 (86) 692.}
\REF\NSW{K. Narain \journal Phys. Lett. &B169 (86) 41;\hfill\break
K. Narain, M. Sarmadi and E. Witten
\journal Nucl. Phys. &B279 (87) 369.}
\REF\NSSW{V. P. Nair, A. Shapere, A. Strominger and F. Wilczek
\journal Nucl. Phys. &B287 (87) 402;\hfill\break
A. Shapere and F. Wilczek
\journal Nucl. Phys. &B320 (89) 669;\hfill\break
B. Sathiapalan \journal Phys. Rev. Lett. &58 (87) 1597.}
\REF\GRV{A. Giveon, E. Rabinovici and G. Veneziano
\journal Nucl. Phys. &B322 (89) 167.}
\REF\DVV{R. Dijkgraaf, E. Verlinde and H. Verlinde
\journal Comm. Math. Phys. &115 (88) 649.}
\REF\DHS{M. Dine, P. Huet and N. Seiberg
\journal Nucl. Phys. &B322 (89) 301.}
\REF\GMR{A. Giveon, N. Malkin and E. Rabinovici
\journal Phys. Lett. &B238 (90) 57.}
\REF\FLST{S. Ferrara, D. Lust, A. Shapere and S. Theisen
\journal Phys. Lett. &B225 (89) 363;\hfill\break
S. Ferrara, D. Lust and S. Theisen
\journal Phys. Lett. &B233 (89) 147.}
\REF\GiPo{A. Giveon and M. Porrati
\journal Nucl. Phys. &B355 (91) 422.}
\REF\hikko{H. Hata, K. Itoh, H. Kunitomo, and K. Ogawa,
\journal Phys. Lett. &172B (86) 186; 195;
\andvol &175B (86) 138; \andjournal Nucl. Phys. &B283 (87) 433;
\andjournal Prog. Theor. Phys. &77 (87) 443.}
\REF\hikkoi{H. Hata, K. Itoh, H. Kunitomo, and K. Ogawa
\journal Phys. Rev. &D34 (86) 2360; \andvol &D35 (87) 1318; 1356}

\REF\Yo{T. Yoneya \journal Phys. Lett. &197B (87) 76.}
\REF\itoh{K. Itoh, Ph. D. Doctoral Thesis, Kyoto University;
\journal Soryusiron Kenkyu &75 (87) 134.}
\REF\MaTa{M. Maeno and H. Takano
\journal Prog. Theor. Phys. &82 (89) 829.}
\REF\HaNa{H. Hata and Y. Nagoshi
\journal Prog. Theor. Phys. &80 (88) 1088.}
\REF\KaKi{M. Kaku and K. Kikkawa
\journal Phys. Rev. &D10 (74) 1110; 1823.}
\REF\SZKKS{M. Saadi and B. Zwiebach
\journal Ann. Phys. &192 (89) 213;
\hfill\break
T. Kugo, H. Kunitomo and K. Suehiro
\journal Phys. Lett. &B226 (89) 48.}
\REF\KuSu{T. Kugo and M. Suehiro
\journal Nucl. Phys. &B337 (90) 434.}
\REF\Zwie{B. Zwiebach
\journal Mod. Phys. Lett. &A5 (90) 2753;\hfill\break
B. Zwiebach \journal Comm. Math. Phys. &136 (91) 83;\hfill\break
H. Sonoda and B. Zwiebach \journal Nucl. Phys. &B331 (90) 592.}
\REF\Se{A. Sen \journal Nucl. Phys. &B345 (90) 551;
\andvol &B347 (90) 270.}
\REF\MuSe{S. Mukherji and A. Sen
\journal Nucl. Phys. &B363 (91) 639.}
\REF\Sei{A. Sen, ``Some Applications of String Field Theory",
TIFR/TH/91-39, September 1991.}
\REF\VaGi{
C. Vafa and P. Ginsparg \journal Nucl. Phys. &B289 (87) 414;
\hfill\break E. Alvarez and M. Osorio
\journal Phys. Rev. &D40 (89) 1150;
\hfill\break A. Giveon, N. Malkin and E. Rabinovici
\journal Phys. Lett. &B220 (89) 551;
\hfill\break D.J.~Gross and I.~Klebanov
\journal Nucl. Phys. &B344 (90) 475.}
\REF\Tseytlindual{A.A.~Tseytlin \journal
Phys. Lett. &B242 (90) 163.}
\REF\GIRO{A. Giveon and M. Rocek, ``Generalized Duality in Curved
String-Backgrounds", IAS preprint, IASSNS-HEP-91/84, December 1991}
\REF\ASENII{A. Sen, ``$O(d) \times O(d)$ Symmetry of the Space of
Cosmological Solutions in String Theory, Scale Factor Duality
and Two-Dimensional Black Holes", TIFR/TH/91-35, 1991;\hfill\break
A. Sen, ``Twisted Black p-Brane Solutions in String Theory",
TIFR/TH/91-37, 1991;\hfill\break
S. F. Hassan and A. Sen, ``Twisting Classical Solutions in
Heterotic String Theory", TIFR/TH/91-40, 1991}
\REF\Venezia{G. Veneziano
\journal Phys. Lett. &B265 (91) 287;\hfill\break
K. A. Meissner and G. Veneziano
\journal Phys. Lett. &B267 (91) 33.}

\chapter{Introduction and Summary}
One of the most interesting properties of string theory
is target space duality invariance [\KIYA--\DVV].
It means that strings cannot
tell the difference between backgrounds that appear to be
quite different as far as particle field theory is concerned.
The simplest example is that of compactification of a single
space-like dimension into a circle of radius $R$. The physics
of strings remains unchanged if the circle becomes of radius
$1/R$. Given such a striking behavior it is only natural
that there has been much discussion about the way in which
duality would manifest itself in the context of a field theory
of strings [\DHS ,\GMR ].
There have been also proposals for low energy
effective actions for string theory having duality invariance
[\FLST , \GiPo ].

We have found that our current understanding of string field
theory (SFT) is sufficient to discuss quite effectively the issue
of target space duality. Since the first priority in the present paper
has been the clarification of the physics of duality, we have used
the form of string field theory that, at the present time,
appears to be easiest to use for our purposes.  This is a
variant of the original closed string field theory formulated
by the Kyoto group [\hikko ,\hikkoi ],
in which the string length parameter, which
was an unphysical parameter, is now taken to be equal to the
$+$ component of the momentum. The string length thus becomes
physical, and gives no problems at the loop level. The price one
pays is that the theory, while gauge covariant, is not fully
Lorentz covariant. This, however, is no serious problem for us
since we will always consider the case where at least two coordinates
$X^0,X^1$ (the first of them time) are not curled up.
The $+$ component of momentum
refers to $X^+ = (X^0+X^1)/\sqrt2$, and all the curled up coordinates
can be treated in the same footing.
The simplicity of using this theory,
refered to as the \ahikko  theory, is that one can find
explicitly exact classical solutions of string field theory.
A large part of the technical tools
necessary for our analysis have been developed in studies
by Yoneya [\Yo ], Itoh [\itoh ], Maeno and Takano [\MaTa ] and
Hata and Nagoshi [\HaNa ].
The light-cone string field theory [\KaKi ]
could have been used for
everything except discussing gauge invariance.
Much of our discussion applies directly to the nonpolynomial
closed string field theory [\SZKKS--\Zwie].
Possibly the hardest point there
is to find exact classical solutions. The methods developed by
A. Sen [\Se--\Sei]
in his study of background independence of closed string
field theory may help give a self-contained explicit discussion
of duality in the context of the nonpolynomial string field theory.

The present work explains in very general grounds how duality
transformations turn out to be discrete symmetries of string
field theory. In making this clear we have had to address
some of the issues of background independence of string field
theory. Toroidal backgrounds are an ideal setting since
the analysis can be done very explicitly. The basic points
we have understood will be summarized now.
\medskip
\noindent
$\underline{\hbox{Aspects of Background Independence}}$.
For a space of backgrounds that corresponds to sigma models
with a fixed number of two dimensional fields we suggest the
idea of universal coordinates and background dependent
representations. For our case, namely toroidal backgrounds,
we think of the coordinate $X^i(\sigma )$ as a universal object,
in fact, for all compactifications we take it to be periodic
with period $2\pi$. The momentum operator $P_i(\sigma )$ is also
universal, since it just represents functional
differentiation $-i\delta /\delta X^i(\sigma)$.
Oscillator expansions simply furnish convenient
background dependent representations of these objects.  Thus we
think of the oscillators $[\alpha, \bar{\alpha}]$ as background
dependent objects and we write this explicitly as
$[\alpha (E), \bar{\alpha}(E)]$. The vacuum state is also
background dependent. Oscillators for different backgrounds
can therefore be related to each other, and these relations
are Bogoliubov type transformations. This viewpoint gives
us a way to relate operators in different conformal field theories
(cft's).
We believe this is a consistent viewpoint since in this context
one can verify explicitly the background independence of
string field vertices defined by overlap conditions. This
is something we would expect to be true on intuitive grounds.
Vertices look background dependent in oscillator form, but this
is only an appearance. Two vertices written for two different
backgrounds are verified to be identical when the oscillators
and vacuum of one background are expressed in terms of those
of the second background.
\medskip
\noindent
$\underline{\hbox{Classical Solutions of String Field Theory}}$.
Using the above ideas operators defined in string field theories
at nearby backgrounds can be related and one can find the
infinitesimal string field classical solution that shifts the
background. While such shift was found earlier by Sen [\Se ]
in the context of the nonpolynomial closed string field theory,
we are able to show explicitly that the new string field theory
after shifting is of the same form as the original one. This
is due to our use of the light-cone type vertex. We then give
an expression for a classical solution corresponding to a finite
shift of background. While this solution has some shortcomings
arising from the singular nature of the light cone vertex
(which we expect would dissapear in the nonpolynomial SFT), it
suggests quite strongly that the classical solution does not
live in the Hilbert space of the original string field theory,
and thus the classical string field cannot be thought in terms
of the component fields of the string field theory. The classical
solution is a superposition (integral) of Fock space states around
different vacua (cft's) and due to the infinite number of oscillators
these vacua are orthogonal making it apparently impossible to
describe the solution in terms of a single Fock space.
This point requires far more investigation, and as a first step
we give an alternative form for the string field classical
solution corresponding to an exactly marginal perturbation.
This solution is found solving the string field theory perturbatively
and applies to any covariant string field theory, in particular to
the nonpolynomial closed string field theory. The solution is given
as an infinite series, each term defined by an off-shell string
amplitude. The convergence of the series is controlled
by the off-shell behavior of the theory, and it may be a tractable
problem.
\medskip
\noindent
$\underline{\hbox{Duality Implies Discrete String Field Symmetries}}$.
The main question we discuss is why and how target space
duality turns out to be a string field symmetry.  A priori
duality relates two different looking theories that turn out
to be physically equivalent. This relation is clearly not in the
form of an invariance. The strategy we follow is quite general and
might have further applications. If two different background field
configurations correspond to the same conformal field theory,
it is clear that the corresponding string field theories must
describe the same physics. We show that this implies the existence
of a homogeneous string field redefinition relating these two string
field actions. Moreover, the two string field theories can also be
related via condensation, or a classical solution that takes us from
one background to the other. If we start from one of the string field
theories the composition of a string field
shift plus the field redefinition brings us back to the
original theory and gives rise to a field transformation corresponding
to an invariance. Thus the string action for
{\it any} fixed background has a discrete symmetry corresponding to
{\it each} possible duality transformation. The symmetries are exact,
but are spontaneously broken unless we are at a background
invariant under duality. This analysis is carried out explicitly
for the complete discrete group of dualities \Oddz.
The discrete duality symmetries turn out to correspond in general
to finite global gauge transformations, as was predicted earlier
on the basis of conformal field theory arguments [\DHS , \GMR ].
We show how to identify them in the context of the string field
theory. We find it quite interesting both that string field theory
is essentially manifestly dual, and that the string field
gauge transformations contain already target space duality.
\medskip
\noindent
$\underline{\hbox{The String Field Coupling Constant Does not Change
Under Duality}}$.
A curious fact
about the discussion of duality from the
viewpoint of first quantization has been the understanding
that duality invariance requires a
shift in the dilaton field [\VaGi ].
We emphasize here that this shift does not involve a
change of the physical string coupling constant.
In string field theory it is manifest that two string
theories formulated around dual backgrounds could not possibly
be equivalent unless the string field coupling constants are
identical. String field theory necessarily uses two
zero modes $(x,q)$ for the compactified space coordinates,
conjugate to the momentum and winding numbers $(p,w)$.
The volumes in which the zero modes $(x,q)$ live
are inversely related, and their product is a constant. This makes
it unnecessary to rescale the coupling constant as we exchange
winding and momentum modes.
We define the string field dilaton $\Phi_s$
to be the field whose condensation changes the string field
coupling constant.
We can then show that the string field condensation that takes
us from one background to the physically equivalent dual background
does not involve the string field dilaton.
There is no actual disagreement with the results of first quantization,
and we explain how this happens by using path integral
methods. The extra factors
necessary for duality, which are introduced in first quantization
by giving the sigma model dilaton $\Phi_\sigma$ a background dependent
expectation value, are seen to arise {\it automatically} when the string
field theory amplitudes are rewritten as first quantized path integrals.
The string field theory is essentially manifestly dual, and
one can choose whether to give a path integral expression
by Fourier transforming either the momentum modes, to obtain the
usual sigma model action for the coordinates $X$, or the winding modes,
to obtain a sigma model action for the dual coordinate $Q$.
If we Fourier transform {\it both} the momentum and winding modes
we obtain an interesting dual-symmetric first quantized
action involving $X$ and $Q$, closely related, though not identical
to the one proposed
by Tseytlin [\Tseytlindual ].
All this implies that the sigma model dilaton $\Phi_\sigma$,
defined as the field coupling to the Euler number of the
surface, and the string field dilaton $\Phi_s$ are not the same.
One must have $\Phi_\sigma = \Phi_s + \ln \det G$, where $G$ is
the background metric for the compactified dimensions.
The necessary shift of the sigma model
dilaton simply reflects the change of the background metric.
Relations between alternative definitions of the dilaton field
in somewhat more general backgrounds have been discussed recently
[\GIRO ].

\bigskip
Let us now give a brief description of the contents of the
various sections of this paper. In Sect.~2 we set up
our conventions for toroidal compactification, discuss the
universal objects and their oscillator expansions, and
give the expressions for the BRST operator and its variation
due to a change in background.  In Sect.~3 we give the
$\alpha = p^+$ HIKKO string field theory for toroidal
backgrounds. We show explicitly the background independence of
string field overlap vertices.
We observe the interesting fact that the string vertex fails
to be dual only due to a phase factor involving a product of
momenta and winding. In Sect.~4  we discuss the relations
between string field theories
formulated around backgrounds related by a discrete duality
transformation $g$ in the complete group of dualities \Oddz.
We give the explicit form of the operator $\U_g$ that defines
the homogeneous redefinition relating dual string field theories.
These operators are seen to give a projective representation
of the group of dualities. The operators satisfy the group
multiplication rules up to extra parity like operators that
turn out to be symmetries of the string field theory.
In Sect.~5 we discuss the condensation of the dilaton extending
the light cone results of Yoneya [\Yo ] to the \ahikko  string field
theory. We then give the classical solutions
for infinitesimal backgrounds shifts, and their form for finite
shifts. Finally we derive the series form of the classical
solution, applicable to the nonpolynomial
closed string field theory. In Sect.~6 we show explicitly how
the asymmetric looking first quantized path integrals
(satisfying duality) arise from manifestly dual symmetric string
field theory in the passage to path integrals.
Sect.~7 deals with the string field symmetry of duality,
its algebra, and its relation with gauge transformations.
In Sect.~8 we offer some comments related to background
independence of string field theory and summarize the main
open questions.

We include three Appendices.  In Appendix A we give some notations
and definitions for the ingredients entering the three string vertex
of the theory.  In Appendix B we show that the \ahikko theory
reproduces the light-cone string field theory amplitudes for processes
involving physical states, thus establishing the correctness of the
theory at the loop level.  Finally in Appendix C we show how to
calculate the overlap of exactly marginal states with the three string
vertex, thus deriving the shift of the BRST operator under an
infinitesimal string condensation. We also include in this Appendix C
a derivation of the SU(2) current algebra at the selfdual radius.
The charges are found by contracting a BRST cohomology class of ghost
number one, representing a suitable global gauge parameter,
against the three string vertex. This shows how global unbroken
symmetries arise in string field theory.

Closed string field theory has been used recently [\ASENII ]
as a tool
to understand cosmological solutions to string theory, scale
factor duality [\Venezia ],
and to generate new classical solutions.

\chapter{String Theory in Toroidal Backgrounds and Universal Objects}

Our objective in this section is to set up the formalism
that will enable us to discuss the string field interpretation
of duality.  We begin by giving the first quantized action
describing bosonic string propagation in a general toroidal
background (we follow the conventions of Ref.\GRV . ):
$$
S = -{1\over 4\pi} \int_0^{2\pi} d\sigma
\int d\tau \left( \sqrt{\gamma} \gamma^{\alpha\beta}
\partial_\alpha X^i \partial_\beta X^j G_{ij}
+ \epsilon^{\alpha\beta} \partial_\alpha X^i
\partial_\beta X^j B_{ij} \right)\eqn\fqi
$$
where we take the world sheet metric to be of signature
$(-,+)$, $\epsilon^{01}=-1$, and we have given just the
part of the action corresponding to the compactified
dimensions. There are $d$ such dimensions, and thus the
indices $i,j$ run from $1$ to $d$. The $d\times d$
constant matrices $G_{ij}$ and $B_{ij}$ represent the background
metric and antisymmetric tensor respectively.
Note that the strings are parameterized by $\sigma \in [0,2\pi ]$.
Our compactification
hypothesis reads
$$X^i \equiv X^i + 2\pi,\eqn\fqii$$
and will not be background dependent.  All of the background
dependence (such as the radii of the tori, etc.) is encoded
in $G$ and $B$; so, in particular, the physical length of the
period in the $i$-th direction is $2\pi R_i$ with
$R_i=\sqrt{G_{ii}}$.  It will be convenient to define the
matrices $E_{\pm} \equiv G \pm B$. The matrix $E_+$ alone
(or, $E_-$ alone) contains the full information about the
background fields, $G$ is the symmetric part of $E_+$
and $B$ the antisymmetric part. Whenever we write $E$
without a subscript, we mean $E_+$:  namely,
$$
E\equiv G+B, \qquad E^t = G-B,
\ee
$$
with superscript $t$ denoting transposed matrix. From the action
one finds that the momentum conjugate
to $X$ is given by
$$2\pi P = G_{ij}{\dot X}^j + B_{ij}X^{j\prime}.\eqn\fqiii$$
The Hamiltonian density $H(\sigma , \tau )$ is given by
$$4\pi H = ({\dot X}^i {\dot X}^j + X^{i\prime}
X^{j\prime}) G_{ij},\eqn\fqiv$$
which, written in terms of proper canonical variables, takes the
form
$$4\pi H = (2\pi)^2 P_i G^{ij} P_j + X^{i\prime} (G-BG^{-1}B)_{ij}
X^{j\prime} + 4\pi X^{i\prime}B_{ik} G^{kj} P_j .\eqn\fqv$$
It is convenient to use matrix notation for this equation, one
writes
$$4\pi H = (X',2\pi P)\, {\cal R} (E) \pmatrix{X'
\cr 2\pi P \cr}\eqn\hmat$$
where the matrix ${\cal R} (E)$ is defined by
$${\cal R}(E) =\pmatrix{G-BG^{-1}B & BG^{-1} \cr
-G^{-1}B & G^{-1}\cr}.\eqn\eqmatrix
$$

Another convenient rewriting of the Hamiltonian is
obtained by defining left and
right components of the momentum ($(-,+)$ respectively):
$$P_{\mp i} = {1\over 2} \left( P_i \pm {1 \over 4\pi}
E_{\mp ij}X^{j\prime} \right) ,\eqn\fqvi$$
one then finds
$$H= 2\pi (P_{-i}G^{ij}P_{-j}
+P_{+i}G^{ij}P_{+j}),\eqn\fqvii$$

The Hamiltonian equations of motion can be solved as usual
to give oscillator expansions:
$$\eqalign{X^i(\sigma ,\tau ) &=
x^i + w^i \sigma + \tau G^{ij}(p_j -B_{jk} w^k)\cr
{}&+ {i\over \sqrt 2 } \sum_{n \not= 0} {1 \over n}
[\alpha^i_n e^{-in(\tau -\sigma )}
+ \bar{\alpha}^i_n  e^{-in(\tau +\sigma )} ]\cr}\eqn\fqviii$$
$$2\pi P_i (\sigma ,\tau ) = p_i
+{1\over \sqrt 2} \sum_{n \not= 0}
[E^t_{ij} \alpha^j_n e^{-in(\tau - \sigma )}
+E_{ij} {\bar \alpha}^j_n e^{-in(\tau + \sigma )}],\eqn\fqix
$$
It should be noted that from the periodicity of $x^i$,
$x^i=x^i+2\pi$, both the momentum
$p_i$ and the winding number $w^i$ take integer eigenvalues.

\section{Universal Objects and Oscillator Expansions}

As is necessary for field theory of strings we look at functionals
of the string coordinates at $\tau =0$.  The coordinates for the
string field are just $X^i(\sigma ) \equiv X^i (\sigma ,\tau =0)$.
Corresponding to these coordinates, we have the operation of
functional differentiation, which is realized by
$P_i (\sigma ) \equiv P_i (\sigma , \tau =0) = -i \delta
/\delta X^i (\sigma )$. We must think of $X^i(\sigma )$ and
$P_i(\sigma )$ as {\it background independent} notions. In
string field
theory the background dependence comes in when in constructing
a kinetic operator and the vertices, one uses the background
field $E=G+B$. Schematically, a string field theory looks like
$$\eqalign{S &= \int dX_1 dX_2 \Psi (X_1) {\cal V}_2 (E,X_r,P_r) \Psi
(X_2)
\cr {}&+ \int (\prod_{r=1}^3 dX_r \Psi (X_r ))
{\cal V}_3 (E,X_r,P_r) + \dots \cr} \eqn\uoi$$
and all the background dependence is concentrated on the vertices
$\cal V$. In a string field theory the background field
$E$ is fixed.
Oscillator expansions are a convenience to study
such actions. As it turns out the expansions of equations
\fqviii\ and \fqix\ , restricted
to $\tau =0$ are extremely convenient. These read:
$$
\eqalign{
X^i(\sigma ) &=
x^i + w^i \sigma
+ {i\over \sqrt 2 } \sum_{n \not= 0} {1 \over n}
[\alpha^i_n (E) e^{in\sigma}
+ \bar{\alpha}^i_n (E)  e^{-in\sigma} ], \cr
2\pi P_i (\sigma) &= p_i
 +{1\over \sqrt 2} \sum_{n \not= 0}
[E^t_{ij} \alpha^j_n (E)e^{in\sigma }
+E_{ij} {\bar \alpha}^j_n (E) e^{-in \sigma}]. \cr
}\eqn\uoii
$$
Here we have introduced the $E$ label to the oscillators
to emphazise that they
{\it depend} on background field $E=G+B$,
since they appear in an explicitly background dependent
expansion of the background independent
objects $X(\sigma ),P(\sigma )$.
The commutation relations arise from
$[X^i (\sigma ) , P_j (\sigma^\prime )]
= i \delta ^i_j \delta (\sigma - \sigma^\prime )$
$$\eqalign{[x^i,p_j] &= i\delta^i_j,\cr
[\alpha^i_m (E), \alpha^j_n
(E)] &= [{\bar \alpha}^i_m (E),{\bar \alpha}^j_n (E)]
= m G^{ij} \delta_{m+n,0} .\cr}\eqn\uoiii$$
As expected, the commutation relations for the oscillators
are background dependent.
We have not introduced a background label for the zero modes
since they will actually turn out to be background independent,
as their commutation relation suggests.

For discussions of target space duality, it is convenient to
introduce another coordinate $Q_i(\sigma )$, dual to $X^i(\sigma )$,
by the relation $Q'_i(\sigma )=2\pi P_i(\sigma )$: that is,
$$
\eqalign{
Q_i(\sigma ) &\equiv  \hbox{constant}
 + \int _0^\sigma  d\sigma ' 2\pi P_i(\sigma ') \cr
 &= q_i + p_i\sigma
 +{i\over \sqrt 2} \sum_{n \not= 0}{1\over n}
[ - E^t_{ij} \alpha^j_n (E)e^{in\sigma }
 + E_{ij} {\bar \alpha}^j_n (E) e^{-in \sigma}]. \cr
}\eqn\eqdefq
$$
The new zero-mode variable $q_i$ is introduced as a
CM coordinate conjugate to $w^i$:
$$
[q_i,w^j] = i\delta _i^j, \qquad
[q_i,x^j] = [q_i,p_j] = 0.
\ee
$$
Since $w^i$ takes integer values, the $q_i$, and hence
$Q_i(\sigma )$ also, must be periodic variables with period $2\pi$
just like $x^i$. The coordinate $Q_i(\sigma )$ is also
background independent.

If we have a string field theory
written for a fixed background $E_0$, it is most convenient to
expand the background independent objects $X(\sigma )$ and
$P(\sigma )$, and the string field, using oscillators
$ \alpha (E_0), \bar \alpha (E_0)$
corresponding to that
particular background.
This leads to a kinetic operator with a diagonal mass operator,
enabling one to read the spectrum easily.
It is important to realize, however, that this is not
required, one may expand a string field theory written around some
background using  oscillators that correspond to a {\it different}
background. This possibility is essential to understand explicitly
the meaning of a shift of background in string field theory.

The universality of $X(\sigma )$ and $P(\sigma )$ allows us to
define a relation between oscillators that correspond to
different backgrounds. We simply equate the two different expansions
of the universal coordinates. It should be noted that we have
precisely the right number of conditions to determine the relations
between the oscillators uniquely. We have two sets of Fourier
coefficients $(\alpha, \bar \alpha )$ and precisely two functions
of $\sigma$, namely, $X$ and $P$. One easily finds that the zero
modes must be identical; this is the reason we did not introduce
a background label for them. For the oscillators we get
the following relations
$$
\eqalign{\alpha_n (E) -{\bar \alpha}_{-n} (E)
&= \alpha_n (E') -{\bar \alpha}_{-n} (E') ,\cr
E^t \alpha_n (E) + E {\bar \alpha}_{-n} (E)
&= E'^t \alpha_n (E') +E'{\bar \alpha}_{-n} (E'),\cr}
\eqn\uoiv
$$
where we have ommitted, for brevity, the indices $i,j...$ both
in the oscillators and in the backgrounds.
In fact, the above relations hold for $n=0$ too. Indeed,
with our normalization convention for the oscillators one
must have
$$
\eqalign{
\alpha _0(E) &\equiv {1\over \sqrt 2}G^{-1}(p-Ew) \cr
\bar \alpha _0(E) &\equiv {1\over \sqrt 2}G^{-1}(p+E^tw). \cr
}\eqn\exi
$$
Inverting the above relations one has
$$
\eqalign{
w &= {1\over \sqrt 2} (\bar{\alpha}_0 - \alpha_0 ),\cr
p &= {1\over \sqrt 2} (E^t \alpha_0 + E\bar{\alpha}_0 ),\cr}
\eqn\wpalpha
$$
and one now sees that \uoiv, for $n=0$ simply says that
$p$ and $w$ are background independent.

Solving equations \uoiv\ for the $\alpha (E') ,\bar\alpha (E')$
in terms of the $\alpha (E) ,\bar\alpha (E)$ oscillators one
finds
$$
\eqalign{
2G'\alpha_n(E') &= (E^t + E')\alpha_n (E)
+(E - E')\bar{\alpha}_{-n} (E) ,\cr
2G'\bar{\alpha}_n(E') &=
(E^t - E'^t)\alpha_{-n} (E)
+(E +E'^t)\bar{\alpha}_n (E).\cr}
\eqn\solv$$

As a first
application of the above results, let us find
the change in the oscillators under a small fluctuation
of the background fields from $E$ to $E'=E + \delta E$.
Defining
$\delta \alpha_n \equiv \alpha_n (E+\delta E) -\alpha_n (E)$
and similarly for $\delta \bar{\alpha}_n$, we find
$$
\eqalign{
2G\delta \alpha_n = - \left( \delta E^t \alpha_n(E) + \delta E\,
\bar{\alpha}_{-n}(E) \right) ,\cr
2G\delta \bar{\alpha}_n = - \left( \delta E^t \alpha_{-n}(E) +
\delta E\, \bar{\alpha}_{n}(E) \right) .\cr}
\eqn\smf
$$
These equations will allow us to relate operators at
nearby values for the background fields.

\section{The BRST Operator}

We will use indices
$\mu ,\nu ,\cdots $
to label the noncompact $(D-d)$ dimensions and  indices $i,j,\cdots $
to label the $(d)$ dimensions that have been curled up into tori
(of course, $D=26$).
The BRST operator on the constant background is given by
$(\eta _{\mu \nu },G_{ij},B_{ij})$:
$$
\QB = -\sum_n\ :\,c_{-n}\left(L_n^X(E) +{1\over 2}L_n^{\FP}
   -\alpha (0)\delta _{n,0}\right)\,: + {\rm a.h.},
\eqn\brst
$$
where the Virasoro operators are:
$$
\eqalign{
L_n^X(E)&= \sum_m {1\over 2}\,:\,\left(\alpha _{n-m}^\mu \eta _{\mu
 \nu }\alpha _m^\nu  +
 \alpha _{n-m}^i(E) G_{ij}\alpha _m^j(E)\right)\,:   \cr
L_n^{\FP} &= \sum_m (n+m)\,:b_{n-m}c_m\,: \cr
}\eqn\eix
$$
The zero-modes $p,w$ appear in the BRST operator
in the form
$$
\QB = -\cp\Big[{1\over 2}(w,p){\cal R}(E) \pmatrix{w \cr p\cr}
 +\cdots\Big] -{1\over 2}\cm \big[-pw +  \cdots\big] +\cdots ,
\eqn\exii$$
where and henceforth
we use the following notation for ghost zero modes:
$$
\eqalign{
&c_0^+\equiv {1\over 2}(c_0+\bar c_0),
\qquad  c_0^-\equiv c_0-\bar c_0, \cr
&b_0^+\equiv b_0+\bar b_0, \qquad
b_0^-\equiv {1\over 2}(b_0-\bar b_0). \cr
}\ee
$$

Let us understand how the BRST operator changes under a shift
of background. It follows from \brst\ that the only change
comes from the contribution of the compact dimensions
to the matter Virasoro generators. Consider the Virasoro
operator $L_0^X$
$$\eqalign{\delta L_0^X &\equiv L_0^X (E+\delta E) - L_0^X (E)\cr
&={1\over 2} \alpha_{-n}(E+\delta E)
(G+\delta G) \alpha_n(E+\delta E)
-{1\over 2}\alpha_{-n}(E) G \alpha_n (E)
\cr
&= -{1\over 2} \alpha_n(E) \delta E\, \bar{\alpha}_n(E) .\cr
}\eqn\varl
$$
A small identical calculation shows that $\delta \bar{L}_0^X
=\delta L_0^X$, and this is quite essential since it implies
that the operator $\Delta N \equiv L_0 -\bar{L}_0$ is unchanged
under a shift
of background, and therefore it is background independent.
Note that the hamiltonian $L_0 + \bar{L}_0$ is not background
independent.
Since we have that the BRST operator commutes with $\Delta N$,
namely $[Q_B(E), \Delta N] =0$, under variation
we must find that $[\delta Q_B , \Delta N ]=0$. This is indeed
what one finds after a small calculation.
$$
\delta \QB = {1\over 2}
\sum_{\ell+n+m=0}(c_\ell+\bar c_{-\ell})
\left(\alpha _n^i(E) \delta E_{ij} \bar\alpha_{-m}^j(E)\right) .
\eqn\varqb
$$
This will be of utility later.

\chapter{String Field Theory in Toroidal Backgrounds}

In this section we set up completely the \ahikko string field
theory in toroidal backgrounds. We then turn to the explicit
analysis of the background independence of the three string
vertex. Much of our discussion below applies to the nonpolynomial
closed string field theory.

\section{The $\alpha =p^+$ HIKKO String Field Theory}

The string field is denoted in our notation by
$$
\ket{\Psi } = c_0^-(\ket{\phi } +c_0^+\ket{\psi }) + ( i\ket{\chi }
 +c_0^+\ket{\eta }),
\eqn\eqi
$$
and it is a Grassmann odd object with ghost number $+3$ (with
respect to the $SL(2;C)$-invariant vacuum).
Actually, in writing an action, only the
field $b_0^-\ket{\Psi}$, which is even and of ghost number
$+2$ appears. The above component string fields
$\phi ,\psi ,\chi ,\eta $ are constructed on the ``down-down''
ghost zero-mode vacuum $\ket{-,-}$ defined by
$$
b_0^+\ket{-,-}=0,\quad
b_0^-\ket{-,-}=0,\quad
\bra{-,-} c_0^-c_0^+ \ket{-,-} = 1,
\eqn\eqiii
$$
Let us henceforth denote $\ket{-,-}$ simply
$\vac $, we then have
$$\bravac c_0^-c_0^+ \vac = 1.
\eqn\eqv
$$
The $\vac$ vacuum is related with the SL(2;C)-invariant vacuum
$\ket{{\bf 1}}$ via
$\ket{{\bf 1}}= -b_{-1}\bar b_{-1}\vac $, and
$\vac = c_1\bar c_1 \ket{{\bf 1}}$.\footnote{*}{There is an
apparent inconsistency concerning this
notation. Take the hermitian conjugate of \eqv, to get
$\bravac c_0^+c_0^- \vac = 1$, in contradiction with \eqv .
An easy way out of this difficulty is
to adopt a coordinate representation for $c_0^+$ and $c_0^-$.
Then we regard $\bravac c_0^-c_0^+ \vac $
as an abbreviation standing for
$\int dc_0^+dc_0^- \, c_0^-c_0^+$. The above difficulty
is avoided by the presence of the integration measure $\int
dc_0^+dc_0^-$
which changes sign under hermitian conjugation since the order of
$dc_0^+$ and $dc_0^-$ is interchanged. The operators $b_0^+$
and $b_0^-$ are then the differential operators
$\partial /\partial c_0^+$
and $\partial /\partial c_0^-$, respectively.
This interpretation is only necessary when dealing with
hermitian conjugation.}
\bigskip
The action of $\alpha =p^+$ HIKKO SFT is given by
$$
S = \bsub{12}\bra{R}\sfd_1 \QB^{(2)} b_0^{-(2)}\sfd_2 +
    {g\over 3}\,\bsub{123}\bra{V}\sfd_1\sfd_2\sfd_3 ,
\eqn\evi
$$
which is invariant under the following gauge transformation:
$$
\eqalign{
\delta (\bm\sfd)& = \QB \bm\ket{\Lambda } + g\ket{\Psi *\Lambda } \cr
  {\rm with}&\ \ \ket{\Psi *\Lambda }_1
    \equiv \bsub{1'23}\bra{V} \ket{R}_{11'}\sfd_2\ket{\Lambda }_3. \cr
}\eqn\evii
$$
The inner-product of states implied by the repeated string labels
$1, 2, \cdots $ also implies integration over the noncompact
zero-modes and summation over the compact ones, as follows:
$$
\int {d^{D-d}p\over (2\pi )^{D-d}}\sum_{p_i}\sum_{w^i}.
$$
\medskip
\noindent
{\it i})\  {\bf 2-point vertex}. This state, in the direct product
of two Hilbert spaces, and denoted as the reflector
$\bsub{12}\bra{R}$, is given by
$$
\bsub{12}\bra{R} = \delta (1,2)\bsub{12}\bravac
\exp{(E_{12})}(c_0^{+(1)}+c_0^{+(2)})
(c_0^{-(1)}+c_0^{-(2)}) e^{-i\pi p_2w_2}{\cal P}_{12},
\eqn\twopt$$
where the exponent $E_{12}$ is defined by
$$E_{12}=
(-)^{n+1}\sum_{n\geq 1}({1\over n}
  \alpha _n^{(1)}\cdot \alpha _n^{(2)}+
   \con\btn -\bon\ctn ) +\ah
\eqn\twopti$$
with
$$ \alpha _n^{(1)}\cdot \alpha _n^{(2)} \equiv
\alpha _n^{\mu (1)}\eta _{\mu \nu } \alpha _n^{\nu (2)} +
\alpha_n^{i(1)}(E) \G \alpha_n^{j(2)}(E) ,\eqn\twoptii$$
and the delta functions and projectors defined by
$$
\eqalign{
&\delta (1,2,\cdots ,n)\equiv
       (2\pi )^{D-d}\delta ^D(\sum_{r=1}^n p^\mu _r)
       \cdot \delta ^d(\sum_{r=1}^n p_{ri})
       \cdot \delta ^d(\sum_{r=1}^n w_r^i),\cr
&{\cal P}_{12\cdots n}\equiv  \prod_{r=1}^n{\cal P}^{(r)},
 \ \ \ {\cal P} \equiv\int _0^{2\pi }{d\theta \over 2\pi }\exp i\theta
 (L-\bar L) \cr}\eqn\exiii
$$
where the two $\delta ^d$'s are Kronecker deltas,
$L=L_0^X(E)+L_0^{\FP}-1$ and $\bar L$ is its
antiholomorphic counterpart.
The hermitian conjugate of $\bra{R}$ coincides with the minus of the
ket reflector $\ket{\tilde R}$:
$$
(\bra{R})^\dagger \equiv  \ket{R} = -\ket{\tilde R},
\eqn\exiv
$$
where the ket reflector $\ket{\tilde R}$ is defined by the property
$$
\bsub{12}\VEV{R\vert \tilde R}_{23}\ket{\Phi }_1 =\ket{\Phi }_3
\eqn\exv
$$
for arbitrary $\ket{\Phi }$.
The string field $\sfd $ satisfies
the following {\it reality condition}:
$$
\bsub{12}\VEV{R\vert \Psi }_1=\bsub{2}\bra{\Psi }
\quad {\rm or}\quad  \bsub{1}\VEV{\Psi \vert \tilde R}_{12} = \sfd_2
\eqn\exvi
$$

The reflector has been written in momentum representation, as
we can see from the fact that $p$ and $w$ appear, instead of
$x$ and $q$. The $p$'s and $w$'s are $c$-numbers and they
are taken to have the value of the momenta of the states that
eventually appear to the right. The vacuum, appearing in
the reflector has nothing to do with momenta.
An alternative notation, preferred by some physicists
would be to let the $p$'s and $w$'s that appear on the
reflector to be operators, and to replace
$$\delta (1,2) {}_{12}\bra{0}\,
\rightarrow \,
\sum_{\bfp_1 ; \bfp_2} \delta(1,2) \bra{\bfp_1,\bfp_2},
\eqn\otherview$$
where the sum extends over all possible values of the
momenta $\bfp = (p_{\mu} , p_i, w^i)$ for each of the strings,
the delta function constraining the sum to the momentum
conserving combinations, and the vacua representing
the corresponding momentum eigenstates.
\medskip
Note the presence in the reflector of the phase factor
$e^{-i\pi p_2 w_2}$. It is the unique possible sign factor of
the form $pw$, as can be checked using momentum conservation.
It is important to note that the phase factor is invariant under
the exchange of $p$ and $w$. This implies that the reflector
treats in the same way the coordinates $X(\sigma )$
and $Q(\sigma )$.
Let us understand why this phase factor is essential in
getting the expected type of connection conditions from
the reflector.  A straightforward calculation
gives the following continuity conditions on the reflector
$$
 \bsub{12}\bra{R}  \left(
\alpha _n^{(1)} + (-)^n\alpha_{-n}^{(2)};\
c_n^{(1)} + (-)^nc_{-n}^{(2)};\
b_n^{(1)} - (-)^nb_{-n}^{(2)} \right) \ = \ 0 ,
\eqn\refleccon
$$
and the same ones for the anti-holomorphic oscillators.
The above hold for all $n$ different from zero. Consider now
the expansion for $X(\sigma )$ in \uoii\ written as
$$X(\sigma ) = x + w \sigma + \tilde{X} (\sigma),\eqn\separx$$
where we explicitly separate out the oscillators. It follows from
\refleccon\ that
$$ \bsub{12}\bra{R}  \left(
\tilde{X}_{(1)} (\sigma ) - \tilde{X}_{(2)} (\pi - \sigma )
\right) = 0.\eqn\connecosc$$
It is clear that the full coordinate must be connected in
a similar fashion. Let us therefore consider the zero modes
$x$ and $w$ for the compactified coordinates.
We must be careful since the zero mode operator $\hat x$ is
not a well defined operator, due to the periodicity condition
on the torus.
Due to the periodicity of $x$, the momentum $p$ takes the integer
eigenvalues $\opp \ket{n} = n\ket{n}$ (here $n$ is a vector of
integers). Rather than trying to
define a coordinate operator, we define a coordinate eigenstate
via
$$
\ket{x} \equiv  \sum_n {e^{-inx}\over \sqrt{(2\pi )^d}}\, \ket{n}.
\eqn\eqxeigen
$$
Then, as desired, the state label $x$ becomes the label of
the point on the torus since
$\ket{x} = \ket{x+2\pi e^{(i)}}$ where
$e^{(i)}$ is a unit vector in the $i$ direction.
The inner product of two coordinate eigenstates is given by
$$
\VEV{x\vert y} = \sum_n{1\over (2\pi )^d}\,e^{in(x-y)} \,
=\sum_m\delta (x-y+2\pi m) \equiv  \bfdelta (x-y),
\eqn\eqperodic
$$
where $\bfdelta$ is a periodic delta function. In order to
understand what type connection the reflector gives we must
evaluate the overlap
$$\bsub{12}\bra{R} \ket{x_1, w_1}_1 \ket{x_2,w_2}_2,$$
where the second label on the kets is the winding eigenvalue.
As far as the zero modes are concerned the above is equals
$$ \sum_{n_1,n_2} \delta^d (n_1 + n_2) \delta^d (w_1 + w_2)
e^{-i\pi n_2 w_2} \,
{e^{-in_1x_1}\over \sqrt{(2\pi )^d}}\,
{e^{-in_2x_2}\over \sqrt{(2\pi )^d}}$$
where the delta functions and the phase factor $e^{-i\pi n_2 w_2}$
came from the reflector. It then follows that
$$\bsub{12}\bra{R} \ket{x_1, w_1}_1 \ket{x_2,w_2}_2
= \bfdelta (x_1 - [x_2 + \pi w_2]) \delta (w_1 + w_2), \ee$$
which means that the vertex ``connects'' zero modes as
$$ x_1 \approx x_2 + \pi w_2 ,\quad w_1 \approx -w_2, \ee$$
where the first one is modulo $2\pi$. Actually the connection
of the windings is a true operator relation when acting on
the reflector, since the winding
operator is well defined. The above implies that
$$x_1 + w_1 \sigma \,\,\approx \,\, x_2 + \pi w_2 - w_2 \sigma
\,\, \approx \,\, x_2 + w_2 (\pi - \sigma ).\eqn\connectzm$$
This fits nicely with equations
\separx\ and \connecosc\ to give the connection condition
$$X_1 (\sigma ) \,\approx \,  X_2 (\pi - \sigma ),  \ee$$
for the full coordinate. This is the expected result, it
shows the relevance of the phase factor. For the dual
coordinate $Q (\sigma)$ one finds a similar result,
namely
$$Q_1 (\sigma ) \, \approx \,  Q_2 (\pi - \sigma ) ,\ee$$
due to the symmetry of the reflector under the exchange
of $p$ and $w$. We also use the $\approx$ symbol for
this coordinate because the zero mode $q$ is not a
well defined operator.
\medskip
\noindent
{\it ii})\ {\bf 3-point vertex}. The three string vertex can be
given in two useful forms:
$$
\eqalign{
\bsub{123}\bra{V} =& \mu ^2_{123}\delta (1,2,3)\bsub{123}\bravac
  (\prod_{r=1}^3 \cpr )\exp(E_{123})  \cr
&\ \ \ \ \times \left(\sum_{r=1}^3 {\bpr\over p^+_r}\right)
 G(\sigma _I) e^{-i\pi (p_3 w_2-p_1 w_1)}{\cal P}_{123},
\cr}
\eqn\vertex
$$
or, equivalently, by
$$
\eqalign{
\bsub{123}\bra{V} =& \mu ^2_{123}\delta (1,2,3)\bsub{123}\bravac
  \exp(F_{123})  \cr
&\ \ \ \ \times [\prod_{r=1}^3(\cpr +{1\over \sqrt 2}W^{(r)}_I)]
  e^{-i\pi (p_3 w_2-p_1 w_1)}{\cal P}_{123},
\cr}
\eqn\exvii
$$
where
$$
\eqalign{
E_{123}&= E_{123}^{\rm ordinary} + E_{123}^{\rm compact}(E) ,
\cr
F_{123}&= F_{123}^{\rm ordinary} + E_{123}^{\rm compact}(E) .
\cr}\eqn\eqexp$$
The expressions for $E,F$ (ordinary) correspond to the noncompact
directions and since they will not be essential in the following
discussion we have relegated them to Appendix A. We give, however,
the expression for $E$(compact):
$$E_{123}^{\rm compact}(E) =
 {1\over 2}\sum_{r,s}\sum_{n,m\geq 0}\bar N^{rs}_{nm}
 \alpha_n^{i(r)}(E) G_{ij}\alpha_m^{j(s)}(E) + \ \ah \ .
\eqn\messy$$
It is interesting to expand this expression out. In what
follows the repeated indices $r,s$ are summed over the
three strings, and the repeated indices $n,m$ are summed
over the {\it positive} integers $1,2,\cdots $.
$$\eqalign{
E_{123}^{\rm compact}(E) &=
{1\over 2}\bar N^{rs}_{nm}
 \alpha_n^{i(r)}(E) G_{ij}\alpha_m^{j(s)}(E) + \ \ah
\cr
{}&\ \ \ +{1\over \sqrt 2}\bar N^{rs}_{n0}
 \left( \alpha_n^{i(r)}(E)(p_{is}-E_{ij}w^j_s)
+ \bar \alpha_n^{i(r)} (E) (p_{is}+E^t_{ij}w^j_s) \right)
\cr
{}&\ \ \ +{1\over 2} \bar{N}_{00}^{rs}
(w_r,p_r) {\cal R} (E) \pmatrix{w_s \cr p_s \cr}
\cr}\eqn\hell$$
where ${\cal R} (E)$ is the matrix introduced before.

The sign factor $e^{-i\pi (p_3 w_2-p_1 w_1)}$ in the above 3-string
vertex, \vertex\ or \exvii, (or its remnant $e^{i\pi p_2 w_2}$ in the
2-point vertex) should be noted. This time, in contrast with the
case of the two-point vertex, the phase factor is not symmetric under
the exchange of $p$ and $w$. It is therefore not dual symmetric.
In fact, this phase factor, which is either one or minus one,
is the only factor that prevents the
string field theory from being completely dual symmetric. It is not
hard to check that the factor cannot be made dual symmetric by
a redefinition of the string field. We therefore expect the
connection conditions on the vertex not to be dual symmetric.
This expression for the sign factor was first given by Maeno
and Takano [\MaTa ]. We refer to this sign factor as a vertex cocycle
factor henceforth.

The following symmetry and Grassmann even-odd properties are
worth remembering:
$$
\eqalign{
\bsub{12}\bra{R}&: \hbox{Grassmann even, symmetric under}
        \ 1 \leftrightarrow 2 \cr
\bsub{123}\bra{V}&:
\hbox{Grassmann odd, anti-symmetric under interchange of 1,2,3}
\cr
\ket{\Psi }&: \hbox{Grassmann odd}  \cr
\ket{\Lambda }&: \hbox{Grassmann even}  \cr
}\eqn\exx
$$

\section{Universality of the Three String Vertex}

The above expression for the 3-string vertex
was obtained by Maeno and
Takano [\MaTa ] starting from the following
naive delta functional expression for the vertex:
$$
\eqalign{
&V[X^{(1)},X^{(2)},X^{(3)}] \ \ \sim  \!\!\!
 \prod_{-\pi \abs{p^+_3}\leq \sigma \leq \pi \abs{p^+_3}}\!\!\!
 \bfdelta \left( \Theta _1X^{(1)}(\sigma _1)
 +\Theta _2X^{(2)}(\sigma _2)-X^{(3)}(\sigma _3) \right),\cr
\noalign{\vskip 5mm}
& \Theta _1(\sigma )\equiv \theta (\pi p^+_1-\abs{\sigma }),\qquad
  \Theta _2(\sigma )\equiv 1-\Theta _1(\sigma ), \cr
& \sigma _1(\sigma )\equiv  {\sigma \over p^+_1},\quad
 \sigma _2(\sigma )\equiv
  {\sigma -\pi p^+_1{\rm sgn}(\sigma )\over p^+_2},\quad
 \sigma _3(\sigma )\equiv
 {\pi \abs{p^+_3}{\rm sgn}(\sigma )-\sigma \over \abs{p^+_3}}.
\cr}\ee
$$
[This overlapping pattern is for the case $p_1^+,p^+_2>0, p^+_3<0$
($p^+_1+p^+_2=\abs{p^+_3}$); other cases are similar.] \
The delta functions for compact coordinates are the periodic ones
defined in  \eqperodic. As expected from this
derivation, they proved that the following Goto-Naka type
connection conditions are satisfied by the above vertex:
$$
\eqalign{
&\bsub{123}\bra{V} \left(
\Theta _1X^{(1)}(\sigma _1)
 +\Theta _2X^{(2)}(\sigma _2)-X^{(3)}(\sigma _3)\right)
\approx  0 \ \ ({\rm mod}\ 2\pi ), \cr
&\bsub{123}\bra{V} \left(
\Theta _1P^{(1)}(\sigma _1)
 +\Theta _2P^{(2)}(\sigma _2)+P^{(3)}(\sigma _3)\right)
=  0 \ . \cr}\eqn\gotonaka$$
The dual coordinate $Q$ (in \eqdefq ), however, does not connect the
way the $X$ coordinate does, one finds
$$
\bsub{123}\bra{V} \left(
\Theta _1Q^{(1)}(\sigma _1)
 +\Theta _2Q^{(2)}(\sigma _2)-Q^{(3)}(\sigma _3)\right)
\, \approx  \bsub{123}\bra{V}
\ \pi(\Theta _1p_2 + \Theta _2p_1)\ \  ({\rm mod}\ 2\pi ).
\eqn\eqGN
$$

We should note that the expressions \vertex\ or \exvii\
for the 3-string vertex $\bsub{123}\bra{V}$ apparently depend on
the background fields $E$, but always satisfy the Goto-Naka
equations \gotonaka\ and \eqGN\ irrespectively of $E$. Note also that
the coordinates $X^{(r)}(\sigma )$ and $Q^{(r)}(\sigma )$
give a complete set of operators in the three string Hilbert space.
Namely there is no operator which commutes with all the
$X^{(r)}(\sigma )$ and $Q^{(r)}(\sigma )$, and hence
the Goto-Naka connection equations \gotonaka\ and \eqGN\
uniquely specify the 3-string vertex up to an overall normalization.
Therefore, despite its appearance, the 3-string vertex \vertex\
or \exvii\
gives in fact a unique object that does not depend on the
background at all. [The coincidence of the normalization will be
checked explicitly.]

This argument proves the universality of the 3-string
vertex. But it is also very illuminating to confirm it directly for
the explicit expression given in \vertex\ or \exvii . The apparently
background dependent part of the vertex is given by
$$
 \bsub{E}\bravac \exp(E_{123}) ,
\eqn\eqVERTEX
$$
where $E_{123}$ here is the
$E_{123}^{\rm compact}(E)$ given above in \hell, which may be
rewritten more concisely as
$$
E_{123} = \ {1\over 2}\Big(\bfalpha ^t, \bfalpha_0^t \Big)
\mymatrix{N}{N_0}{N_0^t}{N_{00}} G
\myvector{\bfalpha }{\bfalpha _0} + \ah\ ,
\eqn\eqEEE
$$
using the following condensed vector- and matrix-notations:
$$
\eqalign{
&\myvector{\bfalpha }{\bfalpha_0} \equiv
\myvector{\arn \ \ (n\geq 1)}{\arz}\,,\quad
\myvector{\bar\bfalpha }{\bar\bfalpha_0} \equiv
\myvector{\bar\arn \ \ (n\geq 1)}{\bar\arz}\,, \cr
\noalign{\vskip 3mm}
&N \equiv  \Bigg[\ \ \Nrsnm\ \ \Bigg],\quad
N_0 \equiv  \Bigg[\ \ \Nrsnz \ \ \Bigg],\quad
N_{00} \equiv  \Bigg[\ \ \Nrszz \ \ \Bigg].\cr
}\ee
$$
Note the suffix $E$ on the vacuum $\bravac$ in
\eqVERTEX\  to emphasize that it is the vacuum of the background
dependent oscillators $\alpha _n(E), \ \bar\alpha _n(E)$.

Now let us show that \eqVERTEX\ is indeed independent of the
background $E$. Under an arbitrary infinitesimal change
of $E$ to $E+\delta E$, the
oscillators $\alpha _n(E), \ \bar\alpha _n(E)$ change, from \smf, by
$$
\eqalign{
\delta \alpha_n &= -{1\over 2}G^{-1} \left( \delta E^t \alpha_n
+ \delta E\,\bar{\alpha}_{-n} \right) ,\cr
\delta \bar{\alpha}_n &= -{1\over 2}G^{-1} \left(
\delta E\, \bar{\alpha}_{n} +
   \delta E^t \alpha_{-n}  \right) .\cr}
\eqn\eqsmf
$$
Here we have omitted the background label $E$ from $\alpha _n(E)$
for brevity. The vacuum corresponding to the changed oscillators
$\alpha '_n\equiv \alpha _n(E+\delta E)=\alpha _n+\delta \alpha _n$
and $\bar\alpha '_n=\bar\alpha _n+\delta \bar\alpha _n$
is also infinitesimally shifted from the original one
$\vac_{E}$:
$$
\vac_{(E+\delta E)} = \vac_{E}
 - \calB \vac_{E}.
\eqn\eqprimevac
$$
$\calB$ is easily found to be given by
$$
\calB = {1\over 2}\bigg(
\bfalpha ^t {\delta E\over n}\,\bar\bfalpha
-\bfalpha ^{\dagger\, T} {\delta E\over n}\,\bar\bfalpha^{\dagger}
\bigg) ,
\eqn\eqcalB
$$
with condensed notation again:
$$
{\delta E\over n} \equiv  \Bigg[ \delta E_{ij}{\delta _{nm}\over n}
\delta ^{rs}
\Bigg] \ .
\ee
$$
Indeed  $\calB$ is an anti-hermitian generator of the Bogoliubov
transformation
$$
[\calB , \alpha_n ] = {1\over 2}G^{-1}
 \delta E\,\bar{\alpha}_{-n} \qquad
[\calB , \bar\alpha_n ] =
 {1\over 2}G^{-1} \delta E^t \alpha_{-n} \ ,
\eqn\eqBogoliubov
$$
for all $n\not=0$, from which, together with \eqsmf, one can see that
the vacuum \eqprimevac\ is really annihilated by the changed
oscillators
$\alpha '_n =\alpha _n+\delta \alpha _n$
and $\bar\alpha '_n=\bar\alpha _n+\delta \bar\alpha _n$ with
$n\geq 1$.

Now we can evaluate the change of the vertex \eqVERTEX\ under the
change of $E$. Working with ket state representation for convenience
of writing, and noting that
$\delta (\vac_{E})= - \calB\,\vac_{E}$, we have
$$
\delta \left(e^{E_{123}^{\dagger}}\vac_{E}\right) =
\delta \left(e^{E_{123}^{\dagger}}\right)\,\vac_{E} -
 e^{E_{123}^{\dagger}}\,\calB\vac_{E}\ .
\eqn\eqverchange
$$
To evaluate the first term we need know the change of
$E_{123}^{\dagger}$, which is calculated by using \eqsmf, \eqEEE\
and the property
$\bfalpha^{\dagger\,T}\delta B\,N\,\bfalpha^\dagger =
\bfalpha_0^t\delta B\,N_{00}\bfalpha_0  = 0 $
owing to the antisymmetry of the $\delta B$ matrix.
One finds
$$
\delta E_{123}^{\dagger} =
 -{1\over 2}\Big(\bfalpha ^{\dagger\,T}, \bfalpha_0^{T} \Big)
\mymatrix{N}{N_0}{N_0^t}{N_{00}}
\myvector{\delta E\,\bar\bfalpha}
{\delta E\,\bar{\bfalpha_0}}
 + \ah(E\rightarrow E^t)\ .
\eqn\eqabove
$$
Here $\ah(E\rightarrow E^t)$ denotes the anti-holomorphic term
which is obtained by making substitutions
$\bfalpha  \rightarrow  \bar\bfalpha$  and $E \rightarrow  E^t$
in the first term. Since the variation
$\delta E\,\bar{\bfalpha}$ consists of
annihilation operators, it does not commute with
$E_{123}^{\dagger}$ and could make the calculation of
$\delta \left(\exp(E_{123}^{\dagger})\right)$ complicated.
Fortunately that
part of the change is just identical with the one given by the
Bogoliubov transformation \eqBogoliubov, so that \eqabove\
can be written in the form
$$
\delta E_{123}^{\dagger} =
- [ \calB, E_{123}^{\dagger} ] +
 \delta _0 E_{123}^{\dagger} ,
\ee$$
with $\delta _0 E_{123}^{\dagger}$
denoting the change in the zero-mode part:
$$
 \delta _0 E_{123}^{\dagger} =
 -{1\over 2}\Big(
(\bfalpha ^{\dagger\,T}N_0 + \bfalpha_0^{T}N_{00})
 \delta E\,\bar{\bfalpha_0}
+ \ah(E\rightarrow E^t)\
\Big)\ .
\eqn\eqzerochange
$$
Since $\delta _0 E_{123}^{\dagger}$ commutes with
$E_{123}^{\dagger}$, we have
$$
\delta \left(e^{E_{123}^{\dagger}}\right) =
 - [ \calB, e^{E_{123}^{\dagger}} ]
 +\delta _0 E_{123}^{\dagger}\,e^{E_{123}^{\dagger}}  \ ,
\ee
$$
and hence the vertex change \eqverchange\ becomes
$$
\delta \left(e^{E_{123}^{\dagger}}\vac_{E}\right) =
 - \calB\,e^{E_{123}^{\dagger}}\,\vac_{E}
 +\delta _0 E_{123}^{\dagger}\,e^{E_{123}^{\dagger}}\,
\vac_{E}\ .
\eqn\eqsaigo
$$
We can now evaluate the first term and show that it cancells exactly
the second term.  Using the expression \eqcalB\ of $\calB$ and making
the annihilation operators in $\calB$ act on
$e^{E_{123}^{\dagger}}\,\vac_{E}$,
we evaluate the first term and find
$$
\eqalign{
&-\calB\,e^{E_{123}^{\dagger}}\,\vac_{E} \cr
& \ \ \  = -{1\over 2}\Bigg\{
\bfalpha^{\dagger\,T}\big(-{1\over n} + N^tn\,N\big)
 \delta E\,\bar\bfalpha^{\dagger}
+ \bfalpha^{\dagger\,T}(N^tn\,N_0) \delta E\,\bar\bfalpha _0 \cr
& \ \ \ \ \quad \ \
 + \bar\bfalpha^{\dagger\,T}(N^tn\,N_0) \delta E^t\bfalpha _0
 + \bfalpha_0^t(N_0^tn\,N_0) \delta E\,\bar\bfalpha _0
\Bigg\}e^{E_{123}^{\dagger}}\vac_{E}
\cr} . \eqn\eqmojiki
$$
Here we can use the following identities [\Yo ] for the Neumann
coefficients of the light-cone type three-string vertex
$$
\eqalign{
N^tn\,N &= {1\over n} ,\cr
N^tn\,N_0 &= -N_0 ,\cr
N^t_0n\,N_0 &= -2N_{00} ,\cr
}\ee
$$
where the last two equalities hold in the presence of conservation
delta-functions (or Kronecker deltas) for the zero-modes.
Using these identities,
we find that \eqmojiki\  becomes
$$
\eqalign{
&-\calB\,e^{E_{123}^{\dagger}}\,\vac_{E} \cr
&\ \ \ \  = {1\over 2}\Big(
 \bfalpha^{\dagger\,T} N_0\delta E\,\bar\bfalpha _0
+ \bar\bfalpha^{\dagger\,T} N_0\delta E^t\bfalpha _0
 + 2\bfalpha_0^t N_{00}\delta E\,\bar\bfalpha _0
\Big)e^{E_{123}^{\dagger}}\vac_{E}\ .\cr}\ee
$$
But we immediately see that the quantity in parenthesis
equals $-\delta _0 E_{123}^{\dagger}$ (see
\eqzerochange ), and therefore the first term in \eqsaigo\
cancels the second term as desired.  We thus have shown
directly that the
3-string vertex is actually independent of the background $E$ despite
its apparent dependence.

\chapter{Equivalence of String Field Theories around Dual Backgrounds}

In order to begin our study of duality in string field theory
we need to understand why string field theory formulated around
backgrounds related by duality transformations describe physically
equivalent theories. Duality transformations are discrete
transformations, and for the case of toroidal compactification
of $d$ space dimensions, they form the group \Oddz. Consider
two backgrounds $E$ and $E'$ related by a duality transformation.
As we have discussed in a previous section we can write a string
field theory $S_E(\Psi )$ around the background $E$, and a string
field theory
$S_{E'}(\Psi)$ around the background $E'$.
We can also choose arbitrarily the string field
coupling constant. Let $g_0$ denote the coupling constant
for the $E$-theory and ${g'}_0$ denote the coupling constant
for the $E'$-theory. These string field
actions are manifestly different, in particular, the kinetic
terms are defined by $Q_B(E)$ and $Q_B(E')$ respectively, and
these two BRST operators are different. The purpose of the
present section is to show that these two actions describe the
same physics if and only if $g_0 = {g'}_0$.

As we will see, in string field theory it is manifest that two
theories written around dual backgrounds could only be equivalent
if their string field coupling constants are identical. The
string field coupling constant is defined from the three-point
couplings of states of the theory. If the spectra of two theories
are identical, and the perturbative S-matrix is identical, the
three point couplings ought to be the same, thus the coupling
constants must be identical. Duality therefore does not involve
a shift in the string field dilaton. This result is in agreement
with first quantization analysis, if this analysis is properly
interpreted. That will be the subject of section 6.

The concrete way of proving the physical equivalence of the
two theories will pave the way for our writing of the
symmetry transformations that leave the string action invariant
(Sect.~7). In the present section we will find, for each
discrete symmetry transformation $g\in$ \Oddz a unitary operator
$\U_g$ that will have the following fundamental property
$$S_E (\U_g \ket{\Psi} ) = S_{g(E)} (\ket{\Psi} ),\eqn\fund$$
for any background $E$, where $g(E)$ denotes the background obtained
by acting with the transformation $g$ on the background $E$.
Equation \fund\ shows that these two string field theories are
related by the homogeneous invertible field redefinition
$\ket{\Psi} \rightarrow \U_g \ket{\Psi}$, and therefore
the field theories are physically equivalent.

It should be emphazised that finding this operator $\U_g$ relating
two apparently different theories does not yet give us a symmetry
transformation. A symmetry transformation corresponds to an invariance
of an action, and the above operator does not yet give us any
such invariance. The operator $\U_g$, however, will be a key element
in the symmetry transformation to be constructed in Sect.~7.

This section is divided into three parts. In the first one
we review the necessary properties of the group \Oddz
and the definition of its action on the backgrounds.
In the second part we construct the operator $\U_g$,
show it is universal, prove \fund\  and
show that the operators $\U_g$ form a representation
of the discrete group \Oddz.

\section{\Odd and Background Fields}

The group \Odd is defined by its elements,
real matrices $g$ of size
$2d \times 2d$, such that $g^t J g = J$, where $J=\pmatrix{0 & I\cr
I& 0\cr}$.  We can spell out explicitly the conditions for a matrix
to belong to \Odd.  Denote $g$ by
$$g = \pmatrix{a&b\cr c&d\cr} \,
\rightarrow
g^t = \pmatrix{a^t & c^t \cr b^t & d^t}.\eqn\soddr$$
where $a,b,c,d$ are $d\times d$ matrices. The conditions
for $g\in$ \Odd are
$$a^tc + c^ta = b^td + d^tb = 0, \quad \hbox{and} \quad
a^t d + c^t b = I.\eqn\condg$$
These relations tell, in particular, that $(a^tc)$ and
$(b^td)$ are
antisymmetric matrices. Useful consequences of the above
conditions are derived next.
If $g\in $\Odd, then $g^t \in$\Odd. This
is proven as follows:
begin with $g^t J g = J$, then
take the inverse in both sides to find $g^{-1} J g^{t-1} = J$
(since $J^{-1} = J$). Now multiply from the left by $g$,
and from the right by $g^t$ to find $J= gJg^t$, which
shows that $g^t \in$ \Odd. If we now apply the conditions
in \condg\ to $g^t$ we find
$$ab^t + ba^t = cd^t + dc^t = 0,
\quad \hbox{and} \quad ad^t + bc^t = I,\eqn\condgadd$$
which means that $(a^tb)$ and $(c^td)$ are also antisymmetric
matrices. With all this information, it is possible now
to check that
$$g^{-1} = \pmatrix{d^t & b^t \cr c^t & a^t \cr}.\eqn\ginv$$
In fact, this follows directly from
$g^tJg = J \, \rightarrow \, (Jg^tJ)g= J^2 = I \, \rightarrow
\, g^{-1} = Jg^tJ$,
which is the result quoted above.
\bigskip
Let us now review the action of \Odd on the background field
$E=G+B$.
In order to have $2d\times 2d$ matrices in \Odd act on the
$d\times d$ matrix $E$ one uses linear fractional transformations.
Let $g \in$ \Odd be given by \soddr\ .
We denote by $E' = g(E)$ the
new background obtained by acting with $g$ on the background $E$.
The background $E'$ is given by
$$E' = g(E) \equiv  (aE + b) (cE + d)^{-1}. \eqn\actback$$
This definition can be checked to be consistent with the group
property: $g(g'(E))$
$= gg'(E)$.
Let us now derive a few useful relations that arise from
\actback, which we now denote as
$$E' = \pmatrix{a&b\cr c&d \cr} E .\eqn\brform$$
Solving for $E$ from \actback\ and transposing, one finds
$$E^t = \pmatrix{d^t & -b^t \cr -c^t & a^t\cr} {E'}^t,\eqn\acti$$
where the matrix above is readily verified to belong to
\Odd. Taking inverses to the previous two equations one
also finds the useful relations (using \ginv\ )
$$
E = \pmatrix{d^t & b^t \cr c^t & a^t \cr} E',\quad
{E'}^t = \pmatrix{a&-b \cr -c&d \cr} E^t.\eqn\actii
$$
An extra pair of relations will be useful:
$$\eqalign{
(d + cE)^t \, G'\, (d+cE) &= G,\cr
(d-cE^t)^t \, G'\, (d-cE^t) &=G.\cr}\eqn\transfg$$
The first relation is derived by writing $G' = (E' + {E'}^t)/2$,
using the expression for $E'$ from \actback\
and evaluating the left hand side. The second equation is derived
similarly beginning with the expression for ${E'}^t$ given
in \actii .
\medskip
Backgrounds related by generic \Odd do
not give equivalent physics.
We need to restrict ourselves to \Oddz. At the level of the
spectrum this follows from the form of the
first quantized hamiltonian $H(E)$
$$H(E) =  {1\over 2} {\vec Z}^t {\cal R}(E) \vec Z  + N + \bar{N}
+ \cdots \eqn\hamor$$
where $N$ and $\bar{N}$ denote the number operators,
the dots represent terms irrelevant for our discussion,
and $\vec Z$ denotes a $2d$-column vector with integer entries
$\vec Z = \pmatrix{\vec m \cr \vec n \cr}$,
where the integers $n_i$ and $m^j$ (with $i,j = 1, \cdots ,d$)
represent momentum and winding quantum mumbers respectively.
The matrix ${\cal R}$ was defined in \eqmatrix\ and has the
property that ${\cal R}(E') = g {\cal R}(E) g^t$ if $E'=g(E)$.
There is a further condition on the spectrum, one must have
$${1\over 2}{\vec Z}^t J \vec Z = N- \bar{N}.\eqn\extraco$$
Consider now a background
$E'= g(E)$, with $g\in$\Odd. One then has
$$
H(E') = {1\over 2} {\vec{Z'}}^t {\cal R}(E') \vec{Z'}
+ N' + {\bar N}'
+\cdots \, =\,{1\over 2}
{\vec{Z'}}^t g {\cal R}(E) g^t \vec{Z'}
+ N' + {\bar N}'
+\cdots .
\eqn\modham$$
Equations \hamor\ and \modham\ can define the same spectrum if we
can consistently set $\vec Z = g^t \vec{Z'}$, and thus
think of the two spectra as identical, although labeled
by different momentum and winding quantum numbers. Two
requirements are enough, $g^t$ must be invertible (it is so),
and all its entries must be integer (otherwise there would
exist some integer vectors $\vec{Z'}$ that would be mapped into
non integer vectors).
There is one extra condition coming from eqn. \extraco, we need
${\vec Z}^t J \vec Z = {\vec{Z'}}^t J \vec{Z'}$
which guarantees that an allowed state remains allowed after
the relabeling of the quantum numbers (and without change of the
oscillator excitations). This requires  $gJg^t = J$, which is
satisfied since $g\in$\Odd. Thus, all our discussion just
shows that backgrounds related by \Odd transformations
with integer entries, give an identical spectrum.

There is one extra discrete symmetry beyond \Oddz. It
corresponds to taking $B \rightarrow -B$. From the form of the
hamiltonian this is seen to be a symmetry of the spectrum
which is taken care by letting $m^i \rightarrow -m^i$
(or $n_i \rightarrow -n_i$). Since this change alters the sign
in the constraint \extraco, one must also have
$N \leftrightarrow \bar N$, by exchanging the right moving
and left moving oscillators. This clearly does not change
the contribution of $N+\bar N$ to the hamiltonian.

\section{The Unitary Operator $\U_g$}

We have seen that the definition of the universal objects
$X,P$ in term of background dependent oscillators led
to definite relations between any two sets of oscillators
corresponding to two different backgrounds. Those relations
were given in equation \uoiv\ . Such relations, of course,
are consistent with the background dependent commutation
relations of the oscillators. They always mix mode numbers,
in particular, oscillators of mode number $+n$ are related
to oscillators of mode numbers $+n$ and $-n$. The only
way to avoid mode number mixing is to have identical backgrounds.
This is sensible because different backgrounds correspond to
different string field theory vacua.

The above arguments do not rule out the possibility that
the physics of different vacua is the same. This is
actually a well known fact in first quantization analysis of
toroidal compactification.  In our language
the key idea is that we can define maps (not equalities)
between sets of oscillators, and these maps will respect
the commutation relations. These maps will be realized
by the operators we are after. Let us begin by recalling that
the commutation relations of the oscillators
$\{\alpha ,\bar{\alpha}\}$
are conveniently summarized by
$$[ {X'}^i (\sigma ), P_j(\sigma' )] = i \delta^i_j
\,{d\over d\sigma} \delta (\sigma - \sigma').\eqn\commuxp$$
These commutation relations, however, are left unchanged
under the following replacement
$$\pmatrix{X' \cr 2\pi P \cr} \, \rightarrow
\, \pmatrix{a^t & c^t \cr b^t & d^t\cr} \,
\pmatrix{X' \cr 2\pi P \cr},\eqn\replxp$$
if the above matrix performing the map belongs to \Odd.
Since this map implies that the zero modes ($w,p$) are transformed as
$$\pmatrix{w \cr p \cr} \, \rightarrow
\, \pmatrix{a^t & c^t \cr b^t & d^t\cr} \,
\pmatrix{w \cr p \cr},\eqn\transfzm$$
and the eigenvalues of $(w,p)$ are integers, the matrix
performing the map must actually belong to \Oddz.

Note that the map, which may be labeled by the
group element, is by definition background
independent (since $X'$ and $P$ are).
Since the map preserves commutation relations it must
be possible to obtain via a unitary operator.
In order to find the implications of the map for oscillators
we have a choice of backgrounds to make on the left
and on the right of the arrow in \replxp\ .
Let us choose the background $E$ for the left hand
side and the background $E'$ (which may or may not
be different for the right hand side), the map
\replxp\ then implies
$$\eqalign{
[{\bar \alpha}_{-n} - \alpha_n](E) \, &\rightarrow
\, a^t [{\bar \alpha}_{-n} - \alpha_n](E')
+ c^t [{E'}^t \alpha_n + E' {\bar \alpha}_{-n}](E'),\cr
[{E}^t \alpha_n + E {\bar \alpha}_{-n}](E)\, &\rightarrow \,
b^t [{\bar \alpha}_{-n} - \alpha_n](E')
+ d^t[{E'}^t \alpha_n + E' {\bar \alpha}_{-n}](E').
\cr}\eqn\maposc
$$
{}From the above maps, a small calculation gives
$$\eqalign{
2G\alpha_n(E) \, \rightarrow \,
&\left[ E(-c^t{E'}^t+a^t) +( d^t {E'}^t -b^t) \right]\,\alpha_n(E')
\cr
{}& +\left[ -E(c^tE'+a^t) +(d^t E'+b^t) \right]\,\bar{\alpha}_{-n}(E'),
\cr
2G\bar{\alpha}_{-n}(E) \, \rightarrow \,
&\left[ E^t(c^t{E'}^t-a^t) +( d^t {E'}^t -b^t) \right]\,\alpha_n(E')
\cr
{}&+\left[ E^t(c^tE'+a^t) +(d^t E'+b^t) \right]\,\bar{\alpha}_{-n}(E').
\cr}\eqn\ostransf$$

We now want to think of the \Oddz matrix as fixed, and find
if the above maps become diagonal in mode number for a particular
choice of $E'$. This requires the following conditions
$$\eqalign{
-E(c^tE'+a^t) +( d^t E'+b^t) &= 0 ,\cr
E^t(c^t{E'}^t-a^t) +( d^t {E'}^t-b^t) &=0.\cr}
\eqn\codiag$$
It follows from \acti\ and \actii,
that the above conditions are simultaneously satisfied if
$$ E' = \pmatrix{a&b\cr c &d\cr} \, E\, =\, g (E),\eqn\relback$$
and therefore for this choice of $E'$ the
maps do not change mode number. A little  calculation gives
$$\eqalign{
G\alpha_n(E)\,&\rightarrow \, (d^t-Ec^t)G' \,\alpha_n(E'),\cr
G\bar{\alpha}_n(E)\,&\rightarrow \,
(d+cE)^tG' \,\bar{\alpha}_n(E'),\cr}
\eqn\almost$$
and using \transfg\ we obtain the simplest form of the map
$$\eqalign{
\alpha_n(E) \, &\rightarrow \, (d-cE^t)^{-1} \alpha_n (E'),\cr
{\bar \alpha}_n(E) \, &\rightarrow \,
(d+cE)^{-1} {\bar \alpha}_n (E').\cr}
\eqn\fform$$
One can verify explicitly that the above maps hold also for
$n=0$, namely, their action on $\alpha_0 ,\bar{\alpha_0}$,
is consistent with \transfzm\ via \wpalpha .
We want to define now the unitary operator that
performs the above map. We will denote this operator by
$U_g$. Rather than trying to construct the operator explicitly
in terms of the operators $X,P$, we will define the operator
by describing how it acts on states (this defines completely
the operator). We therefore write
$$\eqalign{
U_g^{\dagger}\, \alpha_n(E)\, U_g  &= (d-cE^t)^{-1} \alpha_n (E'),\cr
U_g^{\dagger}\, {\bar \alpha}_n(E)\, U_g
&=(d+cE)^{-1} {\bar \alpha}_n (E'),\cr}\eqn\defuop$$

Since we require that the operator be unitary, the above
relations determine  $U_g$ up to phases.
We fix those phases now:
$$U_g \ket{w,p}_{E'} = \ket{a^tw+c^tp,b^tw+d^tp}_E.\eqn\actst$$
One can verify that the state on the right hand side must
be the one shown (up to a phase) by acting on the left hand side
with various operators, for example
$$\eqalign{
{\hat p} U_g \ket{w,p}_{E'}
&= U_g U^\dagger_g {\hat p} U_g \ket{w,p}_{E'}\cr
{}&= U_g (b^t {\hat w} + d^t {\hat p})\ket{w,p}_{E'}\cr
{}&= (b^t w + d^t p) U_g \ket{w,p}_{E'},\cr}\eqn\derpact$$
where from the quoted result follows. The fact that the
operator $U_g$ turns the $E'$ vacuum into the $E$ vacuum
follows from
$$\alpha_n(E)U_g \ket{0}_{E'} = U_g U^\dagger_g \alpha_n (E) U_g
\ket{0}_{E'}
\prop  U_g\alpha_n(E') \ket{0}_{E'} = 0,\eqn\uvacuum$$
which holds for all positive $n$. We therefore have
$$U_g \ket{0}_{E'} = \ket{0}_E \quad \leftrightarrow \quad
U_g^{\dagger}\ket{0}_E = \ket{0}_{E'}.\eqn\uactvac$$

Note that the action of $U_g$ on operators was defined in
a background independent way via \replxp\ . It should be
emphasized that $U_g$ is an operator relating states in
different Hilbert spaces, unless the original state is
in a Hilbert space corresponding to a background that is
invariant under the group element $g$.

Let us now find the action of $U_g$
on the BRST operator and on the vertex.
Most results will follow from the action
of $U_g$ on oscillator bilinears
$$\eqalign{
 U_g^{\dagger} \alpha_n (E)\, G\, \alpha_m (E) \, U_g
&= \alpha_n (E')(d-cE^t)^{-1t}\, G\,(d-cE^t)^{-1}\, \alpha_m (E')\cr
{}&= \alpha_n (E')\,G' \,\alpha_m (E'),\cr}\eqn\actbil$$
where use was made of \transfg\ .  The same equation
holds for the antiholomorphic oscillators. It therefore follows that
we have a very simple action on the Virasoro generators:
$$U_g^{\dagger} \pmatrix{L^X(E)\cr {\bar L}^X(E)\cr}   U_g =
\pmatrix{L^X(E')\cr {\bar L}^X(E')\cr},
\eqn\virasact
$$
and this result implies that
$$U_g^{\dagger} Q_B(E) U_g = Q_B(E'),\eqn\brstact$$
namely, that the operator $U_g$ changes the BRST operator from
that corresponding to the original background $E$ into that
corresponding to the background $g(E)$.

Let us now consider the three string vertex. Recall it is built
of a vacuum, oscillator bilinears and a cocycle factor.
Up to the cocycle factor, equations \actbil\ and \uactvac\ imply that
$${}_{123}\bra{V(E)} U_g^{(1)} U_g^{(2)} U_g^{(3)} =
{}_{123}\bra{V(E')}.\eqn\actvert$$

The cocycle factor is conveniently written as follows
$$\exp (i\pi [p_3^tw_2 - p_1^tw_1]) =
\exp (i\pi[\bfp_3^t P \bfp_2 -\bfp_1^t J \bfp_1 ])
\eqn\nicerfm$$
where the matrix $P$ and the vector $\bfp$ are defined by
$$P = \pmatrix{0 & 0\cr 1& 0\cr},\quad \bfp =
\pmatrix{w \cr p\cr},\eqn\defpp$$
and $J$ is the \Odd metric matrix. It follows that
$$
\eqalign{
\exp (i\pi[\bfp_3^t P \bfp_2 - \bfp_1^t J \bfp_1 ])
U_g^{(1)} U_g^{(2)}&U_g^{(3)}  \cr
= U_g^{(1)} U_g^{(2)} U_g^{(3)} &
\exp (i\pi[\bfp_3^t gPg^t \bfp_2 -\bfp_1^t J \bfp_1 ]),
\cr} \eqn\dercoc
$$
which shows that the form of the cocycle factor is not
preserved. The solution to this is to modify the operator
$U_g$ by including an extra phase factor
$$\U_g = U_g \Upsilon (g,\bfp)\eqn\moduuu$$
In order to give a simple description of $\Upsilon$ let us introduce
some notation. For any matrix $A$, \ \ $A_u$ and $A_l$
are defined to be the upper and lower triangular part matrices of $A$,
 respectively; namely,  for $A=(a_{ij})$
$$
\eqalign{
(A_u)_{ij} &= \cases{ a_{ij} &for $i<j$ \cr
                     0      &for $i\geq j$ \cr} \cr
(A_l)_{ij} &= \cases{ 0  &for $i\leq j$ \cr
                     a_{ij}  &for $i>j$ \cr} \cr
}$$
Then, clearly, $A_l = \big( ( A^t )_u \big)^t$,
and, if $A$ is an antisymmetric matrix,
$A = A_u + A_l$ and  $A_l = -(A_u)^t$.
Let us now give the form for the $\Upsilon$ factor:
$$\Upsilon (g,\bfp) = \exp ( i\pi \bfp^t {\cal A}_u(g)
\bfp ),\eqn\upsildef$$
where ${\cal A}$ is an antisymmetric matrix given by
$$
{\cal A}(g) = gPg^t-P =\pmatrix {ba^t & bc^t \cr -cb^t & dc^t \cr}.
\eqn\cbfdef$$
We now verify that the product
$\Upsilon^{(1)}\Upsilon^{(2)}\Upsilon^{(3)}$, in the presence of
the momentum conservation Kronecker deltas (for $\bfp$)
of the vertex, restores the cocycle to its original form. Begin with
$$\eqalign{
\Upsilon^{(1)}\Upsilon^{(2)}\Upsilon^{(3)}
&= \exp (i\pi \bfp_3^t {\cal A}_u \bfp_2
+ i\pi \bfp_2^t {\cal A}_u \bfp_3)\cr
{}& =\exp (i\pi \bfp_3^t {\cal A}_u \bfp_2
- i\pi \bfp_3^t {\cal A}_l \bfp_2)\cr
{}& = \exp (i\pi \bfp_3^t [{\cal A}_u +{\cal A}_l ] \bfp_2)\cr
{}& = \exp (i\pi \bfp_3^t {\cal A} \bfp_2), \cr}\eqn\workcoc$$
where use was made of momentum conservation to eliminate
$\bfp_1$, and of the identities given above \upsildef\ .
It therefore follows that indeed the vertex cocycle is restored to its
original form:
$$\exp (i\pi\bfp_3^t gPg^t \bfp_2 )
\Upsilon^{(1)}\Upsilon^{(2)}\Upsilon^{(3)}
=\exp (i\pi \bfp_3^t P \bfp_2 ).\eqn\fixcoc$$
Thus, we have finally obtained the proper unitary operator
$\U_g$ that leaves invariant the three string vertex
$${}_{123}\bra{V(E)} \U_g^{(1)} \U_g^{(2)} \U_g^{(3)} =
{}_{123}\bra{V(E')}.\eqn\actvertf$$
[We say ``invariant" because the vertex $\bra{V(E)}$ is actually
independent of $E$ as shown in Sect.~3.2.]
This proves that the interaction term of the action does
not change under the homogeneous redefinition of the
string field induced by $\U$. We must now check that
the kinetic term is changed from that corresponding to
the background $E$ to that corresponding to the background
$E'=g(E)$. For this purpose one first verifies that
the operator $\U$ acts on the
reflector state ${}_{12}\bra{R(E)}$ as expected
$${}_{12}\bra{R(E)} \U_g^{(1)} \U_g^{(2)} = {}_{12}\bra{R(E')},
\eqn\uonref$$
since the vertex cocycle of $\bra{R}$
is not changed by the $U$ operators, and the $\Upsilon$ factors
vanish by momentum conservation. Equation \uonref\ implies that
$${}_{12}\bra{R(E)} \U_g^{(1)} =
{}_{12}\bra{R(E')}\U_g^{(2)\dagger},
\qquad
\U_g^{(1)}\ket{R(E)}_{12}
=\U_g^{(2)\dagger} \ket{R(E')}_{12}.\eqn\uonrefi$$
With this information, we immediately find that the string
kinetic term becomes
$$\eqalign{
{}_{12}\bra{R(E)} \ket{\Psi}_1\, Q_B^{(2)}(E) b_0^{-(2)} \ket{\Psi}_2
\, \rightarrow \,
&{}_{12}\bra{R(E)}\U_g^{(1)} \ket{\Psi}_1\, Q_B^{(2)}(E) b_0^{-(2)}
\,\U_g^{(2)}\ket{\Psi}_2\cr
=&{}_{12}\bra{R(E')} \ket{\Psi}_1\, \U_g^{(2)\dagger}
Q_B^{(2)}(E) \U_g^{(2)}\, b_0^{-(2)} \ket{\Psi}_2 \cr
=&{}_{12}\bra{R(E')} \ket{\Psi}_1 \,Q_B^{(2)}(g(E))
 b_0^{-(2)} \ket{\Psi}_2\cr}
\eqn\uactonkin$$
as desired. This completes our proof of
$S_E(\U \ket{\Psi} ) = S_{g(E)}(\ket{\Psi} )$.

The action of $\U$ on the star product will be of utility later;
using \actvertf\ and \uonrefi\ one finds
$$\U \ket{A \star B} = \ket{\U A \star \U B},\eqn\actstar$$
where $A$ and $B$ are arbitrary string fields.

\section{Group Properties of $\U$}

Having constructed the operator $\U$ and established both its
background independence, and how it relates string field theories
on dual backgrounds, we now establish
the group properties of the operators $\U$. We will show that
$$
\U_{g'} \U_g = \U_{gg'}\,\exp [i\pi {\cal C}(\bfp,g',g) ],
\eqn\eqcocycle
$$
namely, that the operators $\U$ form a projective representation
of the discrete group of dualities, with $\exp(i\pi{\cal C})$
a nontrivial cocycle factor.
It follows directly from our expression for $\U$ that
$$\exp [i\pi {\cal C}(\bfp,g',g) ]=
\exp\left( i\pi \bfp^t
[\calA_u(g) + g\calA_u(g')g^t -\calA_u(gg') ] \bfp \right) .
\eqn\eqthree
$$
We will concentrate on the cocycle factor only since it is clear from
sect. 4.2 that the operators
$U$ satisfy $U_{g'}U_g = U_{gg'}$. It is also clear that
the above cocycle factor satisfies the conditions that
arise from the associativity of the $\U$ operators, since
this cocycle was derived from operators that associate.

Let us show that only the diagonal piece of ${\cal C}$ is relevant.
It follows from \cbfdef\ that
$$
\eqalign{
\calA(gg') &= gg'P (gg')^t - P  \cr
 &= g[g'Pg^{\prime t}]g^t - P \cr
 &= g[\calA(g') + P ]g^t - P \cr
 &= g\calA(g')g^t + \calA(g), \cr
}$$
and taking the upper triangular part of this matrix equation we have
$$
\calA_u(gg') = (g\calA(g')g^t)_u + \calA_u(g).
$$
Using this in \eqthree, we have now
$$\exp [i\pi {\cal C}(\bfp,g',g) ]=
\exp\left( i\pi \bfp^t
[g\calA_u(g')g^t - (g\calA (g') g^t)_u ] \bfp \right) .
\eqn\eqthree
$$
Note that inside the square bracket $[ \cdots ]$ above,
any matrix can be transposed freely
and its sign can be changed
at will (the matrices are integer-valued).
We denote equality in this sense by $\sim $. Then, for the first
term, we have
$$
\eqalign{
g\calA_ug^t
 &= (g\calA_ug^t)_u + (g\calA_ug^t)_l + (g\calA_u g^t)_d  \cr
 &= (g\calA_ug^t)_u + \big([(g\calA_ug^t)^t]_u\big)^t
+(g\calA_u g^t)_d\cr
 &\sim  (g\calA_ug^t)_u - [(g\calA_ug^t)^t]_u
+(g\calA_u g^t)_d\cr
}$$
where $\calA_u$ here means $\calA_u(g')$ and
$A_d$ denotes a diagonal matrix obtained by setting all the
entries of $A$ other than the diagonal elements equal to zero.
For the second term in \eqthree\
since $\calA$ is an antisymmetric matrix, we have
$$
\eqalign{
(g\calA g^t)_u
&= \left[g ( \calA_u - (\calA_u)^{\ t} ) g^t \right]_u \cr
&= (g\calA_ug^t)_u - [(g\calA_ug^t)^t]_u . \cr
}$$
This shows that only the diagonal matrix part
$(g\calA_u(g') g^t)_d$ survives, and this gives us the
simplest form for the cocycle factor:
$$
\exp [i\pi {\cal C}(\bfp,g',g) ] =
\exp\left( i\pi \bfp^t (g\calA_u(g') g^t)_d\,
 \bfp \right), \eqn\finalcoc$$

If we write $(g\calA_u (g') g^t)_d \equiv {\rm diag}(a_i ,b_i)$ the
above expression becomes
$$
\exp [i\pi {\cal C}(\bfp,g',g) ] =
\exp\left( i\pi \sum_{i=1}^d (a_i {w^i}^2 + b_i p_i^2) \right)
\eqn\eqthreeprime
$$
We can see now why the cocycle appearing in the composition
of two $\U$ operators is irrelevant when acting on the
three string vertex: momentum conservation shows that
$\sum_{r=1}^3 {w^i_r}^2$ and $\sum_{r=1}^3 p_{ir}^2$
are necessarily even integers and hence
$\prod_{r=1}^3 \exp(i\pi{\cal C}^{(r)}) = 1$.
Incidentally, this also shows that the string field theory is
invariant under the following field transformations:
$$
\Psi \, \rightarrow \Psi'=
\exp\left( i\pi \sum_{i=1}^d (a_i {w^i}^2 + b_i p_i^2) \right) \Psi,
\ee
$$
with $a_i$ and $b_i$ taking the values of $0$ or $1$.
These are parity-like
transformations on the component fields; for instance, in the case of
$a_i=\delta_{i1}$ and $b_i=0$,
the component fields corresponding to odd $w_1$ eigenvalues change
their signs
and those corresponding to even eigenvalues remain unchanged.
The complete set of such parity transformations form a discrete
closed group with $2^{2d}$ elements in all.

Before closing this section let address the issue of the
nontriviality of the cocycle in \eqcocycle\ .
If the cocycle were trivial, a suitable
redefinition of the $\Upsilon $ factor would eliminate it.
So consider redefining our $\Upsilon $ factor \upsildef\ into
$$
\Upsilon (g,\bfp) = \exp \big[
i\pi \bfp^t \big({\cal A}_u(g)+M(g)\big)\bfp \big]
\ee
$$
by adding an integer-valued matrix function $M(g)$ of $g$.
This new factor $\Upsilon$ must
still preserve the form of the three string vertex
(otherwise we could just take $M=-{\cal A}_u$). Following
our previous analysis we conclude that $M$ must be diagonal.
If this were not the case, then in addition to the diagonal
piece, which preserves the vertex cocycle, $M$ would have an extra
piece $M=A_u$ for some antisymmetric integer
matrix $A$. Then the vertex cocycle would acquire the extra factor
$\exp ( i\pi \bfp_3^t A \bfp_2 )$.  Since $\bfp_3$ and
$\bfp_2$ are independent, and this factor must be unity always,
$A$ must be zero. This shows $M$ is diagonal.
Then, in view of Eq.(4.48),
the condition for the cocycle factor to vanish for this $\Upsilon $
is given by the equation
$$
M(gg') \sim  M(g) + gM(g')g^t - (g\calA_u(g') g^t)_d
\eqn\iscoctr
$$
where $\sim$ means equality when used inside the brackets in
$\exp({i\pi \bfp^t [{\cdots} ]\bfp})$.
In fact, inside the brackets any symmetric integer
matrix can be replaced
by its diagonal part, so the condition for the triviality of
the cocycle reduces to
$$
M(gg') \sim  M(g) + (gM(g')g^t)_d - (g\calA_u(g') g^t)_d
\eqn\iscoctrf
$$
We do not know if a diagonal matrix $M(g)$ satisfying this
equation exists.

\chapter{Condensation of States and Classical Solutions}

In this section we will discuss classical solutions of
string field theory, in particular, classical solutions that
correspond to changes of backgrounds. It is important to
emphasize that the classical solutions that change backgrounds
need not involve change of coupling constant of the theory.
To make that clear we will first consider the condensation
of the dilaton, and explain how it induces a change of
coupling constant. In this respect, our work is an extension
of that of Refs. [\Yo ,\HaNa ].

Consider a state $\ket{S}$ which condensates or acquires a
vacuum expectation value. Assume that when contracted with
the three-string vertex it gives the following result
$$
\bsub{123}\bra{V}\ket{S}_3
 = \bsub{12}\bra{R} [\O{2}_S,\QB^{(2)}] \bz{2} ,
\eqn\ebi
$$
namely, it can be written as a commutator of the BRST
operator with a (Grassman even) differential ${\cal O}_S$.
Such property implies that the condensation
of $\ket{S}$ converts the three-string interaction
term into kinetic-like term:
$$
\eqalign{
\bsub{123}\bra{V}\sfd_1\sfd_2\ket{S}_3
 &= \bsub{12}\bra{R} \sfd_1 [\O{2}_S,\QB^{(2)}] \bz{2} \sfd_2  \cr
 &= \bra{\Psi } [{\cal O}_S, \QB] \bm \sfd  \cr
}\eqn\ebii
$$

If the operator $\calO_S$ is anti-hermitian, this change of
kinetic term can be cancelled by making a homogeneous
field redefinition by $\calO_S$.  Indeed, the following
(infinitesimal)  inhomogeneous field transformation
$$
\delta \sfd = {1\over g}\ket{S} + \calO_S\sfd
\eqn\etransf
$$
gives the following change of the action \evi:
$$
\eqalign{
\delta S =& {2\over g}\bra{\Psi }\QB \bm \ket{S} \cr
 &+g\cdot {1\over g}\bsub{123}\bra{V}\sfd_1\sfd_2\ket{S}_3
 +\bra{\Psi }[\QB, \calO_S] \bm\sfd
\cr
&+ {g\over 3}\bsub{123}\bra{V}
(\sum_{r=1}^3\calO_S^{(r)})\sfd_1\sfd_2\sfd_3 .
\cr
}\eqn\eds
$$
The first term in the right hand side vanishes if the equation of motion
$$
\QB\bm\ket{S}=0
\eqn\eqmotion
$$
is satisfied, and the two terms in the second line cancel each other
out because of
eq.\ebii. Therefore,  if the vertex $\bra{V}$ is an eigenstate of the
operator $(\sum_{r=1}^3\calO_S^{(r)})$,
$$
\bra{V}(\sum_{r=1}^3\calO_S^{(r)})=\lambda _S \bra{V},
\eqn\eeigen
$$
with eigenvalue $\lambda_S$, then the total change of the action
is simply given by
$$
\delta S = {\lambda _Sg\over 3}\bsub{123}\bra{V}\sfd_1\sfd_2\sfd_3 ,
\eqn\eSchange
$$
implying that the string field coupling constant $g$ is changed by the
field condensation $g^{-1}\ket{S}$ by the amount
$$
\delta g = \lambda _S g.
\eqn\egchange
$$
It should be noted that the role of the homogeneous part of the
field transformation was that of a field redefinition useful
to bring the action to a form where one could read directly the
fact that the string coupling constant had changed.

\section{Dilaton condensation}

We now apply the above discussion to the condensation of
the dilaton state
$$
\eqalign{
&\ket{D} = \cm \left[ \alpha _{-1}^\mu \eta _{\mu \nu }
\bar \alpha _{-1}^\nu  + c_{-1}\bar b_{-1} -b_{-1}\bar c_{-1}
\right] \vac \delta _\epsilon (p,w) ,
\cr
&\ \delta _\epsilon (p,w) =
\lim_{\epsilon \rightarrow 0}{1\over 2}[\delta (p^+-\epsilon )+\delta
 (p^++\epsilon )]
(2\pi )^{D-d}
\big(\prod_{\mu \not=+}\delta(p^\mu )\big)\cdot
\delta ^d(p_i) \delta ^d(w^i)\ , \cr
}\eqn\ebiii
$$
in the $\alpha =p^+$ HIKKO theory.\footnote{*}{Dilaton condensation
is difficult to analyze in the original HIKKO string field theory
due to the fact that changes in the unphysical string length
parameter change the effective coupling constant of the theory.}
A calculation similar to
that of Hata and Nagoshi [\HaNa ] shows that
condensation of this dilaton
yields the following differential operator:
$$
\calO_D = {1\over 2}\left[
\calD_{X}+\calD_{\FP}+N_{\FP}-\eulerder{p^+} \right] .
\eqn\ebiv
$$
where the dilation operators $\calD_{X},
\calD_{\FP}$ and ghost number operator are defined as:
$$
\eqalign{
\calD_{X} &= {1\over 2}\eulerder{p_\mu } +
\sum_{n\not=0} {1\over n}\alpha _n^\mu  \eta _{\mu \nu } \bar \alpha
 _n^\nu
\cr
\calD_{\FP} &= {1\over 2}[\bp,\cp] +
\sum_{n\not=0} b_n \bar c_n + \bar b_n c_n
\cr
N_{\FP} &= {1\over 2}[\cp,\bp] +
\sum_{n\not=0} :\,c_{-n} b_n + \bar c_{-n} \bar b_n\,:
\cr
}\eqn\ebv
$$
Since the operator $\calO_D$ is anti-hermitian and the dilaton state
$\ket{D}$ satisfies the on-shell equation \eqmotion, the dilaton
condensation
$$
\delta \sfd = {1\over g}\ket{D} + \calO_D\sfd ,
\eqn\ediltransf
$$
yields a change of the coupling constant (\egchange ):
$$
\delta g = \lambda _D g,\qquad {\rm where}\quad
\bra{V}\calO_D =\lambda _D \bra{V},
\eqn\elambda
$$
where $\calO_D$ stands for the sum
$\sum_{r=1}^3\calO_D^{(r)}$. The eigenvalue $\lambda _D$ of
the operator $\calO_D$ defined in \ebiv\ is obtained as
follows.  Noting that $\calD_X$ and $\calD_{\FP}$
act on oscillators as
$$
\eqalign{
\calD_X &: \qquad \alpha _n\rightarrow \bar\alpha _{-n},\quad \bar\alpha
 _n\rightarrow \alpha _{-n},
\cr
\calD_{\FP}&: \qquad c_n \leftrightarrow\ -\bar c_{-n},
 \quad b_n \leftrightarrow\ \bar b_{-n},
\cr
}\eqn\exx
$$
we find the following eigenvalues for each factor
appearing in the vertex \exvii\ under the action of the
operators
$\calD_X, \calD_{\FP}, N_{\FP}$ and $-\eulerder{p^+}$:
\vskip 5mm


$$\hbox{\vbox{\offinterlineskip
\def\strut{\hbox{\vrule height 15pt depth 10pt width 0pt}}
\hrule
\halign{
\strut\vrule#\tabskip 0.1in&
\hfil$#$\hfil &
\vrule#&
\hfil$#$\hfil &
\vrule#&
\hfil$#$\hfil &
\vrule#&
\hfil$#$\hfil &
\vrule#\tabskip 0.0in\cr
& &&  \delta (123) && \Pi _r(\cp+W/\sqrt2)
  &&      \exp(F^\dagger_{123})\vac   & \cr\noalign{\hrule}
& \calD_X={d_0\over 2}+p^\mu \pder{p^\mu }+(\hbox{oscillators}) &&
       -d_0 && 0 && {d_0\over 2}\times 3 & \cr\noalign{\hrule}
& \calD_{\FP}={1\over 2}-\cp\bp +(\hbox{oscillators}) &&
       0  && -3  &&      {1\over 2}\times 3  & \cr\noalign{\hrule}
& N_{\FP}=-{1\over 2}+\cp\bp +(\hbox{oscillators}) &&
       0 && +3  &&  -{1\over 2}\times 3  & \cr\noalign{\hrule}
& -\eulerder{p^+}=-1-2p^+\pder{p^+} &&
  +2 && 0  &&   -1\times 3  & \cr\noalign{\hrule}}}}$$
\vskip 3mm
\noindent
where $d_0 \equiv  D-d$ is the number of uncompactified coordinates.
Note that these are quantum numbers for the {\it ket-state} vertex
$\ket{V}$.  For the desired bra $\bra{V}$, they change signs
because of anti-hermiticity of those operators. The factor of three
in the last column comes from the fact that these operators actually
stand for sums over the three strings. So from
 eq.\ebiv\  for $\calO_D$,
we have
$$
\bra{V}\calO_D =\lambda _D \bra{V},
\qquad \lambda _D = -{1\over 4}(d_0-2).
\eqn\ebxx
$$
Note that the coupling constant change we have obtained is proportional
to the \nextline
transversal dimensions $(d_0-2)$ and the $-2$ contribution
came from $-\eulerder{p^+}$.

It may be interesting to note that the condensation of the following
``transversal dilaton" $\ket{trD}$ also gives the same change of
coupling constant:
$$
\eqalign{
&\ket{trD} =  \cm \bigg[ \sum_{\mu ,\nu \not=\pm }
\alpha _{-1}^\mu \eta _{\mu \nu }\bar \alpha _{-1}^\nu
\bigg] \vac \delta _\epsilon (p,w) , \cr
&\ \ \longrightarrow \ \
\calO_{trD}= {1\over 2}\calD_X^{\rm transverse}
\ \ \longrightarrow \ \
\delta g = -{1\over 4}(d_0-2)g ,
\cr}\eqn\ebxxi
$$
where
$$
\calD_{X}^{\rm transverse} = {1\over 2}\sum_{\mu \not=\pm
 }\eulerder{p_\mu } +
\SUM{n\not=0}{\mu ,\nu \not=\pm } {1\over n}\alpha _n^\mu  \eta _{\mu
 \nu } \bar \alpha _n^\nu  .
\ee
$$

Having understood how dilaton condensation changes the coupling
constant we now turn to condensation of states that do not change
the coupling constant but rather the string background.

\section{Condensation of Exactly Marginal States}

Our discussion of target space duality as a string field symmetry
relies on the existence of suitable classical solutions that
shift the backgrounds. Even more, the statement of background
independence in the space of toroidal compactifications is essentially
the statement that there are classical solutions that move us in
this space.

Our discussion of condensation of states in this section begins
at the infinitesimal level. Here the convenience of using the light
cone style vertex will be manifest. In contrast with the less
direct calculation which is possible in the covariant closed string
field theory [\Se ], we will be completely explicit. We will show
how the change of the BRST operator comes about directly by
the condensation. Moreover, after the infinitesimal condensation
the action has the correct form at the new background.
Our analysis benefitted from the earlier discussion of Itoh [\itoh ].
We then turn to finite condensation. We give an expression for the
classical solution that corresponds to a finite change of background.
Here several issues arise, having to do with the space where the
solution lives, and with the singular nature of the light cone vertex.
These will be discussed explicitly.

Let us begin by considering the condensation of the
following exactly marginal state:
$$
\ket{\hbox{EM}} = \cm \left[ \alpha _{-1}^i(E)a_{ij}
\bar \alpha _{-1}^j(E)
\right] \vac \delta _\epsilon (p,w),
\eqn\ebxxii
$$
which is a state with a tensor coefficient $a_{ij}$
in the curled up dimensions.
Let us denote $\alpha _n^i(E)$ simply as $\alpha _n^i$ by
omitting the background label $(E)$, whenever confusion would not
occur. The condensation of this state yields a contribution to
the kinetic term of the theory coming from the contraction of
$\ket{\hbox{EM}}$ against the three string vertex. This is actually
an interesting calculation, and therefore we give its complete
details in Appendix C. The result of the condensation is
$$
\bsub{123}\bra{V}\ket{\hbox{EM}}_3 =
-{1\over 2}
\bsub{12}\bra{R}\sum_{\ell+n+m=0}(c_\ell+\bar c_{-\ell})^{(2)}
(\alpha _n^i a_{ij} \bar\alpha _{-m}^j)^{(2)} b_0^{-(2)}.
\eqn\ebxxiii
$$
Therefore,
under the following infinitesimal trasformation ($a_{ij}\ll 1$):
$$
\sfd\ \ \longrightarrow \ \ \sfd'= {1\over g}\ket{\hbox{EM}} + \sfd .
\eqn\ebxxv
$$
it follows from  eq.\ebxxiii\ that the kinetic operator
is now changed into
$$
\QB' = \QB(E) -{1\over 2}
\sum_{\ell+n+m=0}(c_\ell+\bar c_{-\ell})
(\alpha _n^i a_{ij} \bar\alpha_{-m}^j) .
\eqn\ebxxvi
$$
Comparing this with \varqb,
we see that this transformed BRST operator $\QB'$
is just the BRST operator on the new background
$ E' = E -a $; i.e., $\delta E=-a$.
$$
\QB' = \QB(E' = E -a) .
\eqn\ebxxx
$$
Conversely, the change of the background metric by an
amount $\delta E$ is realized by the following string field
 condensation
$$
\delta \ket{\Psi _0} = -{1\over g}
 \cm \left[ \alpha _{-1}^i(E)\,(\delta E)_{ij}
\bar \alpha _{-1}^j(E)
\right] \vac_{\alpha (E)} \delta _\epsilon (p,w).
\eqn\ebxxxi
$$
This is the form of the infinitesimal string field condensation
we were looking for.
Let us see what has happened with the complete action after the
transformation indicated in \ebxxv . The kinetic term has become
that of the new background and the three-string vertex now couples
three $\ket{\Psi}$'s. Since the vertex is background independent
the whole result is simply the string field theory around the
new background, namely
$$S_E (\delta \ket{\Psi_0} + \ket{\Psi} ) =
S_{E+\delta E} (\ket{\Psi})
+ {\cal O}((\delta \ket{\Psi_0} )^2) .\eqn\shiftact$$

Let us now consider a finite change of the background fields from
$E_0=G_0+B_0$ to $E_1=G_1+B_1$.  We take an arbitrary
interpolating path
$E(t)=G(t)+B(t)$ satisfying $E(0)=E_0$ and $E(1)=E_1$.
Then we simply integrate the infinitesimal string field
condensation along the path to obtain the following state:
$$
\ket{\Psi _0} \equiv \ket{E_0;E_1} =  -{1\over g}
 \cm \int _0^1 \hbox{dt}\, \alpha _{-1}^i(t)\,
{\hbox{d}E_{ij} \over \hbox{dt}}(t) \,
\bar \alpha _{-1}^j(t)
 \vac_{\alpha (t)} \delta _\epsilon (p,w),
\eqn\ebxxxii
$$
where $\alpha _n^i(t) \equiv  \alpha _n^i(E(t))$.
Formally, this state is expected, as a consequence of
\shiftact\ to shift the string action from the background
$E_0$ to the background $E_1$, namely
$$S_{E_0} ( \ket{\Psi_0} + \ket{\Psi} ) =
S_{E_1} (\ket{\Psi}) .\eqn\shiftaction$$
Note that this equation does not involve the value of the
original action $S_{E_0} (\ket{\Psi_0})$ at the classical
solution because this constant is zero. Indeed, let
$S(t) \equiv S(\ket{\Psi} =\ket{\Psi_0 (t)})$, where
$\ket{\Psi_0 (t)}$ denotes the state indicated in equation
\ebxxxii\ integrated only up to $t$. The constant in question
is $S(t=1)$. It follows that
$${dS\over dt}(t) = \left( {\delta S \over \delta \Psi}
\right)_{\Psi = \Psi_0(t)} \cdot {d\Psi_0 \over dt} = 0,
\eqn\nocons$$
since $\Psi_0(t)$ is a classical solution for all $t$.
Since $S(t=0)= 0$, then $S(t) \equiv 0$, and we verify that
there is no constant term in \shiftaction .

Why didn't we consider the
differential operator $\calO_{\hbox{EM}}$ in analogy to the
dilaton case ?   There is an ``operator'' $\calO_{\hbox{EM}}$
for which eq.\ebi\ holds.
It is given by
$$
\calO_{\hbox{EM}}={1\over 2}\left(
 \pder{\alpha _0^i}(G^{-1}a)^i_{\ j}\bar\alpha _0^j
 + \alpha _0^i(aG^{-1})_i^{\ j}\pder{\bar\alpha _0^j}
 + \sum_{n\not=0}{1\over n}\alpha _n^i a_{ij} \bar\alpha _n^j
\right)\ , \eqn\edaop
$$
but this operator is completely ill-defined because
it contains the differential
operators $\pder{\alpha _0^i}$
and $\pder{\bar\alpha _0^j}$ that correspond to the
zero-mode coordinate operators $x^i$ and $q_i$ in the
compactified directions.
(It is, however, interesting to note the similarity of the non-zero
mode part of $\calO_{\hbox{EM}}$
and the generator $\calB$
of the Bogoliubov transformation in \eqBogoliubov.
Furthermore, when the zero modes take continuous values,
$\calO_{\hbox{EM}}$ becomes a generator of $O(d,d;R)$.)

\REF\Hataetal{H.~Hata, K.~Itoh, T.~Kugo, H.~Kunitomo and
K.~Ogawa \journal Phys. Lett. &175 (86) 138.}

\medskip
There is actually one difficulty with the above classical
solution arising from our use of the light-cone vertex.
It can be verified that this solution is not path-independent.
The simplest way to check this is to perform two infinitesimal
string field condensations $\delta E_1$ and $\delta E_2$
successively, but in two different orders. These would read
$$\ket{\Psi} = \alpha_{-1}(E)\, \delta E_1 \,
\bar{\alpha}_{-1}(E)\ket{0}_E +
\alpha_{-1}(E+\delta E_1) \,\delta E_2
\,\bar{\alpha}_{-1}(E+\delta E_1)\ket{0}_{E+\delta E_1},
\eqn\nonindep$$
and a similar one with the labels 1 and 2 exchanged.
A simple calculation, using the Bogoliubov transformed
vacua to first order, shows they are not equal. By performing
the shifts in the order $\delta E_1$ ,$\delta E_2$,
$-\delta E_1$, $-\delta E_2$,  this
gives us a nonvanishing string field corresponding to a
condensation that should not change the background. This would
seem impossible on account that the BRST operator should be
shifted as $Q \rightarrow Q + \Psi_0 \star$ , where $\Psi_0$
is the classical solution. But in the light cone field theory
there exist nonvanishing string fields whose product with any
ordinary string field (of nonzero $p_+$) is zero
(see Ref. [\Hataetal ]).
Such pathology is not expected to occur for a string field
theory whose star product does not admit singular configurations,
as is the case of the nonpolynomial closed string field theory.

Let us now discuss the most important issue, that of the space
where the classical solutions are expected to live.
At face value one may think that our finite classical solution
should live in the Hilbert space of the original string theory.
We could then speak of the string classical solution as a collection
of classical solutions for the component fields of the theory.
This does not seem to be possible, according to \ebxxxii .
The classical solution is a sum (actually integral) of
very simple Fock space vectors, but each on a different vacuum.
Since we are dealing with a system with infinite number of
degrees of freedom (oscillators) it turns out that the
different vacua, as related formally by Bogoliubov transformations
are actually orthogonal. Their inner product is always zero!
There is no way we can perform the sum in a single Hilbert
space, unless we cut-off the number of oscillators. As the
cutoff is removed the difficulties reappear.
We may be forced to admit that nontrivial classical solutions
must live outside the Hilbert space of the original background,
but this will demand that we learn how to define string field
theory beyond the usual methods based on oscillator expansions.
The natural language for string field theory, at any rate,
is likely to be that of functionals, and we may be learning that
restricting ourselves to functionals corresponding to a
single vacuum is a very unnatural thing to do. Understanding
the implications of this fact is possibly the most important
issue that we face in string field theory.

\REF\KADA{L. Kadanoff
\journal Ann. Phys. &120 (79) 39;\hfill\break
J. Cardy \journal J. Phys. &A (87) L891.}
\REF\DVVV{R. Dijkgraaf, E. Verlinde and H. Verlinde in :
{\it Perspectives of String Theory}, eds. P. Di Vecchia and
J. L. Petersen (World Scientific, Singapore), 1988.}
\REF\DAKU{D. Kutasov \journal Phys. Lett. &B220, (89) 153.}
\REF\ChSc{S. Chaudhuri and J. A. Schwartz \journal Phys. Lett.
&B219 (89) 291.}
\REF\CBT{C. B. Thorn \journal Nucl. Phys. &B286 (87) 61.}
\REF\POL{J. Polchinski \journal Nucl. Phys. &B307 (88) 61.}

\section{Solving the Classical Equations Recursively}

In this section we find an explicit expression for the
classical string field solution corresponding to an
exactly marginal operator. The solution will be expressed
as a series, and if the series converges, it will define
a classical solution corresponding to a finite shift of
background. This solution applies to any form of closed
string field theory using symmetric vertices. We have in
mind, of course, the nonpolynomial closed string field theory.
The convergence of the series depends on the off-shell behavior
of the theory.

String field condensation that changes the toroidal backgrounds
corresponds to exactly marginal operators of the conformal field
theory. Such operators are dimension (1,1) primary fields
of the form current-current $J\bar{J}$. The corresponding BRST
invariant states, which are used in the string field theory,
are the states created by the dimension (0,0) operator
$cJ \bar{c}\bar{J}$. Such states, for our case can be chosen
to have zero momentum and zero winding.

Consider a conformal field theory, and denote by $\phi_i$,
$i=1,2,\cdots $ the dimension (1,1) primary fields of the
theory, and by $\lambda_i$ their corresponding couplings.
We thus consider deformation of the conformal field theory via the
perturbations
$$S_{\hbox{cft}}(\lambda ) = S_{\hbox{cft}} +\sum_i \lambda_i
\int d^2z \phi_i (z,\bar{z})\eqn\deformcft$$
If one of the dimension (1,1) operators above, say $\phi_A$, is
exactly marginal then it must happen that the operator product
coefficient $c_{iAA}$ must vanish for all $i$ (including $A$)
[\KADA--\ChSc].
As we will see, the recursive solution of the string field equations
without obstructions will demand the above condition and, in
addition, higher order requirements. These state that
the string scattering
amplitude, in genus zero, of {\it any} dimension (1,1) operator
$\phi_i$ with $n \geq 2$ copies of the exactly
marginal operator $\phi_A$ vanishes:
$$\int_{M_{0,n+1}} <\phi_i
\underbrace{\phi_A \phi_A \cdots \phi_A}_{n} > = 0\eqn\cftmarg$$
We will not attempt to give a conformal field theory derivation
of this statement. We will simply assume that the obstructions
vanish and find the string field solution. The work of Mukherji and
Sen [\MuSe ] provides evidence that string field theory
obstructions correspond to conformal field theory
nonzero beta functions. Our analysis in this section parallels
that of [\MuSe ] in general strategy.

Consider now the equations of motion of the nonpolynomial
closed string field theory:
$$Q b_0^- \ket{\Psi} + {1\over 2!} \ket{\Psi^2}
+ {1\over 3!} \ket{\Psi^3} + \cdots = 0 .\eqn\sfteqnm$$
We now attempt a perturbative solution of this field equation
via the expansion
$$\ket{\Psi}= \sum_{n=1}^\infty \epsilon^n \ket{\Psi_n}
= \epsilon \ket{\Psi_1} + \epsilon^2 \ket{\Psi_2} + \cdots .
\eqn\ansfield$$
The equations
that we must solve recursively read:
$$\eqalign{
Qb_0^- \ket{\Psi_1} &= 0\cr
Qb_0^- \ket{\Psi_2} &= -{1\over 2} \ket{\Psi_1^2} ,\cr
Qb_0^- \ket{\Psi_3} &= - \ket{\Psi_1 \Psi_2}
-{1\over 3!} \ket{\Psi_1^3},\cr}\eqn\recurseq$$
and so on.
Note that the structure of the equations is such that
the right hand sides must correspond to BRST trivial
states. Therefore the  BRST operator must annihilate every
right hand side. The identities relating the BRST operator
and the string products guarantee that this condition
is satisfied automatically to every order if the lower
order equations are satisfied. The only obstruction to
solving these equations is that a state corresponding to
a BRST cohomology class may appear in the right hand side
(a state annihilated by $Q$ which is not of the form
$Q\ket{\alpha}$ for any state $\ket{\alpha}$). Since the
string field has ghost number $+3$ and $Qb_0^-$ has
ghost number zero, the terms in the right hand sides must
have ghost number $+3$. Thus the
{\it obstructions are the BRST cohomology classes at ghost
number +3}. In critical string theory, a full copy of
the physical cohomology appears at ghost number $+3$ [\CBT ],
so there exist potential obstructions.

Our ansatz is that to leading order the string field is
the BRST invariant version of the marginal operator
$\phi_A$, namely
$$\ket{\Psi_1} = \ket{\phi_A} \, \quad
\rightarrow \, Qb_0^- \ket{\phi_A} = 0\eqn\anzatzfi$$

We will now try to solve all the higher order equations.
Note that all the higher order corrections $\ket{\Psi_n}$
($n\geq 2$) to the string field correspond to
{\it unphysical} states.
This is so because they must not be annihilated
by $Q$, as is seen in \recurseq . The effect of this is
that string field condensation of the massless fields is
not sufficient to change the background, we must also
give expectation values to unphysical fields, namely,
to the zero momentum components of massive fields in the
string field theory. While at each stage we
could add some BRST invariant physical field to
$\ket{\Psi_n}$ we will not do so. The states in the right
hand side must be annihilated by $b_0^-$ and by
$L_0^-$, and these conditions are guaranteed by the structure
of the string field theory. In order to solve the equations
we use the following Lemma.

\noindent
$\underline{\hbox{Lemma}}$. Consider a state $\ket{A}$ such
that
$$Q\ket{A} = 0, \quad L_0^- \ket{A} = 0, \quad
L_0^+ |A> \not= 0.\eqn\lemmcond$$
If the state is a linear superposition of
Fock space states, we require that all those states have nonzero
$L_0^+$ eigenvalue. Then one can solve
$$Qb_0^- \ket{\psi} = b_0^-\ket{A}\eqn\solveee$$
with
$$b_0^- \ket{\psi} =
- {b_0^+\over L_0^+}\, b_0^- \ket{A}.\eqn\solvedee$$

\noindent
$\underline{\hbox{Proof}}$.  This is just proven by calculation.
We use
$$Q= -c_0^+ L_0^+ + c_0^- L_0^- + b_0^+ M^+ + b_0^- M^- + \hat{Q}$$
and the generic expression for the field $b_0^- \ket{A}$
$$b_0^- \ket{A}= A_0 \ket{+-} + A_1 \ket{--}$$
which given the odd statistics of the string field $\ket{\Psi}$,
and the even statistics of the SL(2,C) vacuum (convention),
we have that $A_0$ is even and $A_1$ odd. One then finds
$$0= Qb_0^- \ket{A} = (L_0^+ A_1 + \hat{Q} A_0)\ket{+-}
+ (M^+ A_0 + \hat{Q} A_1)\ket{--}\eqn\qona$$
Now verify that the solution given in \solvedee\ is correct
$$b_0^- \ket{\psi} = - {b_0^+\over L_0^+}\, b_0^- \ket{A}
= -{b_0^+ \over L_0^+} \, A_0 \ket{+-} =
-{1 \over L_0^+} \, A_0 \ket{--},$$
and upon acting with the BRST operator one gets
$$\eqalign{
Qb_0^- \ket{\Psi} &= - Q {1\over L_0^+} A_0 \ket{--}\cr
{}&= (c_0^+ L_0^+ - \hat{Q}) {1\over L_0^+} A_0 \ket{--}\cr
{}&= A_0 \ket{+-} -  {1\over L_0^+}\hat{Q} A_0 \ket{--}\cr
{}&= A_0 \ket{+-} + A_1\ket{--} = b_0^-\ket{A},\cr}\eqn\checksol$$
where use was made of \qona . This proves the lemma. Note that
the reason we had to do an explicit check was that $b_0^+$
annihilates part of $\ket{A}$.
\medskip
Before beginning to consider the recursive solution, let us
establish one more useful result. We are going to solve
equations of the type
$$Qb_0^- \ket{\Psi} = A_0 \ket{+-} + A_1\ket{--}\eqn\ruleout$$
as indicated before. The right hand side is annihilated
by $Q$ and conformal field theory should imply that
the right hand side must not contain a nontrivial
BRST class, since otherwise the recursive procedure falls
flat on its face. Thus it must only contain BRST trivial
states. Let us show that it {\it cannot} contain
BRST trivial states of dimension (0,0). While such
states would present no obstruction to the recursive
procedure, it is useful to show they are not present since
this will simplify considerably our results and enable
us to use the lemma. Consider all possible type of
states that can appear for the right hand side of
\ruleout\ . Since they must have dimension (0,0) and
the momentum must be zero, they must be made by one
holomorphic and one antiholomorphic oscillator.
Taking into account ghost number the only possible states are
$$c_{-1} \bar{\alpha}_{-1}^\mu \ket{--}, \quad
\bar{c}_{-1} \alpha_{-1}^\mu \ket{--},\eqn\firstgr$$
and
$$c_{-1}\bar{b}_{-1} \ket{+-},\quad
\bar{c}_{-1}b_{-1} \ket{+-},\quad
\alpha_{-1}^\mu a_{\mu\nu} \bar{\alpha}_{-1}^\nu \ket{+-}.
\eqn\lastgr$$
{}From the latter group the combination
$(\bar{c}_{-1}b_{-1}+ c_{-1}\bar{b}_{-1})\ket{+-}$ is BRST
invariant, but also nontrivial, as is checked in a straighforward
way by writing states of suitable ghost number. The last
state in \lastgr\ is also BRST invariant and nontrivial.
\footnote{*}{The ``graviton trace state'' which is unphysical
for nonzero momentum, becomes physical at zero momentum.
Even though it can then be written as $Q \ket{\alpha}$
[\POL ],
it should not be considered a trivial state because the ket
$\ket{\alpha}$ involves the $X$ field, which is not
a conformal field.}
The states in \firstgr\ are more delicate. They are BRST
invariant and actually BRST trivial (the first one is
$-Q b_{-1} \bar{\alpha}_{-1}^\mu \ket{+-}$, for example).
They cannot arise in the right hand side of \ruleout\
because of Lorentz invariance of the string field theory.
Since the right hand side is built of string products, which
are manifestly Lorentz invariant, an $\alpha^\mu$ oscillator
can only appear contracted with a momenta. Since all momenta
are zero, it cannot appear at all. This concludes our proof
that all states appearing in the right hand side of \ruleout\
will satisfy the conditions of the lemma.
\medskip
The first nontrivial equation to solve is
$$Qb_0^- \ket{\Psi_2} = -{1\over 2} \ket{\phi_A^2}\eqn\firstnt$$
where $\ket{\phi_A^2} = \ket{\phi_A \star \phi_A}$. Indeed,
$Q$ acting on the right hand side vanishes on the account that
$Q$ acts as a derivation on the star product, and the fact
that $\ket{\phi_A}$ is BRST invariant. Now this is solved by
$$\eqalign{
b_0^- \ket{\Psi_2} &={b_0^+ \over 2L_0^+}
\ket{\phi_A \star \phi_A},\cr
{}&= {b_0^+ \over 2L_0^+}
\sum_r < f_1 \circ \Phi_r^c(0)\,
f_2 \circ (b_0^- \phi_A(0))\,
f_3 \circ (b_0^- \phi_A(0))\, > \,b_0^-\ket{\Phi_r},\cr
{}&= b_0^+ \sum_r
{1 \over 2L_{0r}^+}{\cal A}\,(\Phi_r^c , \phi_A ,\phi_A) \,
b_0^- \ket{\Phi_r}.\cr}\eqn\secsol$$
Here we have used the conformal field theory definition
of the string product (see [\Se ]).
The bra $\bra{\Phi_r^c}$ denotes the state conjugate to
$\ket{\Phi_r}$. This means that
$\VEV{\Phi_r^c \vert \Phi_s}=\delta_{rs}$.
Note how ghost number works.  The star product of two fields of
ghost number three must give a field of ghost number three; indeed,
it follows from the expression that $\Phi^c$ must have ghost number
two (to add up to six in the correlator), and thus $\Phi$ has ghost
number four, and finally $b_0^- \Phi$ has ghost number three.
Note that while our earlier arguments show that
$b_0^-\ket{\Phi_r}$ must be BRST trivial, that does not imply
that $\Phi_r^c$ is a BRST trivial operator, thus the correlator
above does not vanish. The correlator, with the string field
theoretic functions $f_i$ telling us how to insert the states,
is giving us the off-shell amplitude for scattering of the
marginal operators into the $\Phi^c$ operator, and we have
denoted, for simplicity, the correlator, by the letter ${\cal A}$.
The sum over $r$ runs over all states of the Hilbert space,
satisfying $L_0^- = 0$, but  only states $\ket{\Phi_r}$ of
ghost number four, that are not annihilated by $b_0^{\pm}$
will contribute. If $\ket{\Phi_r}$
$= \alpha_{-1}^\mu a_{\mu\nu} \bar{\alpha}_{-1}^\nu \ket{++}$,
which is the BRST nontrivial marginal perturbation, it better
be that the correlator with $\Phi^c_r$ vanish. Here
$\ket{\Phi_r^c}$
$=\alpha_{-1}^\mu a_{\mu\nu} \bar{\alpha}_{-1}^\nu \ket{--}$
and the correlator is simply (since all fields are on shell)
$C_{rAA}$, which is required to vanish in the conformal field
theory.

We now generalize. Consider the equation for $\ket{\Psi_3}$.
We solve it as
$$\eqalign{
b_0^- \ket{\Psi_3} &= {b_0^+\over L_0^+}
\ket{\Psi_1 \star \Psi_2} + {b_0^+\over 3! L_0^+}\ket{\Psi_1^3},\cr
{}&= {b_0^+\over L_0^+} {1\over 3!} \left(
3\ket{\phi_A \star c_0^- {b_0^+\over L_0^+}
\ket{\phi_A \star \phi_A}} + \ket{\phi_A^3} \right).\cr}\eqn\threee$$
But it is clear now that the quantity inside parenthesis is
building a four point amplitude, the first term corresponding to
the three Feynman diagrams with an intermediate propagator,
as the presence of $(b_0^+/L_0^+)$ indicates, and the last term
being the product that defines the four point function. Thus
the above result can be written as
$$
b_0^- \ket{\Psi_3} = b_0^+ \sum_r {1\over 3!\,L_{0r}^+}
{\cal A}\,(\Phi_r^c, \phi_A , \phi_A ,\phi_A) \,
b_0^- \ket{\Phi_r},
\ee$$
where ${\cal A}$ here denotes the off-shell four external state
amplitude (with the integral over moduli space understood) calculated
using the string diagrams of the corresponding string field theory.
It is clear what is now the complete generalization. The final
solution is therefore
$$b_0^- \ket{\Psi}
= b_0^- \epsilon\ket{\phi_A} +
b_0^+ \sum_{n\geq 2 , r} {\epsilon^n\over n!}
{1\over L_{0r}^+} {\cal A}\, (\Phi_r^c ,
\underbrace{\phi_A, \phi_A, \cdots \phi_A}_{n})\,b_0^-\ket{\Phi_r} .
\eqn\ufffinal$$
where ${\cal A}$ denotes the off-shell scattering amplitude
(summed over moduli space) for the field $\Phi_c^r$ with
$n$ marginal fields.
This formula suggests that to every order in $\epsilon$
the string field components of the solution are finite.
It is nice that classical solutions and off-shell amplitudes
are related like this, it indeed indicates that a good
string field theory must have
nice off-shell structure. The question of whether or not the finite
classical solution is in the Hilbert space of the theory becomes
just the issue of convergence of the whole series.

\REF\VeVe{E. Verlinde and H. Verlinde,
``Lectures In String Perturbation Theory", Trieste School of
Superstrings, April 1988.}

\chapter{Relation with First Quantization}

\def\ubar#1{\underline{\ \ }\!\!\!\!#1}
\def\XX{{\ubar{X}}}
\def\QQ{{\underline Q}}
\def\PP{{\ubar{P}}}
\def\DXX{{\cal D}\!\XX}
\def\DQQ{{\cal D}\!\QQ}
\def\DPP{{\cal D}\!\PP}

It is well-known that the usual
path-integral expression for the partition function of
a free scalar living on a circle of radius $R$ ($X \equiv X+2\pi R$)
must be multiplied by a radius dependent factor, if one
wishes to have a duality invariant expression [\VeVe ].
Namely, the amplitude at the $L$-loop level is
given by
$$
Z_L(R) = g^{-2}(g^2R)^L\int _{x:\ {\rm fixed}}\DX\,\exp(-S[X])\ ,
\eqn\eci
$$
where $S[X]$ is the usual sigma model action
$$
S[X] =
 {1\over \pi }\int d^2z\,\partial X(z,\bar z)\bar\partial X(z,\bar z)
\eqn\ecii
$$
with $ d^2z\equiv dzd\bar z/(-2i)=dtd\sigma $.
The path-integral contains one trivial
integration $\int _0^{2\pi R} dx$ over the zero mode $x$ of
the position $X$ at a time. Due to translational invariance
the integrand does not depend on $x$ and one gets
a factor of $R$ that has been extracted explicitly in \eci
(thus the label ``${x:\ {\rm fixed}}$").

For $d$-dimensional compactification, the formulas
\eci\  and \ecii\ are replaced by
$$
\eqalign{
&Z_L(E) = g^{-2}(g^2\rtG)^L
\int _{x:\ {\rm fixed}}(\rtG\DX) \exp(-S[ X; E ]) \cr
&S[ X; E ] = {1\over \pi }\int d^2z\,\partial X(z,\bar z)\,E^t\,
\bar\partial X(z,\bar z)\,.
\cr}\eqn\esigmaaction
$$
Here $X\equiv X + 2\pi$, and the path-integration
measure $(\rtG\DX)$ means that {\it each} integration has a
volume factor $\rtG$. (This is due to the different
periodicity conventions in  \eci\ and \esigmaaction .)

The purpose of this section is to show that the prefactor $R^L$
or $\rtG^{\,L}$ in these path-integral expressions automatically
appears in the string field theory and does imply neither
coupling constant change nor dilaton condensation.
To show this, we first discuss the path-integral for a point particle
moving on a torus, which contains some of the features of the
string case. We then turn to strings moving on a torus, and to
higher loop amplitudes.

\REF\Kashiwa{T.~Kashiwa \journal Int.~J.~Mod.~Phys. &A5 (89) 375.}

\section{Particle Moving on a Target Space Torus\foot{
{\rm The authors learned the derivation of the path-integral formula
presented in this subsection from T.~Kashiwa, whom they would like to
thank. See also Ref. [\Kashiwa ]. }}}

We consider a particle on a $d$ dimensional target space torus,
namely, the position $x^i$ ($i=1,2,\cdots ,d$) of the particle
obeys the identifications $x^i  \equiv  x^i  + 2\pi$.
The hamiltonian will be given by.
$$
\opH= {1\over 2}\opp_i {G^{-1}}^{ij} \opp_j .
\ee
$$
Henceforth we omit vector indices: e.g.,
$xGx=x^i G_{ij}x^j $.
{}From the periodicity of $x$, the momentum $p$ takes the integer
eigenvalues:
$$
\opp \ket{n} = n\ket{n},\qquad n=(n_i)\in {\bf Z}
\ee
$$
with normalization and
completeness relations
$$
\VEV{n\vert m} = \delta _{n,m}, \qquad \sum_n \ket{n}\bra{n} = 1.
\ee
$$
The coordinate eigenstate, however, cannot be defined by
$\opx \ket{x} = x \ket{x}$, since the eigenvalue $x$ is defined
only modulo the periodicity and hence the operator $\opx$
is not a well-defined operator.
As we explained in \eqxeigen, we define the coordinate
eigenstate via momentum eigenstate as
$$
\ket{x} \equiv  \sum_n {e^{-inx}\over \sqrt{(2\pi )^d}}\, \ket{n}.
\ee
$$
Then the state label $x$ actually becomes the label of
the point on the torus, satisfying
$\ket{x} = \ket{x+2\pi e^{(i)}}$
($e^{(i)}$ : a unit vector in the $i$ direction).
The inner product of the coordinate eigenstates is given by
the periodic delta function,
$$
\VEV{x\vert y} = \sum_n{1\over (2\pi )^d}\,e^{in(x-y)} \,
=\sum_m\delta (x-y+2\pi m) \equiv  \bfdelta (x-y),
\ee
$$
and the completeness relation reads
$$
\int _{C}dx\, \ket{x}\bra{x} = 1, \qquad
({\rm torus}\ {C}: \ 0\leq x^i <2\pi ).
\ee
$$
\medskip
Let us now derive the path-integral formula for the
transition amplitude
$$
{\cal T} = \bra{x_F} e^{-\opH T} \ket{x_I},\qquad
(0\leq x_F,x_I<2\pi ).
\ee
$$
For an infinitesimal time interval $\Delta t\equiv T/(N+1)$
$(N\gg 1)$, we evaluate the
transition amplitude as follows:
$$
\eqalign{
\bra{x_{j+1}} e^{-\opH \Delta t} &\ket{x_j}
= \sum_{n_j}\VEV{x_{j+1}\vert n_j}\bra{n_j}
 e^{-\opH \Delta t} \ket{x_j}  \cr
{}&= \sum_{n_j}{1\over (2\pi )^d}
\exp(-{1\over 2}n_jG^{-1}n_j\Delta t) \,e^{in_j(x_{j+1}-x_j)} \cr
{}&= \int {dp_j\over (2\pi )^d}
\exp\left(-{1\over 2}p_jG^{-1}p_j\Delta t+ip_j(x_{j+1}-x_j)\right)
\sum_{n_j}\delta (p_j-n_j)  \cr
{}&= \int {dp_j\over (2\pi )^d} \sum_{m_j}
\exp\left(-{1\over 2}p_jG^{-1}p_j\Delta t
+ip_j(x_{j+1}-x_j+2\pi m_j)\right) ,
\cr}\ee
$$
where in the last step we used Poisson's formula to replace a sum
of delta functions with a sum of exponentials.
We thus get for the transition amplitude ${\cal T}$,
$$
\eqalign{
{\cal T} &= \lim_{N\rightarrow \infty }
\left(\prod_{j=1}^N \int _{C} {dx_jdp_j\over (2\pi )^d}\sum_{m_j}
 \right)
\left( \int  {dp_0\over (2\pi )^d}\sum_{m_0} \right)  \cr
&\ \ \ \ \ \ \ \ \ \times
\exp\left\{\Delta t\bigg[
\sum_{j=0}^Nip_j{x_{j+1}-x_j+2\pi m_j\over \Delta t}
-{1\over 2}p_jG^{-1}p_j\bigg]\right\} ,
\cr}\ee
$$
where $x_0=x_I$ and $x_{N+1}=x_F$. Now we define the following new
coordinates:
$$
\tilde x_0 = x_0;\quad \tilde x_j \equiv  x_j + 2\pi \ell_j,
\qquad \ell_j\equiv \sum_{k=0}^{j-1} m_k,\quad
(j=1, \cdots , N+1).
\ee
$$
The sums over $m_j$ ($j=0,\cdots N$) can now be traded for
sums over the $\ell_j$ ($j=1, \cdots ,N+1$),
and the $x$-integration region, restricted to the torus,
is combined with the $\ell$ sums so that the restriction to the torus
disappears for $\tilde x$:
$$
\int _{C} dx_j \sum_{\ell_j} = \int _{-\infty }^\infty d\tilde x_j
\qquad {\rm for}\ \ j=1,2,\cdots ,N.
\eqn\einteg
$$
Denoting $\ell_{N+1} \equiv n$, ($\ell_{N+1}$  was not used in
\einteg\ ) the transition amplitude becomes
$$
\eqalign{
{\cal T} &= \lim_{N\rightarrow \infty } \sum_{n=-\infty }^\infty
\left(\prod_{j=1}^N \int _{-\infty }^\infty  {d\tilde x_jdp_j\over
(2\pi
 )^d} \right)
 \int  {dp_0\over (2\pi )^d}  \cr
&\ \ \ \ \ \ \
\times \exp\left\{\Delta t\bigg[\sum_{j=0}^Nip_j{\tilde x_{j+1}-\tilde
 x_j\over \Delta t}
-{1\over 2}p_jG^{-1}p_j\bigg]\right\} ,
\cr}\eqn\eqtramp
$$
with boundary conditions $ \tilde x_0 = x_I$ and
$ \tilde x_{N+1} = x_F + 2\pi n$.  If we path-integrate out the
momentum variables, we obtain
$$
\eqalign{
{\cal T} &= \lim_{N\rightarrow \infty } \sum_{n=-\infty }^\infty
\left(\prod_{j=1}^N \int _{-\infty }^\infty
{\rtG\,d\tilde x_j\over \sqrt{(2\pi \Delta t)^d}} \right)
\ {\rtG\over \sqrt{(2\pi \Delta t)^d}}\  \cr
&\ \ \ \ \ \ \ \times
\exp\left\{-\Delta t\bigg[\sum_{j=0}^N{1\over 2}
\left({\tilde x_{j+1}-\tilde x_j\over \Delta t}\right)\,G\,
\left({\tilde x_{j+1}-\tilde x_j\over \Delta t}\right)\bigg]\right\}\ .
\cr}\eqn\eformula
$$
In the limit $N\rightarrow \infty $, omitting the tilde of $x$,
one writes
$$
{\cal T} = \rtG\,\sum_{n=-\infty }^\infty
\INT{x(0)=x_I}{x(T) = x_F +2\pi n}
(\rtG\,\D{x})
\exp\left\{-\int _0^Tdt\ {1\over 2}\dot x(t)\,G\,\dot x(t)\right\}  \ ,
\eqn\etransition
$$
where the prefactor $\rtG$ came from the factor
$\rtG/\sqrt{(2\pi \Delta t)^d}$
and the singular factor $1/\sqrt{(2\pi \Delta t)^d}$ was
omitted, as usual in the path-integral formulas.
This is the desired formula for the particle moving on a torus. The
prefactor $\rtG$ in this formula will play a key role below.
It is useful to understand in simple terms how the main features of
\etransition\ arise. For every interval $\Delta t$, we introduced
 an integral over $p$ and a sum over $m$. The number of intermediate
$x$ integrations, however, is one less than the number of intervals,
thus, when reassembling the result one is left with an extra sum over
$m$, giving rise to the winding of the particle as it moves in time,
and an extra $p$ integral that gives rise to the $\sqrt{G}$ prefactor.

\section{String Moving on a Target Space Torus}

Now we consider a string moving on a target space $d$ dimensional
torus. The dynamical variable is $X(\tilde{\sigma} )$
with identifications $X \equiv X+ 2\pi$ (the target space index
will be ommitted). This string coordinate is expanded as (see \fqviii\ )
$$
X(\tilde\sigma ) = x + w\tilde\sigma
+ ({\rm oscillators}).
\eqn\expandx
$$
Corresponding to the dynamical variable $x$ we have the momentum
operator $\hat{p}$, and corresponding to the operator $\hat{w}$, whose
eigenvalues $w$ appear above, we have the dynamical variable $q$.
In operator language, the dynamics of the string will be determined
by
$$
\eqalign{
 L + \bar L &= {1\over 2}(\opp-B\opw)G^{-1}(\opp-B\opw) +
{1\over 2}\opw G\opw + ({\rm oscillators}), \cr
L-\bar L &= -\opp\opw + ({\rm oscillators}) \ , \cr
}\ee
$$
where the first term corresponds to the generator of time
translations, and the second is the generator of rotations
of the string. Our expressions will concentrate on the
zero-mode pieces, the full expressions will be written
when necessary. Using the ($p$,$w$) eigenstate, given by
$$
\eqalign{
&\opp \ket{n,m} = n \ket{n,m}, \qquad
\opw \ket{n,m} = m \ket{n,m},  \cr
& \VEV{n,m\vert k,\ell} = \delta _{n,k}\delta _{m,\ell} , \cr
}\ee
$$
the $x$- and $q$-eigenstates are defined by
$$
\eqalign{
\ket{x,m} &\equiv  \sum_n\ {e^{-inx}\over \sqrt{(2\pi)^d}}\,
 \ket{n,m}, \cr
\ket{n,q} &\equiv  \sum_m\ {e^{-imq+i\pi nm}\over\sqrt{(2\pi)^d}}\,
 \ket{n,m}, \cr
\ket{x,q} &\equiv  \sum_{n,m}\ {e^{-inx-imq+i\pi nm}\over (2\pi )^d}\,
 \ket{n,m}, \cr
}\eqn\eqxqeigenstate
$$
and again, $x$ and $q$ become  labels on the points on tori of
unit radii: e.g.,
$\ket{x,q} = \ket{ x+2\pi e^{(i)}, q+2\pi e^{(j)} }$.
Note that the $q$-eigenstate is defined with an additional sign
factor $\exp(i\pi nm)$, in order
to compensate for the asymmetry of the vertex
due to the cocycle factor. [See the end of this section.]
Ommission of this sign factor would not affect the results of
the present subsection.

Let us evaluate the partition function
$$
Z_{T,\theta } \equiv  \tr \left( e^{-(T-i\theta )L-(T+i\theta )\bar L}
 \right) \,
= \tr \,e^{-\calH T},
\ee
$$
where $\calH $ takes the form, omitting the oscillator parts,
$$
\calH  =  {1\over 2}\,
(\opp-B\opw)G^{-1}(\opp-B\opw) +
{1\over 2}\opw G\opw + i\varphi \opp\opw \ ,
\ee
$$
with $\varphi \equiv  \theta /T$.
Evaluation of $Z_{T,\theta }$ can be done in four ways by
using either momentum- or coordinate-representations:
$$
\eqalign{
Z_{T,\theta } &= \sum_{n,m}\bra{n,m}e^{-\calH T}\ket{n,m} \cr
  &= \int _{C}dx\,\sum_{m}\,\bra{x,m}e^{-\calH T}\ket{x,m} \cr
  &= \int _{C}dq\,\sum_{n}\,\bra{n,q}e^{-\calH T}\ket{n,q} \cr
  &= \int _{C}dx\int _{C}dq\,\bra{x,q}e^{-\calH T}\ket{x,q}\ .
 \cr
}\ee
$$
Duality is manifest in the first expression. It is also
manifest in the last expression, which
also leads to an interesting path-integral expression as we shall
see at the end of this section.
To reach the $x$-space sigma model path-integral expression,
it is quickest to start with the second representation.
Following the same procedure as in particle
case, we find
$$
\eqalign{
Z_{T,\theta } =&  \big(\int _{C}\rtG dx \big) \sum_{n=-\infty
 }^\infty
\INT{x(0):\ {\rm fixed}}{x(T) = x(0)+2\pi n}
(\rtG\,\D{x})\sum_{m=-\infty }^\infty  \,  \cr
 &\ \times \exp\left\{-\int _0^Tdt\,\big[
{1\over 2}(\dot x(t)-\varphi m)\,G\,(\dot x(t)-\varphi m)
+ {1\over 2}mGm -i\dot x Bm \big]\,\right\}\ \
\cr}\eqn\estringPI
$$
A few comments are in order. The sum over $m$, corresponding
to the winding of the string at any time, is the same sum we
began with, since the hamiltonian is diagonal in winding
eigenstates (as well as momentum eigenstates). The integral
over $x$ is also the same one we started with; the extra factor
of $\sqrt{G}$ arises because of an unmatched $p$ integration,
as in the particle case. The winding in time, described
by the integer $n$, arises from an unmatched $m_j$ sum, as in
the particle case.

Let us now translate this result into a string path integral
with a sigma model lagrangian.  Note that the coordinate
$\tilde\sigma$ in $X(\tilde\sigma ,\tilde t )$
is a co-moving coordinate fixed to
the string, which is different from the coordinate
$\sigma \equiv  {\rm Re}\,z$ on the Riemann surface with metric
$ds^2 = \abs{dz}^2$. The relation is
$$
\sigma  = \tilde \sigma  + \varphi t \qquad
( \varphi = {\theta \over T}).
\eqn\relcoord$$
(Recall the usual description of a torus with moduli
$2\pi\tau =\theta +iT$ on
the complex $z$-plane as the paralellogram with corners ($0, 2\pi ,
\theta +iT , 2\pi + \theta + iT$))
Therefore, the above together with \expandx\ gives us
$$
X(t,\sigma ) = x(t) + w\,(\sigma -\varphi t) + (\hbox{oscillator
 modes}),
\eqn\epath
$$
so that $\dot X(t,\sigma ) =$
$\dot x(t) - \varphi w +$ oscillator modes; and $X'(t,\sigma ) = w$.
Then we see that,
when the oscillator modes are taken into account, the action
functional in the exponent in eq.\estringPI\ takes the form:
$$
-{1\over 2\pi }\int _0^Tdt\int _0^{2\pi }d\sigma \,\big[
{1\over 2}\dot X\,G\,\dot X + {1\over 2}X'GX' -i\dot X B X' \big]\ .
\eqn\eexponent
$$
This is just identical with $i$ times the sigma model action $S$
given in (2.1) with
$\gamma ^{\alpha \beta } = \eta ^{\alpha \beta }$,
if we go back to the original 2D Minkowski world sheet with
identification $t=i\tau $ ($\tau $: Minkowski time).
If we use the complex coordinate (in Euclidean space),
$z = t + i\sigma , \  \bar z = t -i\sigma$,
then the above \eexponent\ is also seen to agree with
the action given in \esigmaaction:
$$
-S[ X; E ] = -{1\over \pi }\int _{0\leq t\leq T}d^2z\,
\partial X(z,\bar z)\,E^t\,\bar\partial X(z,\bar z)\,.
\ee
$$
\noindent
Noting that the boundary condition $x(T) = x(0) + 2\pi n$ implies,
via \epath,
the condition $X(T,\sigma +\theta ) =$ $X(0,\sigma )+2\pi n$
for the string coordinate,
we finally find that eq.\estringPI\ gives the following
path-integral expression for the string partition function:
$$
Z_{T,\theta } = \big(\int _{C}\rtG dx \big) \sum_{n,m}
\INT{X(T,\sigma +\theta ) = X(0,\sigma )+2\pi n}{X(t,2\pi )
=X(t,0)+2\pi m}(\rtG\,\DX)' \exp(-S[ X; E ]) \ ,
\eqn\etorusPI
$$
where the prime in $(\rtG\,\DX)'$ means that the integration over
the CM coordinate $x(0)=X(0,0)$ is omitted.
Note that the factor $\rtG$ appeared as promised.
This proves the formula
\esigmaaction\ for the $L=1$ case.

\section{Higher Loop Amplitudes}

The amplitudes at any loop level are constructed in SFT by
the products of vertices connected by the string propagators.
Each propagator is written in the form:
$$
{{\cal P}\over L + \bar L} = \int _0^\infty dT\,e^{-T(L+\bar L)}\cdot
\int _0^{2\pi }{d\theta \over 2\pi }\,e^{i\theta (L-\bar L)}\,
= \int _0^\infty dT\int _0^{2\pi }{d\theta \over 2\pi }\,e^{-\calH T},
\eqn\sftprop$$
where ${\cal P}$ is the projector to rotational invariant states,
and $\calH $ was defined before. The amplitude corresponding to
a Riemann surface with definite
moduli is given by the product of
``finite moduli propagators" $e^{-\calH T}$ and
vertices $\bra{V}$.

We now consider a generic loop diagram with no external legs. The
expression for this amplitude given by SFT will be converted into
an $X$-space sigma model path integral. From the
previous two results, \etransition\ for the point particle
transition amplitude and \etorusPI\  for the string partition
function, we see that the ``finite moduli propagator"
$e^{-\calH T}$ is given by
$$
\bra{X_F} e^{-\calH T} \ket{X_I} =  \rtG \,\sum_n
\INT{X(0,\sigma )=X_I(\sigma )}{X(T,\sigma +\theta ) = X_F(\sigma )
+2\pi n}
(\rtG\,\DX) \exp(-S[ X; E ]) \ .
\eqn\ecx
$$
There is also the boundary condition $X(t,2\pi ) = X(t,0) + 2\pi m$,
where $m$ is the winding number of $X_I(\sigma )$ (or $X_F(\sigma )$).
There is no sum $\sum_m$ because the winding must remain fixed.
One should note that this propagator is associated with
the factor $\rtG$.

Now we come to the vertex.
Again the essential part is the zero-modes $p$ and $w$, so let us
concentrate on those modes alone.
Generally any $N$-string
vertex is of the form
$$
\bra{V_N} \ \sim \  \sum_\bfn\sum_\bfm
\delta (\sum_{r=1}^N n_r)\delta (\sum_{r=1}^N m_r)
\prod_{r=1}^N \bsub{r}\bra{n_r,m_r}
\ee
$$
for the relevant zero-mode parts, where $n_r$ and $m_r$ denote
eigenvalues of $p$ and $w$ as before and $\bfn\equiv (n_r),\ \bfm\equiv
 (m_r)$.
In the $x$-representation the basis $\bra{n}$ are
Fourier transformed into the $x$ basis using
$\VEV{n\vert x} = e^{-inx}/\sqrt{(2\pi )^d}$ and the vertex becomes
a vertex function coupling several wavefunctions:
$$
\intT\prod_{r=1}^Ndx_r\VEV{V_N\vert \bfx,\bfm} \sim
\intT\prod_{r=1}^Ndx_r \sum_\bfn \sum_\bfm
\delta (\sum_{r=1}^N n_r)\delta (\sum_{r=1}^N m_r)
\exp(-i\sum_rn_rx_r) .
\ee
$$
The integrations $\intT dx_r$ came from the insertion of
completeness relation $1=$ \nextline $\intT dx\ket{x}\bra{x}$.
The summation over $\bfn$ with conservation factor
$\delta (\sum_{r=1}^N n_r)$ gives \nextline
$\prod_{r=1}^{N-1}\bfdelta(x_r-x_N)$ up to irrelevant factors of
$\sqrt{(2\pi )^d}$.  Multiplying $1=\intT dx\bfdelta(x-x_N)$,
the vertex function becomes
$$
\intT\prod_{r=1}^Ndx_r\VEV{V_N\vert \bfx,\bfm} \sim  \intT dx \cdot
\bigg[\prod_{r=1}^N\intT dx_r \bfdelta(x_r-x)\bigg]
\cdot \sum_\bfm \delta (\sum_{r=1}^N m_r)
\eqn\ecxii
$$
Note that essentially a single integral $\intT dx$ exists
at each vertex,
since all the other integrals over $x_r$ are trivial; they simply
set $x_r=x$. This $x$ is the position of the vertex.

We are now almost finished.  As we saw in \ecx,  each
propagator has a prefactor $\rtG $, so a factor $\rtG^{\,P}$
appears for a Feynman diagram with $P$ propagators.
At each vertex, however, there is an integration
$\intT dx$, which is to be included as
a part of the path-integral over the 2D world
sheet spanned by the diagram. But the integration measure in the
path-integral is $(\rtG\,\D{x})$ and accordingly the integral
$\intT dx$ at each vertex should be multiplied by $\rtG $ so as to
construct the path-integral correctly. Since for each vertex
we need a factor $\rtG$, the overall left over factor
relating the $L$-loop diagram
with $P$ propagators and $V$
vertices to a sigma model $x$-path integral is
$$
(\rtG)^{P-V} =(\rtG)^{L-1},
\ee
$$
and thus we end up with the
$$
(g^2\rtG)^{(L-1)}\int (\rtG\DX) \exp(-S[ X; E ]).
\ee
$$
Note, however, that this path-integral still contains an
integration $\intT\rtG dx$ over the zero mode $x$ of the $X$
coordinate (at a time)
on which the action does not depend. So extracting that
factor we finally obtain the following expression for the general
$L$-loop amplitude
$$
Z_L(E) = g^{-2}(g^2\rtG)^L
\int _{x:\ {\rm fixed}}(\rtG\DX) \exp(-S[ X; E ]),
\ee
$$
and finish the proof of \esigmaaction.

\section{Dual Sigma Models}

In the above we derived the sigma model path-integral
expression \esigmaaction\ from string field theory.
The final expression is very asymmetric from the
viewpoint of duality.  But note that
the starting set up of string field theory in the
$p$-$w$ momentum representation is manifestly
dual-symmetric (aside from the cocycle factor in the vertex).
In particular, the operator $\calH $ in the propagator satisfies
the duality relation
$$
\calH(P,X';E)
= \calH(P_Q, Q'; \tilde E) \ ,
\eqn\eduality
$$
where $\tilde E$ is the dual background $\tilde E = E^{-1}$ and
$$
2\pi P_Q(\sigma ) \equiv  X'(\sigma ), \qquad
Q'(\sigma ) \equiv  2\pi P(\sigma ) \ .
\ee
$$

Therefore the aparent asymmetry in the above path-integral formula
resulted simply because we chose the $x$-coordinate representation.
In fact, we could have chosen the $q$-representation by
Fourier-transforming the $w$-eigenstates $\bra{m}$ but keeping the
momentum representation for $p$-freedom. Then, as is clear from the
duality relation \eduality, we would have obtained the following
$q$-space sigma model path-integral formula for the {\it same}
$L$-loop amplitude:
$$
Z_L(E) = g^{-2}(g^2\sqrt{\tilde G})^L
\int _{q:\ {\rm fixed}}(\sqrt{\tilde G}\,\D{Q})
\exp(-S[ Q; \tilde E ]).
\eqn\eqPIinQ
$$
Note also that $\sqrt{\tilde G}= 1/\sqrt{G-BG^{-1}B}$.

One might notice here that the dual coordinate $Q(\sigma)$ does not
connect smoothly on our vertex, as the Goto-Naka conditions
\eqGN\ shows, and
wonder what happened in obtaining the $q$-sigma model
path-integral formula \eqPIinQ. The asymmetry in the  $X(\sigma)$
and $Q(\sigma)$ connection conditions is a reflection of the
asymmetry in the vertex cocycle factor under
the exchange $p\ \leftrightarrow\ w$.
But this asymmetry is compensated by the additional sign factor
$\exp(i\pi nm)$ put in the definition of $\ket{n,q}$ eigenstate
\eqxqeigenstate, and we can get the same vertex factors for this $q$
case as for the $x$ case and obtain \eqPIinQ.
The reason why this happens is easy to understand: putting the sign
factor $\exp(i\pi nm)$ in \eqxqeigenstate\ is equivalent to giving the
coordinate $q$ the meaning that it stands for the eigenvalue of
the operator $Q(\sigma)+\pi p$ instead of $Q(\sigma)$. But the operator
 $Q(\sigma)+\pi p$ is just the coordinate which is smoothly
connected (mod $2\pi$) on our vertex as is seen in the Goto-Naka
conditions \eqGN.

Finally in this subsection, let us comment on a manifestly
dual-symmetric sigma model which automatically results if we use
coordinate representations both for the $p$ and $w$ degrees of freedom.
Consider the following (Minkowskian)
transition amplitude in the $x,q$-coordinate
representation:
$$
{\cal T} = \bra{x_F,q_F,\XX_F} e^{-i\opH T} \ket{x_I,q_I,\XX_I},\qquad
(0\leq \,x_F,x_I, q_F,q_I\,<2\pi ).
\ee
$$
The $\XX$ denotes  $X(\sigma)$ with the zero-mode parts omitted. We are
considering the $\theta =0$ case, for simplicity, and then $\calH$ reduces
to $\opH$ given in (2.7). Performing the same procedure as in the
particle case to reach \eqtramp\ for the $x,q$, and $\XX$ degrees of
freedom, and
using the expression of $\opH$ in (2.7), we clearly obtain
$$
{\cal T} =  \sum_{n,m=-\infty }^\infty
 \INT{x(T)=x(0)+2\pi n}{q(T)=q(0)+2\pi m}\!
\calD x \calD p \calD q \calD w \DXX \DPP  \ \exp(iS),
\ee$$
with an action functional $S$ given by
$$
S =
\int_0^T dt \int_0^{2\pi}{d\sigma \over 2\pi }\bigg[
2\pi \PP \dot\XX +p\dot x + w\dot q -
 {1\over 2}(X',2\pi P) {\cal R} (E) \pmatrix{X'\cr 2\pi P \cr}\bigg].
\ee$$
It is amusing to note that this action takes
a manifestly dual-symmetric form if we use the $Q(\sigma )$ coordinate
defined in \eqdefq\ instead of $P(\sigma )$ and perform a suitable
partial integration:
$$
\eqalign{
S &=
\int_0^T \!dt \int_0^{2\pi}{d\sigma \over 2\pi }\,\bigg[\
 (w, p) J  \pmatrix{ \dot x \cr \dot q \cr} +
{1\over 2} (\XX',\QQ') J
\pmatrix{ \dot \XX \cr \dot \QQ \cr} \cr
& \qquad \qquad \qquad \ \ \ \ \ \
-{1\over 2} (X',Q') {\cal R} (E) \pmatrix{X'\cr Q'\cr}\bigg] \cr
&\ \quad + \hbox{surface term}\ , \cr}
\eqn\eqsymmeact$$
where $J$ is the \Odd metric matrix
$\mymatrix{0}{1}{1}{0}$. As for the non-zero mode parts,
this action happens to coincide exactly
with the dual-symmetric action which
was proposed by Tseytlin [\Tseytlindual ]
some time ago. But there are some differences
for the zero-mode parts; for instance, the $\sigma $-linear terms
$\dot p\sigma $ in $\dot X$ and $\dot w\sigma $ in $\dot Q$
do not appear here while they did in Ref. [\Tseytlindual ].
The surface term in \eqsymmeact, which appeared as a result of partial
integration, is given by
$$
\int_0^{2\pi }{d\sigma \over 2\pi }\,{1\over 2}
\bigg[\XX\QQ'\bigg]_{t=0}^{t=T}
= \int_0^{2\pi }d\sigma \,{1\over 2}\big[
\PP(\sigma ,T)\XX(\sigma ,T) - \PP(\sigma ,0)\XX(\sigma ,0)
\big]\ .
\ee$$
This is not dual-symmetric but it simply reflects the asymmetry of
the initial and final states, specified by the $\XX$ eigenvalues.
It should be noted that the path-integral measure also takes the
dual-symmetric form $
\calD x \calD q \calD w \calD p \DXX \DQQ$.

\REF\RocekVerlinde{M. Rocek
and E. Verlinde, ``Duality, Quotients and Currents", IAS preprint,
IASSNS-HEP-91/68, October 1991.}
\REF\hikkoii{H. Hata, K. Itoh,
T. Kugo, H. Kunitomo, and K. Ogawa
\journal Prog. Theor. Phys. &77 (87) 443.}

\chapter{String Field Duality Transformations}

In this section we begin by deriving discrete symmetry
transformations of the string field that are invariances
of the string action. They arise due to the physical
equivalence of string field theories written around dual
backgrounds, plus the existence of classical solutions
that connect those dual backgrounds. We verify
that they generate the discrete group of dualities
\Oddz. This full group of symmetry transformations exists
for any possible background $E$, and it leaves the action
invariant. All symmetries, except those corresponding to
group elements $g$ that leave the background invariant
($g(E) = E$), are spontaneously broken.
Dine et. al. [\DHS ] anticipated from conformal field
theory arguments that duality must correspond to global
gauge transformations in a field theory description. This
result was generalized by Giveon et. al. [\GMR ]
for the case of generalized discrete dualities. We will
indeed show that the discrete symmetries we have obtained,
arise mostly from the string field gauge group.

While in string field theory duality turns naturally into
a symmetry transformation of the string field leaving the action
invariant and existing for all backgrounds, in conformal field theory
duality is generically thought as a relation between two
apparently different conformal field theories that are
actually identical. A general way to obtain dual sigma models
corresponding to the same conformal field theory by starting
with a self-dual sigma model and gauging different combinations
of chiral currents has been given recently [\RocekVerlinde ].
\bigskip
Let us now begin our derivation by finding the
discrete global symmetry corresponding
to a generic \Oddz transformation $g$. We have shown that
corresponding to any such group element there is a unitary
operator $\U_g$ such that for any background
$E$ one has $S_E (\U_g \Psi ) = S_{g(E)}(\Psi)$,
or equivalently
$$S_E (\Psi ) = S_{g(E)}(\U_g^{\dagger}\Psi).\eqn\udaggpsi$$
(While we will write, for brevity, the string field as a functional,
it is convenient to think of it as a ket, in order to use the
equations derived earlier.)  Consider now the classical
solution $\Psi (E;g(E))$. We have
established in section 5, equation \shiftaction\ that
$$S_E (\Psi (E;g(E)) + \Psi ) = S_{g(E)} (\Psi ),\eqn\shiftb$$
namely, that the classical solution shifts the theory precisely
to the final background. It follows from the above two equations
that
$$S_E (\Psi (E;g(E)) + \U_g^{\dagger}\Psi)\, =\,
S_{g(E)}(\U_g^{\dagger}\Psi)\, = \, S_E (\Psi),\eqn\dersymm$$
which shows that $S_E$ is invariant under the following
string field discrete transformation
$${\cal D}_g : \, \Psi \,\rightarrow \,{\cal D}_g \Psi
\, \equiv \,\Psi (E;g(E)) +
\U_g^{\dagger} \Psi .\eqn\dsymm$$
The discrete symmetry transformation ${\cal D}_g$ is the
symmetry we were after. It consists of an inhomogeneous term,
given by the classical solution, plus a homogeneous term
in which the operator $\U$ acts on the field. The symmetry
is spontaneously broken unless the first term vanishes,
and this only happens if the background $E$ is invariant
under $g$. Let us derive now the group properties of the
discrete transformations, consider a further discrete
transformation
$${\cal D}_{g'} : \, \Psi \,\rightarrow \,{\cal D}_{g'} \Psi
\, = \,\Psi (E;g'(E)) +
\U_{g'}^{\dagger} \Psi .\eqn\dsymmi$$
and now consider
$$\eqalign{
{\cal D}_{g'}{\cal D}_g \Psi &= \, {\cal D}_{g'}
[\Psi (E;g(E)) + \U_g^{\dagger} \Psi ],\cr
{}&= \Psi (E;g(E)) + \U_g^{\dagger} [\Psi (E;g'(E)) +
\U_{g'}^{\dagger} \Psi ],\cr
{}&= \Psi (E;g(E)) + \U_g^{\dagger} \Psi (E;g'(E))
+\U_g^{\dagger} \U_{g'}^{\dagger} \Psi \cr}\eqn\algebd$$
In order to simplify further we note that classical solutions
have a simple behaviour under the action of $\U$:
$$\U_g^{\dagger} \Psi (E_0;E_1) = \Psi (g(E_0);g(E_1)),
\eqn\uonclas$$
as one easily verifies using equation \ebxxxii\ (note that the
phase factor in $\U^{\dagger}$ is irrelevant because the classical
solution ket has
zero momentum and zero winding). It thus follows
that \algebd\ simplifies to
$$\eqalign{
\quad\quad\quad\quad &=
\Psi (E;g(E)) + \Psi (g(E);gg'(E))
+ \exp (-i\pi {\cal C}(p,g',g)) \,\U_{gg'}^\dagger \Psi \cr
{}&= \Psi (E;gg'(E)) +
+\exp (-i\pi {\cal C}(p,g',g)) \,\U_{gg'}^\dagger \Psi,\cr
{}&=  \exp (-i\pi {\cal C}(p,g',g)) \,{\cal D}_{gg'} \Psi,\cr}
\eqn\algebdi$$
which shows that the second quantized operators
${\cal D}$ satisfy the algebra
$${\cal D}_{g'}\,{\cal D}_g \, = \,
\exp (-i\pi {\cal C}(p,g',g)) {\cal D}_{gg'}.\eqn\algebdii$$
Note that the action of the operators ${\cal D}$
on the string field is background dependent, it
depends on $E$ via the classical solution. Operators
${\cal D}_g$ referring to different backgrounds are simply
related by a shift in the string field. The algebra of the
operators is clearly background independent.

A natural question that comes to mind is whether these
operators commute with gauge transformations of the
string field theory. We represent the gauge transformations
as
$${\cal G}(\Lambda ): \Psi \, \rightarrow \,
{\cal G}(\Lambda ) \Psi \,  \equiv \,
\Psi + Q(E)\Lambda +
g_0 \Psi \star \Lambda ,\eqn\gaugetr$$
(note that $g_0$ is the coupling constant) one can show that
$${\cal D}_{g^{-1}}\, {\cal G}(\Lambda) \, {\cal D}_g
= {\cal G}(\U_g^\dagger \Lambda ),\eqn\gtsym$$
where use was made of \actstar, and of the equation
$$Q(g(E)) + g_0 \Psi (g(E);E)\, \star  = Q(E).\eqn\qactdef$$

Equation \gtsym\ shows
that the discrete symmetries generate automorphisms
of the gauge group.  This suggests strongly that the discrete
symmetries correspond to large gauge transformations.
In the remaining of this section we will show explicitly
how this is obtained in the string field theory for the
case of the standard $R \rightarrow 1/R$ duality. This
will illustrate how the conformal field theory arguments
of Ref. [\DHS ]  apply. For the case of the more general
symmetry transformations one may not have a background
that they leave invariant, and the arguments of [\DHS ]
do not tell us what is the connection with gauge transformations.
For example, the composition of two discrete transformations,
each having a fixed point background, may not have a fixed
background (in the space of backgrounds we are considering).
In this case, however, it is clear that the resulting
transformation is a gauge transformation, which is never
unbroken, but can be identified at any background.
Reference [\GMR ] shows that this is essentially the generic case,
and that all discrete symmetries can be written as products
of symmetry transformations at special backgrounds with extended
symmetry, plus permutations of spacetime coordinates.
These permutations are clearly symmetries of string field
theory, but it is not clear to us if they
belong to the string field gauge group. The complication
arises because we
only know the infinitesimal string field
gauge transformations, and permutations cannot
be built from infinitesimal rotations, due to the
compactification of the extra coordinates.

The standard duality inversion is defined by the
\Oddz matrix $g_D$ given by
$$g_D = \pmatrix{0&I \cr I&0\cr}.\eqn\standdual$$
It follows that det$g_D=(-1)^d$. Acting on backgrounds the
transformation $g_D$ is recognized to give
the well-known action on backgrounds (Ref. [\GRV] ), indeed
$$E' = G' +B' = g_D(E) = [0E+I][IE + 0]^{-1} =
E^{-1} = (G+B)^{-1}.\eqn\dualback$$

The background invariant under the duality trasformation
is $E=I$, and we will therefore discuss string field theory
around it. The oscillators corresponding to this background
will be simply denoted as $\alpha ,\bar{\alpha}$ and the
operator $\U_{g_D}$ will just be denoted as $\U$. It follows from
equations \upsildef\ and \cbfdef\ that
$$\U= U\exp (i\pi p \cdot w )\eqn\udefdual$$
and the action of $\U$ on the oscillators and zero modes
is given by (see \defuop\ )
$$\U^{\dagger} \pmatrix{\alpha_n \cr \bar{\alpha}_n \cr}
\U = \pmatrix{-\alpha_n \cr \bar{\alpha}_n \cr},\quad
\U^{\dagger} \pmatrix{w \cr p \cr}\U =
\pmatrix{p \cr w \cr}\eqn\uactosc$$
which says that all the $\alpha_n$ oscillators, including
$n=0$, are changed sign, and the bar oscillators are left
unchanged. In more geometrical terms
$$\U^{\dagger} \pmatrix{ X(\sigma ) \cr Q(\sigma) \cr}\U
=\pmatrix{Q(\sigma ) \cr X(\sigma ) \cr}.\eqn\uactcoord$$

It is convenient to introduce the general decomposition
$X^i (\sigma ) = X_+^i (\sigma ) + X_-^i(\sigma )$
with
$$\eqalign{
X_+^i (\sigma ) &= x_+^i - G^{ij}p_{+j}\sigma
+ {i\over \sqrt{2}} \sum {1\over n} \alpha^i_n e^{in\sigma}\cr
X_-^i (\sigma ) &= x_-^i + G^{ij}p_{-j}\sigma
+ {i\over \sqrt{2}} \sum {1\over n} {\bar\alpha}^i_n
e^{-in\sigma}\cr}
\eqn\separx$$
where the momentum zero modes $p_+ ,p_-$ are given by
$$\eqalign{
p_{+i} &= {1\over 2} (p_i - E_{ij} w^j),\cr
p_{-i} &= {1\over 2} (p_i +E^t_{ij} w^j),\cr}
\eqn\sparp$$
which for the case at hand ($E= I$) reduce to
$$p_{\pm i} = {1\over 2} (p_i \mp w^i).\eqn\sdspm$$
The mass-shell
conditions read
$$ {1\over 2} M^2 = N + \bar{N} + p_+^2 + p_-^2 -2,
\quad N- \bar{N} = p_-^2 - p_+^2 ,\eqn\masshel$$
and we will denote the momentum eigenstates
by $\ket{p_+ , p_-}$. It is well known that at
this background one has an $SU(2)^d \otimes
SU(2)^d$ symmetry. The gauge bosons for the
$SU(2) \otimes
SU(2)$ that arises from
the $i$-th curled coordinate are associated to the massless states
$$\eqalign{
{}&\alpha_{-1}^\mu \ket{{0}, \pm {k}^i},\quad
\alpha_{-1}^\mu {\bar\alpha}_{-1}^i \ket{0,0}\cr
{}&{\bar\alpha}_{-1}^\mu \ket{\pm {k}^i,{0}},
\quad {\bar\alpha}_{-1}^\mu \alpha_{-1}^i \ket{0,0}\cr},
\eqn\gaugebos$$
where the ${k}^i$ is a $d$-component vector whose $i$-th
entry is $+1$ and all others are zero.
Now we want to find the global tranformations associated with
such gauge bosons. From the standard string field gauge
transformations
$$
\eqalign{
\delta (\bm\sfd)& = \QB \bm\ket{\Lambda } + g_0\ket{\Psi *\Lambda } \cr
  {\rm with}&\ \ \ket{\Psi *\Lambda }_1
    \equiv \bsub{1'23}\bra{V} \ket{R}_{11'}\sfd_2\ket{\Lambda }_3. \cr
}$$
we must require, in order to have an unbroken symmetry, that
$\QB \bm\ket{\Lambda } = 0$, and for the symmetry to be global
the momentum for the open coordinates $p_\mu = 0$, which implies
$M^2 = 0$. Moreover, the ghost number of $\bm\ket{\Lambda}$
must be $-1$ (with respect to the vacuum state $\ket{0}$).
In order to get this ghost number we need an antighost
oscillator, and the only two possibilities are $b_{-1}$
and $\bar{b}_{-1}$ ($b_0^{\pm}$ annihilates $\ket{0}$, and
$b_{-n}$ is ruled out since it cannot give a massless state).
It follows now from the mass-shell conditions that the desired
states are given by
$$\eqalign{
\bm\ket{\Lambda^i_\pm} &=
\bar{b}_{-1}\ket{\pm {k}^i, 0},\quad
\bm\ket{\Lambda^i_3} = \bar{b}_{-1} \alpha_{-1}^i \ket{0,0}\cr
\bm\ket{{\bar\Lambda}^i_\pm} &=
b_{-1}\ket{{0}, \pm {k}^i},\quad
\bm\ket{{\bar\Lambda}^i_3} = b_{-1} {\bar\alpha}_{-1}^i \ket{0,0}\cr}.
\eqn\gaugepar$$
Here we must take the string length $\alpha$ equal to zero.
One easily verifies that the above states are BRST invariant.
The gauge transformations associated with these gauge parameters
are given by
$$
\delta(\bm\sfd) = g_0\ket{\Psi *\Lambda }
= -{g_0\over \sqrt2}\, E\, \bm\ket{\Psi} ,\eqn\globtr$$
where the operator $E$ arises from the contraction of
$\Lambda$ against the vertex. The calculation of the operator
$E$ is familiar from Hata et. al. [\hikkoii ] and is explained
in Appendix C.  One obtains
$$\eqalign{
E_{\pm}^i &= {e^{i\pi pk^i}\over \sqrt2} \int {d\sigma \over 2\pi}
: \exp (\pm 2i {k}^i \cdot X_+ (\sigma )): ,\quad
E^i_3 = p_+ = \alpha_0^i/\sqrt2,\cr
\bar E_{\pm}^i
&= {e^{i\pi pk^i}\over \sqrt2} \int {d\sigma \over 2\pi}
: \exp (\pm 2i {k}^i \cdot X_- (\sigma )): ,\quad
\bar E^i_3 = p_- = \bar\alpha_0^i/\sqrt2,\cr
}\eqn\genstra$$
As is easily confirmed, these operators give generators of the gauge
group $SU(2)^d \otimes SU(2)^d$; e.g.,
$ [E_+^i, -E_-^j] = E_3^i\delta ^{ij},\
 [E_3^i, E_{\pm }^j] = \pm E_{\pm }^j\delta ^{ij}$.
If we define
$$
E_{\pm, n}^i = {e^{i\pi pk^i}\over \sqrt2} \int {d\sigma \over 2\pi}
e^{-in\sigma }: \exp (\pm 2i {k}^i \cdot X_+ (\sigma )): ,
\ee$$
we then have $[ \alpha_n^i/\sqrt2, \,E_\pm^j ]$
$= \pm E_{\pm, n}^i \delta^{ij}$, and this
implies that the operators $(\alpha_n^i/\sqrt2 ,$
 \nextline $E_{\pm, n}^i)$ form
a spin one representation of the $SU(2)$ we are considering.
Thus via a global rotation we can indeed make
$\alpha_n \rightarrow -\alpha_n$. This shows our $\U$
operator performing the duality rotation is just a global $SU$(2)
gauge transformation.

\chapter{Conclusions and Open Questions}

We believe that string field theory, as presently formulated,
is powerful enough to give useful insights into the basic issues
of target space duality. As we have seen it affords a manifestly
dual formulation of the theory, where basic physical facts, such
as the invariance of the string coupling constant are completely
clear. The string field picture explains the origin of the discrete
symmetries as a simple consequence of the facts that two different
backgrounds lead to the same physics, and that there are classical
solutions shifting us from one background to the others.

The most important questions left open by our work have to do
with background independence of string field theory and classical
solutions. Our notion of universal coordinates $X(\sigma )$ and
$P(\sigma )$ is basically the idea that these are field operators
whose existence is independent of the background and whose
(field) algebra is always the same. The various backgrounds
correspond to inequivalent representations of this unique
algebra. In this way we learned how to relate
different theories corresponding to different backgrounds,
and how to write operators in one background in terms
of operators in another background. One feels that there
should be more understanding of how this fits together with
studies of deformations of conformal field theories, and
possibly with geometrical approaches to the study of the
space (or subspaces) of conformal field theories. Our notion
of universal coordinates applies only to conformal field
theories with two-dimensional field theory Lagrangians. It
is not clear to us how to extend these ideas to conformal
field theories described more abstractly in terms of their
operator content.

One of the most puzzling aspects of our results is the
indication that classical solutions correponding to finite
changes in the background may not live in the conventional
Hilbert space of the theory. If this is really the case,
the idea of component fields loses meaning beyond
perturbation theory, and a classical string field solution
will not correspond to a classical solution for the
component fields. It would also mean that we need to learn
how to define string field theory for a class of
functional fields larger than the conventional one, which
corresponds to Fock space states. As a way to test these
ideas we explored a recursive solution of the string field
equations, in the spirit of Ref. [\MuSe ]. The solution
is written as an infinite series of vectors in the Hilbert
space of the original theory. For this finite solution to
make sense the series must converge. Each term of the series
corresponds to an off-shell amplitude of the string field
theory, and we hope it will be possible
to reach a conclusion on the issue of convergence in the
near future.
\bigskip

\ack
We are happy to acknowledge useful conversations with
H.~Hata, T.~Kashiwa, E.~Kiritsis, M.~Maeno,  A.~Giveon, M.~Douglas,
D.~Gross, D.~Kutasov and E.~Verlinde.

B.~Zwiebach wishes to acknowledge the hospitality of the Yukawa
Institute for Theoretical Physics, where most of this work was done,
and the hospitality of the Institute for Advanced Study, where this
work was finished.
T.~K. is supported in part by the Grant-in-Aid for Cooperative
Research (\# 02302020) and the Grant-in-Aid for Scientific Research
(\# 02640225) from the Ministry of Education, Science and Culture.
B.~Z. is supported in part by D.O.E. grant DE-AC02-76ER03069 and
NSF grant \#PHY91-06210.
\endpage

\APPENDIX{A}{A. \ Quantities appearing in the vertex of
the \ahikko theory}

Here we give some explicit expressions for the quantities which
appear in the three-string vertex \vertex\ or \exvii\
of the \ahikko theory:
$$
\eqalign{
E_{123}^{\rm ordinary} &=
 \sum_{r,s}\sum_{n,m\geq 0}\bar N^{rs}_{nm} \left(
  {1\over 2}\alpha _n^{\mu (r)}\eta _{\mu \nu }\alpha _m^{\nu (s)} +
  i\gamma _n^{(r)}\beta _m^{(s)} + \ {\rm a.h.}\ \right)
\cr
\noalign{\vskip 0.6cm}
F_{123}^{\rm ordinary} &=
 {1\over 2}\sum_{r,s}\sum_{n,m\geq 0}
  \bar N^{rs}_{nm} \alpha _n^{\mu (r)}\eta _{\mu \nu }\alpha _m^{\nu
 (s)}  \cr
&\ \ \ \ \ +
 \sum_{r,s}\sum_{n,m\geq 1}\bar N^{rs}_{nm}
  i\gamma _n^{(r)}\beta _m^{(s)} + \ {\rm a.h.parts}
\cr
&= \sum_{r,s}\sum_{n,m\geq 1}\bar N^{rs}_{nm} \left(
  {1\over 2}\alpha _n^{\mu (r)}\eta _{\mu \nu }\alpha _m^{\nu (s)} +
  i\gamma _n^{(r)}\beta _m^{(s)} + \ {\rm a.h.}\ \right)
\cr
&\ \ \ \ \ +
 {1\over \sqrt 2}\sum_r\sum_{n\geq 1}\bar N_n^r(\alpha _n^{\mu (r)}+
\bar \alpha _n^{\mu (r)})
\eta _{\mu \nu }{P}^\nu  + \tau _0 \sum_r {1\over p^+_r}
{p_r^2\over 2}
\cr}\eqn\appi$$

$$
\eqalign{
&\gamma _n^{(r)}=inp_r^+\crn, \qquad \beta _n^{(r)}=\brn/p_r^+  \cr
\noalign{\vskip 0.3cm}
&{P}^\mu =p^+_rp_{r+1}^\mu -p^+_{r+1}p_r^\mu ,  \cr
\noalign{\vskip 0.3cm}
&\mu _{123}=\exp \left(-\tau _0\sum_{r=1}^3(1/p^+_r)\right)
\quad \tau _0=\sum_{r=1}^3p^+_r\ln \abs{p^+_r}, \cr
}\eqn\exix
$$
$$
\eqalign{
&G(\sigma _I) = {p^+_r\over 2}\sum_{n=-\infty }^\infty (\crn +\bar\crn)
\cos n\sigma _I^{(r)}, \quad r= \hbox{1 or 2 or 3}
\cr
\noalign{\vskip 0.5cm}
&W_I^{(r)}= -{i\over \sqrt 2}\sum_s\sum_{n\geq 1}
\left(\chi ^{rs}\bar N^s_n+\sum_{m=1}^{n-1}\bar N^{ss}_{n-m,m}/p^+_r
\right) (\gamma _n^{(s)}+ \bar \gamma _n^{(s)}), \cr
}\eqn\eqghost
$$
The Neumann coefficients $\Nrsnm$ and $\Nrn$ as well as coefficients
 $\chi ^{rs}$ are the same as defined by [\hikkoi ] with the
understanding that $\alpha =p^+$.

When $\alpha _3\equiv \epsilon $ becomes very small compared with
$\alpha _2$ and
$\alpha _1=-(\alpha _2+\epsilon )$, the measure factor $\mu ^2_{123}$
has a singularity
$$
\mu ^2_{123} = \left({e\alpha _2\over \epsilon }\right)^2
\big(1+ O(\epsilon )\big).
\eqn\eqmulimit
$$
In the calculations of string field condensation in Sect.~5,
we need several formulas showing
how the various quantities in the vertex behave in this limit.
Such detailed formulas can be found in Hata and Nagoshi [\HaNa ].
Here we only cite
$$
\eqalign{
\bar N^{3r}_{1n} &= \left(\epsilon {{\rm sgn}(\epsilon \alpha _2)
\over e\alpha _2}\right)
\times \cases{ 1  & for $r=1$ \cr
          (-1)^{n+1}  & for $r=2$ \cr}, \qquad (n\geq 1),
\cr
\bar N^{3r}_{10} &= \left(\epsilon {{\rm sgn}(\epsilon \alpha _2)
\over e\alpha _2}\right)
\times \cases{ 1  & for $r=1$ \cr
          0  & for $r=2$ \cr}, \cr}
\eqn\eqlimit
$$
which will be used in deriving Eq. \ebxxiii . The latter formula for
the $n=0$
case is valid only in the presence of zero-mode conservation factor.

\APPENDIX{B}{B. \
Physical equivalence of $\alpha =p^+$ HIKKO and light-cone SFTs}

In this appendix
we explain why the \ahikko theory correctly reproduces the
light-cone string field theory amplitudes at any loop order.
Of course, this is the case only for processes with external
states of physical polarizations.

The vertex in the {\it gauge-fixed} \ahikko theory takes the form
$$
\bsub{123}\bra{v} = \mu ^2_{123}\delta (1,2,3)\bsub{123}\bravac
  \exp(F_{123})  {\cal P}_{123}
\ee $$
For notational simplicity we consider the case in which all
the coordinates are uncompactified.
Then the exponent $F_{123}$ in the vertex is the same as
$F_{123}^{\rm ordinary}$ given in \appi.
An important fact is that, when $\alpha =p^+$, the momentum
$$
{  P}^\mu =\alpha _rp_{r+1}^\mu -\alpha _{r+1}p_r^\mu
\ee $$
appearing in $F_{123}$ does not contain the + component:
${P}^+ =p_r^+p_{r+1}^+ -p_{r+1}^+p_r^+ = 0$.
Moreover, when $\alpha =p^+$,
the momentum-square term in \eqi\ becomes purely transversal:
$$
 \tau _0 \sum_r {1\over p^+_r}{p_r^2\over 2} =
 \tau _0 \sum_r {1\over p^+_r}{\bfp_r^2\over 2} +
 \tau _0 \sum_r p_r^-
=\tau _0 \sum_r {1\over p^+_r}{\bfp_r^2\over 2} ,
\ee $$
due to the conservation of $p^-$.
We will use boldface letters to denote transverse vectors.
Now the exponent of the vertex takes the form
$$
F_{123} = F_{123}^{\rm LC} +
 F_{123}^{\rm extra}
\ee $$
where the first part $F_{123}^{\rm LC}$ is exactly the same one
as in the light-cone SFT,
$$
\eqalign{
F_{123}^{\rm LC} &=
 \sum_{r,s}\sum_{n,m\geq 1}\bar N^{rs}_{nm} \left(
  {1\over 2}\bfalpha _n^{(r)}\cdot
   \bfalpha _m^{(s)} + \ {\rm a.h.}\ \right)
\cr
&\ \ \ \ \ +
 {1\over \sqrt 2}\sum_r\sum_{n\geq 1}\bar N_n^r
(\bfalpha _n^{(r)}+\bar \bfalpha _n^{(r)})
\cdot {\bf P}  + \tau _0 \sum_r {1\over \alpha _r}
{\bfp_r^2\over 2} \cr
}\ee
$$
and the second part $F_{123}^{\rm extra}$
contains the extra modes
$\alpha _n^+, \alpha _n^-, \gamma _n, \beta _n$
of the covariant theory:
$$
\eqalign{
 F_{123}^{\rm extra} &=
 \sum_{r,s}\sum_{n,m\geq 1}\bar N^{rs}_{nm} \left(
  \alpha _n^{+(r)}\alpha _m^{-(s)} +
  i\gamma _n^{(r)}\beta _m^{(s)} + \ {\rm a.h.}\ \right)
\cr
&\ \ \ \ \ +
 {1\over \sqrt 2}\sum_r\sum_{n\geq 1}\bar N_n^r
(\alpha _n^{+(r)}+\bar \alpha _n^{+(r)}){  P}^- .
\cr}
\ee $$
Writing schematically
$ F_{123}^{\rm extra} =
  \alpha^{+}N\alpha^{-}
 + i\gamma N\beta + \alpha^+P^-
 + \ {\rm a.h.}$,
the vertex takes the form
$$
\bra{v} = \bra{v_{\rm LC}} \otimes \bsub{\rm EX}\bravac
e^{  \alpha^{+}N\alpha^{-}
 + i\gamma N\beta + \alpha^{+}P^-
 + \ {\rm a.h.} }.
\ee
$$
Here $\vac_{\rm EX}$ denotes the vacuum for the modes
$\alpha _n^+, \alpha _n^-, \gamma _n, \beta _n$ and $\bra{v_{\rm LC}}$
is just
the vertex appearing in the light-cone SFT.
It is also important to remember that the
$\alpha^{+}\alpha^{-}
+ i\gamma\beta$ term has $OSp$(1,1$\vert$2) symmetry.

The general (tree or loop) amplitude in this theory is calculated
by evaluating an expression of the form
$$
\calM =  \big(\prod \int d\ell\big)\big(\prod \bra{v}\big)
\big(\prod e^{-L\tau -\bar L \bar \tau }\big)
\big(\prod \ket{R}\big)
\ket{{\rm external}}
\eqn\eqamplitude $$
where $\bra{v}, \ket{R}, \ket{{\rm external}}$ are vertices,
reflectors and external states, respectively, \nextline
$e^{-L\tau -\bar L \bar \tau }$ are propagators with definite moduli
and  $\prod \int d\ell$ stands for the integration over the loop
momenta $\ell$. The physical
external states are constructed by using
the DDF modes $A_n^i$ alone which are given by
$$
A_n^i = \oint {dz\over 2\pi i}z^{n-1}(\sum_m \alpha ^i_m z^{-m})
 \exp\Big(-{n\over p^+}\sum_{\ell\not=0}{1\over \ell}\alpha _{\ell}^+
 z^{-\ell}\Big) .
\ee $$
So we write the physical external state schematically as
$$
\ket{{\rm external}} = \ket{\varphi_{\rm LC}}
 \otimes  e^{-\alpha ^{+\dagger}z}\vac_{\rm EX} ,
\ee $$
where the state
$\ket{\varphi_{\rm LC}}$ is a state written in terms of transverse
modes alone which reduces to the same state as in the light-cone SFT
after $z$-integration if
the factor $e^{-\alpha ^{+\dagger}z}$ can be replaced by 1.
The Klein-Gordon-Virasoro operator $L$ is written as a sum of that
of the light-cone SFT and an extra piece:
$$
\eqalign{
L &= L_{\rm LC} + L_{\rm extra}, \cr
L_{\rm LC} &=
 {1\over 2}\bfp^2 + p^+p^- + \sum_{n\geq 1}
\bfalpha _{-n}\cdot \bfalpha _n ,
\cr
L_{\rm extra}&= \sum_{n\geq 1}
(\alpha ^+_{-n}\alpha ^-_n + \alpha ^-_{-n}\alpha ^+_n +
 i\gamma _{-n}\beta _n  - i\beta _{-n}\gamma _n) .
\cr}
\ee $$
We write again schematically
$$
L_{\rm extra} = \alpha ^{+\dagger}\alpha ^-  + \alpha ^{-\dagger}\alpha ^+
  +i\gamma ^\dagger \beta  + i\gamma \,\beta ^\dagger .
\ee $$
The reflector in the gauge-fixed theory is given by
$$
\eqalign{
\bsub{12}\bra{R} &= \delta (1,2)\bsub{12}\bravac
\exp{(E_{12})}{\cal P}_{12}, \cr
E_{12} &=
(-)^{n+1}\sum_{n\geq 1}({1\over n}
  \alpha _n^{(1)}\cdot \alpha _n^{(2)}+
  i\gamma _n^{(1)}\beta _n^{(2)}+
  i\gamma _n^{(2)}\beta _n^{(1)} ) + \ah \cr
}\ee $$
Again we write the ket reflector
schematically
$$
\ket{R} = \ket{R_{\rm LC}} \otimes
 e^{(\alpha ^{+\dagger}\alpha ^{-\dagger} + i\gamma \beta )}
\vac_{\rm EX}
\ee $$
where $\ket{R_{\rm LC}}$ is the reflector in the light-cone SFT.
Note again that the extra mode parts of $L$ and the reflector are
\Osp invariant.

Now we can evaluate the amplitude \eqamplitude: substituting the above
schematic expressions for the external states, reflectors, vertices and
$L$, we find
$$
\calM = \big(\prod \int d\ell \big) \calM_{\rm LC}\cdot
\calM_{\rm extra}
\ee$$
where
$$
\calM_{\rm LC} =
\big(\prod \bra{v_{\rm LC}}\big)
\big(\prod e^{-L_{\rm LC}\tau -\ah }\big)
\big(\prod \ket{R_{\rm LC}}\big)
\ket{\varphi_{\rm LC}}
\ee$$
is the amplitude in the light-cone SFT before the loop-integration,
and $\calM_{\rm extra}$ is the similar one for the extra modes which
can be schematically written in the following form (omitting the
anti-holomorphic parts):
$$
\eqalign{
\calM_{\rm extra} &=
\big(\prod\bsub{\rm EX}\bravac
e^{  \alpha^{+}N\alpha^{-}
 + i\gamma N\beta + \alpha^{+}P^- }\big)
\big(\prod e^{-(\alpha ^{+\dagger}\alpha ^-
+ \alpha ^{-\dagger}\alpha ^+
  +i\gamma ^\dagger \beta  + i\gamma \,\beta ^\dagger)\tau  } \big)
\cr
&\qquad \qquad \qquad  \times
\big(\prod
 e^{(\alpha ^{+\dagger}\alpha ^{-\dagger} + i\gamma \beta )}
\vac_{\rm EX}\big)
  e^{-\alpha ^{+\dagger}z}\vac_{\rm EX} .
\cr}\eqn\eqEX
$$

Let us evaluate this amplitude $\calM_{\rm extra}$
for extra mode part. We claim that the momentum dependent factor
$\exp(\alpha^{+}P^-)$ in the vertex can be set equal to one.
This is seen as follows: since the $\alpha ^{+}$ oscillators are
contracted
with $P^-$  and have non-zero commutator only with
$\alpha ^{-\dagger}$, which in turn appears in
\eqEX\ contracted only with $\alpha ^+$,
or $\alpha ^{+\dagger}$, the momenta $P^-$ must appear always in the form
$\alpha ^{+}P^-$ or $\alpha ^{+\dagger}P^-$
at any stage of the calculation of \eqEX.  But those
oscillators are eventually elliminated on the bra or ket vacuum.
Thus the terms containing a $P^-$ factor can give no contribution to the
amplitude $\calM_{\rm extra}$, and we can set $P^-$ equal to zero
in \eqEX. (Note that, if there were a term of the form
$\alpha ^{-\dagger}K^+$ or $\alpha ^{-}K^+$ with some momentum $K^+$,
then the term
$\alpha ^{+}P^-$ could have given a finite contribution proportional
to $P^-K^+$.) For the same reason we can set the factor
$\exp(-\alpha ^{+\dagger}z)$
in the external state equal to one. Thus the amplitude
$\calM_{\rm extra}$ becomes
$$
\eqalign{
\calM_{\rm extra} &=  \bsub{\rm EX}\bravac
\big(\prod e^{\alpha^{+}N\alpha^{-} + i\gamma N\beta}\big)
\cr
&\qquad \ \times
\big(\prod e^{-(\alpha ^{+\dagger}\alpha ^-
+ \alpha ^{-\dagger}\alpha ^+
  +i\gamma ^\dagger \beta  + i\gamma \,\beta ^\dagger)\tau  } \big)
\big(\prod
 e^{(\alpha ^{+\dagger}\alpha ^{-\dagger} + i\gamma \beta )}\big)
\vac_{\rm EX}.
\cr}\ee
$$
Note that this is completely \Osp symmetric. Therefore it has to be one,
since whatever factor is given by the $\alpha ^+, \alpha ^-$
oscillators, it is
cancelled by the contribution of the $\gamma , \beta $ oscillators.
We thus find
$$
\calM = \big(\prod \int d\ell \big) \calM_{\rm LC} .
$$
This coincides with the amplitude in the light-cone SFT. (Recall
that the external states also reduced to the light-cone ones since
the factor $\exp(-\alpha ^{+\dagger}z)$ was replaced by one.)
Namely we have proven that the physical amplitudes
in the \ahikko theory indeed agree
 with those in the light-cone SFT.

A comment may be in order. In the light-cone SFT there is only
propagation forward in time due to the structure of the kinetic
term plus the independence of the vertices on $p^-$, which
implies locality in the light cone time. The kinetic term
in the \ahikko theory has the same structure as the light cone
theory, and we have shown that the $P^-$ dependence of the
vertices dissapeared for physical amplitudes.
Therefore the string diagrams for physical amplitudes agree.

\APPENDIX{C}{C. \ Derivation of Eqn. \ebxxiii\ and Eqn. \genstra\ .}

First we briefly explain how Eq. \ebxxiii\ is derived.
We have to evaluate
$$
\bsub{123}\bra{V}\ket{\hbox{EM}}_3 = \lim_{\epsilon \rightarrow 0}
\bsub{123}\bra{V} c_0^{-(3)}
\left[ \alpha _{-1}^i(E)a_{ij}\bar \alpha _{-1}^j(E)
\right]^{(3)} \vac_3 \delta _\epsilon (p_3,w_3),
$$
where integrations (or summations) over $p_3$ and $w_3$ are implied.
We omit the background label $(E)$ from $\alpha _n^i(E)$ henceforth.
Since the ghost oscillator dependence is trivial here, we
first calculate the ghost zero-mode part
substituting the vertex expression \vertex\ and find
$$
\eqalign{
\bsub{123}\bra{V}\ket{\hbox{EM}}_3 &=
\lim_{\epsilon \rightarrow 0} \mu ^2_{123}\delta (1,2,3)\bsub{123}
\bravac
  {1\over \alpha _1\alpha _2}(\alpha _1c_0^{+(1)}-\alpha _2c_0^{+(2)})
\exp(E_{123}) \cr
&\ \ \ \ \times
 G(\sigma _I) e^{-i\pi (p_3 w_2-p_1 w_1)}{\cal P}_{123}
( \alpha _{-1}^ia_{ij}\bar \alpha _{-1}^j )^{(3)}
\vac_3 \delta _\epsilon (p_3,w_3).
\cr}\eqn\eqkk
$$
where use was made of Eqn. \eqv\ and the $b$ zero modes in the
vertex were moved towards the vacuum on the left.
Since $\mu ^2_{123} \sim  \left(e\alpha _2/\epsilon \right)^2 $,
we have to evaluate the rest of the expression to $O(\epsilon ^2)$.
But we have
$$
\bsub{123}\bravac \exp(E_{123})
( \alpha _{-1}^ia_{ij}\bar \alpha _{-1}^j)^{(3)} \vac_3
= \bsub{12}\bravac \exp(E'_{12})\SUM{r,s=1,2}{n,m\geq 0}
\bar N^{3r}_{1n}\bar N^{3s}_{1m}
 \alpha _{n}^{i(r)}a_{ij}\bar \alpha _{m}^{j(s)}
\eqn\eqapiv
$$
where use was made of Eqn. \messy , and with $E'_{12}$
denoting $E_{123}$ with string-three oscillators eliminated.
Since the Neumann coefficient factor $\bar N^{3r}_{1n}\bar N^{3s}_{1m}$
is already of $O(\epsilon ^2)$ as is seen in \eqlimit, we have only to
calculate the $O(1)$ part for all the other quantities in Eq.\eqkk.
Then $\exp(E'_{12})$ becomes the exponent $\exp(E_{12})$ of the 2-point
vertex $\bsub{12}\bra{R}$ in (3.6) and so we have
$$
\lim_{\epsilon \rightarrow 0} \delta (1,2)\bsub{12}\bravac
  {1\over \alpha _1\alpha _2}
(\alpha _1c_0^{+(1)}-\alpha _2c_0^{+(2)})\exp(E'_{12})
 e^{i\pi p_1 w_1}
= -{1\over \alpha _1} \bsub{12}\bra{R'} \,b_0^{-(1)}\ ,
\ee
$$
where $\bra{R'}$ denotes the reflector, but without the
rotational projector ${\cal P}_{12}={\cal P}^{(1)}{\cal P}^{(2)}$.
The ghost prefactor $G(\sigma _I)$ in \eqkk\ yields in this limit
$$
G(\sigma _I) = {\alpha _1\over 2}\sum_{\ell=-\infty }^\infty
(\col +\bar\col) ,\eqn\eqapv
$$
using Eq.(A.3) and the fact that the interaction point
$\sigma _I^{(1)}$ for string one becomes
zero for $\alpha _3\ \rightarrow \ 0$.
Now using Eqs. \eqapiv, \eqapv, \eqmulimit\ and \eqlimit,
we find that Eq.\eqkk\  becomes
$$
\eqalign{
\bsub{123}\bra{V}\ket{\hbox{EM}}_3 &=
 \bsub{12}\bra{R'}\,
{1\over 2}\sum_{\ell=-\infty }^\infty (\col +\bar\col)
\cr
&\quad \times \Big[
\sum_{n,m\geq 0} \alpha _n^{(1)}*\bar\alpha _m^{(1)}
+\sum_{n,m\geq 1} (-)^{n+m}\alpha _n^{(2)}*\bar\alpha _m^{(2)}
\cr & \qquad
+\SUM{n\geq 0}{m\geq 1} (-)^{m+1}\alpha _n^{(1)}*\bar\alpha _m^{(2)}
+\SUM{n\geq 1}{m\geq 0} (-)^{n+1}\alpha _n^{(2)}*\bar\alpha _m^{(1)}
\Big] b_0^{-(1)}{\cal P}_{12} , \cr}
\ee$$
with the abbreviation
$\alpha _n* \bar\alpha _m \equiv
\alpha _n^i a_{ij}\bar\alpha _m^j$.
We can now use the following continuity conditions on
$ \bsub{12}\bra{R'}$,
$$
 \bsub{12}\bra{R'}  \left(
\alpha _n^{(1)} + (-)^n\alpha_{-n}^{(2)},\
c_n^{(1)} + (-)^nc_{-n}^{(2)},\
b_n^{(1)} - (-)^nb_{-n}^{(2)} \right) \ = \ 0 ,
\eqn\eqrefcon
$$
and the analogous ones for the anti-holomorphic oscillators to find
$$
\eqalign{
\bsub{123}\bra{V}\ket{\hbox{EM}}_3
&=  \bsub{12}\bra{R'}\,
{1\over 2}\sum_\ell (\col +\bar c_\ell^{(1)})
\sum_{n,m} \alpha _n^{(1)}*\bar\alpha _m^{(1)}
 b_0^{-(1)}{\cal P}_{12} \cr
&=  \bsub{12}\bra{R}\,
{1\over 2}\sum_{\ell+n+m=0} (\col +\bar c_{-\ell}^{(1)})
(\alpha _n^{(1)}*\bar\alpha _{-m}^{(1)})
 b_0^{-(1)} . \cr}
\ee$$
In going to the second expression we have first moved
the operator ${\cal P}^{(2)}$ all the way up to $\bra{R'}$
and used $\bra{R'} {\cal P}^{(2)}= \bra{R'}{\cal P}_{12}$
$=\bra{R}$ as follows from \eqrefcon. Then the projector
${\cal P}^{(1)}$,
that appears actually both to the left and to the right
of the prefactor picks up only the terms in which the
separate mode number sums of the
holomorphic and anti-holomorphic oscillators are equal. The second
expression is seen to imply Eq. \ebxxiii\ after use of \eqrefcon.

\REF\Hataunp{H. Hata, Private
communication, unpublished.}

The calculation of dilaton condensation (\ebiii )
is somewhat
more complicated because of the presence of ghost oscillators
in the dilaton state.  This time it is easier to use
the vertex expression \exvii\ rather than \vertex .  We then
use the various $\epsilon $-expansion formulas for the Neumann
coefficients $\bar N_{nm}^{rs}$
and the coefficients in $W_I^{(r)}$ which are given in Ref. [\HaNa ].
We here only cite a particularly useful
formula which we learned from Hata [\Hataunp ]:
$$
\eqalign{
\bsub{123}\bravac \exp &(F_{123}) \vac_3\Big\vert_{p_3=w_3=0} \cr
&=\bsub{12}\bravac\exp(E_{12})
 \Big[ 1 -{\epsilon \over 2\alpha _1} {\sum_{n,m}}'
{1\over n+m}\big(\alpha _n^{(1)}\cdot \alpha _m^{(1)}+
2i\gamma _n^{(1)}\beta _m^{(1)} + \ah \big) \cr
&\qquad \ \ \qquad \qquad \qquad \
- {\epsilon \over \alpha _1}\sum_{n\not=0}
\big(\, : c_n^{(1)}b_{-n}^{(1)} : + \ah \big)
\ + O(\epsilon ^2) \Big] ,\cr}
\eqn\eqlast
$$
with
$$
\alpha _n\cdot \alpha _n \equiv
\alpha _n^{\mu }\eta _{\mu \nu } \alpha _n^{\nu} +
\alpha_n^{i}(E) G_{ij} \alpha_n^{j}(E) .
$$
Here $p_3$ is set equal to zero except for the $p_3^+=\alpha _3$
component,
of course, and $E_{12}$ is the exponent of the 2-point vertex
$\bsub{12}\bra{R}$.
The primed summation $\sum'_{n,m}$ means the summation excluding
the $n=m=0$ or $n+m=0$ terms for
$\alpha _n^{(1)}\cdot \alpha _m^{(1)}$ part and the $nm=0$ or $n+m=0$
terms for the
$2i\gamma _n^{(1)}\beta _m^{(1)}$ part. For our case (\ahikko theory)
the terms containing
$\alpha _0^+=\bar \alpha _0^+ = p^+/\sqrt2$ should also
be excluded from the summation.
The second \Osp asymmetric term in \eqlast\ comes from the $\epsilon$
difference between $\alpha_1$ and $-\alpha_2$ contained in
$\gamma^{(1)}\beta^{(2)}+\gamma^{(2)}\beta^{(1)}$.

Finally we explain how \genstra\ is derived.  For the case of the
$E^i_{\pm }$ generators we have to evaluate
$$
\eqalign{
\bsub{123}\bra{V}\ket{\Lambda ^i_{\pm }}_3 &=
\lim_{\epsilon =\alpha _3 \rightarrow  0}\bsub{123}\bra{V}c_0^{-(3)}
\bar b_{-1}^{(3)}\ket{\pm k^i,0}_3 \cr
&=
\lim_{\epsilon \rightarrow 0} \mu ^2_{123}\delta (1,2,3)\bsub{123}
\bravac\exp(F_{123})
(c_0^{+(1)}+{1\over \sqrt2}W_I^{(1)}) \cr
&\qquad \qquad  \times
(c_0^{+(2)}+{1\over \sqrt2}W_I^{(2)})
e^{-i\pi (p_1w_3-p_2w_2)}{\cal P}_{123}
\bar b_{-1}^{(3)}\ket{\pm k^i,0}_3 , \cr
}\eqn\eqfinal
$$
where we have used the vertex expression \exvii\ and the cyclic
symmetry of the vertex cocycle factor
$\exp(-i\pi (p_3w_2-p_1w_1))=\exp(-i\pi (p_1w_3-p_2w_2))$
for later convenience. Since the momentum $p_{+3}=\pm k^i$
is non-zero and the exponent $F_{123}$ contains a singular
$(\hbox{zero-mode})^2$ term
$\tau _0\sum_{r=1}^3(p_{+r}^2+p_{-r}^2)/\alpha _r$,
we have the factor
$$
\mu _{123}^2\exp(\tau _0\sum_{r=1}^3{p_{+r}^2+p_{-r}^2\over \alpha _r})
\ \sim  \  \mu _{123}\
\sim  \ {e\alpha _2\over \epsilon }{\rm sgn}(\epsilon \alpha _2) .
\ee $$
Since this is $O(1/\epsilon )$, we have to evaluate the other terms up
to $O(\epsilon )$. The oscillator $\bar b_{-1}^{(3)}$ can be
contracted with $F_{123}$ or $W_I^{(r)}$. But, since
$\bar b_{-1}^{(3)}=\epsilon \bar \beta _{-1}^{(3)}$ is already of
$O(\epsilon )$, the contraction with $F_{123}$ does not contribute.
The contraction with $W_I^{(r)}$ gives
$$
\eqalign{
\bsub{3}\bravac &
(c_0^{+(1)}+{1\over \sqrt2}W_I^{(1)})
(c_0^{+(2)}+{1\over \sqrt2}W_I^{(2)})
\bar b_{-1}^{(3)}\vac_3  \cr
&\qquad =
\left( c_0^{+(1)}+c_0^{+(2)}+
{1\over \sqrt2}({W'}_I^{(1)} + {W'}_I^{(2)}) \right)\,
{1\over 2}
\left({\epsilon \over e\alpha _2}{\rm sgn}(\epsilon \alpha _2)\right)
\cr}\ee
$$
by the help of expression (A.3) for $ W_I^{(r)}$ and the
related limiting formulas of Ref. [\HaNa ], where ${W'}_I^{(r)}$
denotes $W_I^{(r)}$ with string-three oscillators eliminated.  The second
term $ ({W'}_I^{(1)} + {W'}_I^{(2)})$ vanishes on the reflector
$\bsub{12}\bra{R}$ or on $\bsub{12}\bravac e^{E_{12}}$.
Noting the presence of the term linear in $p_{+3}=\pm k^i$
in $F_{123}$,
we see that the exponent $F'_{123}$ ($F_{123}$ with
the $(\hbox{zero-mode})^2$ term omitted) approaches
$$
\lim_{\epsilon \rightarrow 0}\bsub{123}\bravac\exp(F'_{123})\vac_3
\bigg\vert_{p_{+3}=\pm k^i}
= \bsub{12}\bravac\exp\bigg(E_{12} \pm \sqrt2 \alpha _2
\sum_{r=2,3}\sum_{n\geq 1}\bar N_n^r\,k^i\cdot \alpha _n^{(r)}\bigg) .
\ee $$
Since $\alpha _2\bar N_n^2 =(-)^n\alpha _2\bar N_n^3 = -1/n$
in this limit, we find
$$
\eqalign{
\bsub{123}\bra{V}\ket{\Lambda ^i_{\pm }}_3
&= {1\over 2}\ \delta (1,2,3)
\big\vert_{p_{+3}=\pm k^i} \bsub{12}\bravac e^{E_{12}}
e^{i\pi p_2w_2}
(c_0^{+(1)}+c_0^{+(2)}) \cr
& \qquad \times
\exp\bigg(\mp \sqrt2\sum_{n\geq 1}{1\over n}
(\alpha _n^{(1)}+(-)^n\alpha _n^{(2)})\cdot k^i\bigg)
e^{-i\pi p_1w_3} {\cal P}_{12} . \cr
}\ee
$$
We note that the equality
$$
\delta (p_{+1}+p_{+2}\pm k^i) =
\delta (p_{+1}+p_{+2})\,\exp(\pm 2ik^i\cdot x_{+1})
$$
holds since $[x_+^i, p_{+j}] = (i/2)\delta ^i_j$ (although the $\delta $
here is a Kronecker's delta). Using this and $w_3=\mp k^i$, we find
$$
\eqalign{
\bsub{123}\bra{V}\ket{\Lambda ^i_{\pm }}_3
&= {1\over 2}\bsub{12}\bra{R'}b_0^{-(1)}
e^{\pm 2ik^ix_{+1}} \cr
&\ \ \times
\exp\bigg(\mp \sqrt2\sum_{n\geq 1}{1\over n}
(\alpha _n^{(1)}+(-)^n\alpha _n^{(2)})\cdot k^i\bigg)
e^{\pm i\pi p_1k^i} {\cal P}_{12} . \cr}
\ee
$$
Owing to the connection condition \eqrefcon, this equals
$$
\eqalign{
&={1\over 2}\bsub{12}\bra{R}b_0^{-(1)}
{\cal P}^{(1)}:\exp\big(\pm 2ik^i\cdot X_+^{(1)}(\sigma =0)\big):
{\cal P}^{(1)}e^{\pm i\pi p_1k^i}  \cr
&={1\over 2}e^{\pm i\pi p_1k^i}\int {d\sigma \over 2\pi }
\bsub{12}\bra{R}b_0^{-(1)}
:\exp\big(\pm 2ik^i\cdot X_+^{(1)}(\sigma )\big): \ .  \cr}
\ee
$$
This gives rise to the desired result for $E_{\pm }^i$ in \genstra .
The calculation for the case of $E_3^i$ is much simpler and
can be carried out similarly.

\refout
\bye